\documentclass[onecolumn,english,journal]{IEEEtran}
\usepackage[latin9]{inputenc}
\usepackage{color}
\usepackage{float}
\usepackage{amsmath}
\usepackage{amsthm}
\usepackage{amssymb}
\usepackage{graphicx}
\usepackage{geometry}
\usepackage{xcolor}
\makeatletter

\providecommand{\tabularnewline}{\\}
\floatstyle{ruled}
\newfloat{algorithm}{tbp}{loa}
\providecommand{\algorithmname}{Algorithm}
\floatname{algorithm}{\protect\algorithmname}

\theoremstyle{plain}
\newtheorem{thm}{\protect\theoremname}
\theoremstyle{plain}
\newtheorem{cor}{\protect\corollaryname}
\theoremstyle{remark}
\newtheorem{rem}{\protect\remarkname}

\theoremstyle{plain}
\newtheorem{assum}{\protect\assumptionname}

\theoremstyle{plain}
\newtheorem{cond}{\protect\conditionname}

\@ifundefined{date}{}{\date{}}
\usepackage{algpseudocode}

\usepackage{caption}
\ifdefined\showcaptionsetup
 
\fi

\usepackage{babel}
\providecommand{\corollaryname}{Corollary}
\providecommand{\theoremname}{Theorem}
\providecommand{\assumptionname}{Assumption}
\providecommand{\conditionname}{Condition}

\ifdefined\showcaptionsetup
 
\fi

\ifdefined\showcaptionsetup
 \PassOptionsToPackage{caption=false}{subfig}
\fi
\usepackage{subfig}
\makeatother

\usepackage{babel}
\providecommand{\corollaryname}{Corollary}
\providecommand{\remarkname}{Remark}
\providecommand{\theoremname}{Theorem}

\begin{document}
\title{On the Distribution of Age of Information in Time-varying Updating
Systems}
\author{Jin~Xu, Weiqi~Wang, and~Natarajan~Gautam,~\IEEEmembership{Senior Member,~IEEE}
\thanks{Jin Xu is with School of Management, Huazhong University of Science
and Technology, Wuhan 430074, China.} \thanks{Weiqi Wang is with Department of Physics and Astronomy, University
of Victoria, Victoria, BC V8P 5C2, Canada. \emph{Corresponding author}:
Weiqi Wang. Email: weiqiwang93@gmail.com} \thanks{Natarajan Gautam is with Department of Electrical Engineering and Computer Science,
Syracuse University, Syracuse, NY 13244, U.S.}}

\maketitle
\begin{abstract}
Age of Information (AoI) is a crucial metric for quantifying information
freshness in real-time decision-making systems. In many practical
scenarios, the sampling rate of data packets is time-varying due to
factors such as energy constraints or the use of event-driven sampling
strategies. Evaluating AoI under such conditions is particularly challenging,
as system states become temporally correlated, and traditional analytical
frameworks that rely on stationary assumptions and steady-state analysis
are no longer applicable. In this study, we investigate an $M_{t}/G/1/1$
queueing system with a time-varying sampling rate and probabilistic
preemption. We propose a novel analytical framework based on multi-dimensional
partial differential equations (PDEs) to capture the time evolution
of the system's status distribution. To solve the PDEs, we develop
a decomposition technique that breaks the original high-dimensional
PDE into a set of lower-dimensional subsystems representing the marginal
distributions of the original system. Solving these subsystems allows
us to derive the AoI distribution at arbitrary time instances. Interestingly,
we show that despite AoI being defined as the time elapsed since the
last informative update, it does not exhibit a memoryless property,
even when processing times are negligible, due to its inherent dependence
on the historical sampling process. Furthermore, we demonstrate that
our framework extends naturally to the stationary setting, where we
derive a closed-form expression for the Laplace--Stieltjes Transform
(LST) of the AoI in the steady state. Our numerical experiments further
reveal that the AoI in the time-varying setting exhibits a non-trivial
lag in response to changes in the sampling rate, driven by its correlation
with past sampling rates and processing times. Our results also indicate
that in both time-varying and stationary scenarios, no single preemption
probability or processing time distribution can minimize AoI violation
probability across all violation thresholds. Finally, we formulate
an optimization problem for system parameters in the time-varying
setting and propose a heuristic method to identify sampling and preemption rates
that reduce sampling costs while satisfying AoI constraints across
different time periods. 
\end{abstract}


\section{Introduction}

In real-time decision-making scenarios such as autonomous driving
\cite{kaul2012real,xu2022aoi,shi2024enhancing} and smart manufacturing
systems \cite{xu2021age,wang2025meta,huang2023aoi,li2026condition}, the freshness
of information is paramount. These systems rely on the timely delivery
of sampled data to ensure responsiveness, accuracy, and safety in
rapidly changing environments. However, data collected from the physical
environment such as high-resolution images or audio recordings that
capture the current state of the system (e.g., road conditions, robotic
motion, or manufacturing processes), require non-negligible transmission
and processing times. These delays can make information stale by the
time it reaches decision-makers, leading to inaccurate inferences
and poor decision quality. As a result, it is essential to both quantify
information freshness and develop strategies to improve it, thereby
maintaining high-quality decision-making performance.

AoI is a fundamental metric that quantifies the timeliness of updates
received by decision-makers or system operators. It is defined as
the time elapsed since the most recent reception of fresh information
about the system state. As such, AoI is jointly influenced by the
sampling, transmission, and processing of data packets. A critical
characteristic of real-world sensor networks is that the sampling
rate of data packets is often non-stationary, changing over time due
to varying system constraints and environmental conditions. This \emph{time-varying}
nature of sampling is prominent in various scenarios. One example
is energy-harvesting systems, where the energy collected from ambient
sources (e.g., solar) fluctuates with environmental dynamics. In such
settings, sensors must adapt their sampling rates to align with the
available energy supply \cite{seuaciuc2010energy,wang2021energy}.
Another is the event-triggered sampling scenario, where data is sampled
only upon the occurrence of significant events. Event-triggered sampling
has been widely adopted in sensor networks as an efficient alternative
to periodic sampling \cite{cassandras2014event,zhang2014event,florescu2021event},
since it can notably extend the lifespan of wireless sensor networks
by reducing redundant measurements \cite{malawade2022ecofusion,algabroun2025parametric}
and can also alleviate the communication load in bandwidth-constrained
environments by transmitting only essential data \cite{zhang2019networked,zhang2023sampled}.
Since the monitored physical processes are often inherently non-stationary,
the sampling rate under event-triggered strategies becomes time-varying
by design. Examples of time-varying sampling rates can be found in \cite{jury1959analysis}, and examples of event-triggered sampling are discussed in \cite{li2018stabilization}. In this work, we consider a general time-varying sampling rate framework that can accommodate various types of sampling strategies. It is crucial to understand and evaluate AoI
in systems with time-varying sampling rates, as it enables system
operators to monitor system status, maintain reliability, and support
timely and effective decision-making.

In this research, we aim to characterize the AoI distribution in systems
with time-varying sampling rates. Unlike most existing studies that
primarily focus on the mean AoI \cite{kaul2012real,costa2016age},
we pursue a more general and informative characterization by analyzing
the entire AoI distribution. Understanding the full distribution is
often more valuable than considering only the mean, especially in
time-sensitive and reliability-critical applications. First, the AoI
distribution captures the variability and tail behavior of AoI, which
is crucial in real-time decision-making systems \cite{ayan2020probability,zhong2023stochastic,wang2025meta}.
A high variance or heavy tail in the AoI distribution may indicate
frequent occurrences of extreme AoI values, severely impacting the
timeliness and reliability of decisions. Second, a detailed understanding
of the distribution enables the evaluation of the AoI violation probability,
i.e., the probability that AoI exceeds a predefined threshold. This
metric directly quantifies the risk of outdated information and is
a critical indicator of system performance and reliability \cite{zhou2020risk,hu2021status}.
As we will demonstrate in our analysis, the AoI violation probability
becomes particularly relevant in non-stationary systems, where requirements
for AoI may vary over time. In practical applications like autonomous
driving or industrial automation, ensuring that AoI remains below
specific thresholds with high probability across different time intervals
is essential for maintaining safety and production quality.

However, characterizing the AoI distribution in systems with time-varying
sampling rates poses significant challenges. The temporal variability
in the sampling process introduces non-Markovian and non-stationary
dynamics, and the system state at any given time depends not only
on the current input but also on the entire history of system processes.
In such settings, a steady-state distribution may not exist, which
precludes the use of classical analytical tools developed for stationary
systems such as continuous-time Markov chain (CTMC) analysis \cite{costa2016age},
renewal-type analysis \cite{xu2021age,inoue2019general,champati2019distribution},
and Stochastic Hybrid Systems (SHS) analysis \cite{yates2020age},
which rely on the existence of steady-state distributions. The results
for the stationary scenarios cannot be directly applied to the time-varying
scenarios. Most existing studies on queueing systems with time-varying
parameters use fluid and diffusion approximations to analyze metrics
like waiting time and queue length (e.g., \cite{ko2013critically,yom2014erlang,whitt2018time}).
While these models offer useful insights under high-traffic or many-server
regimes, they only approximate the mean behavior and cannot capture
the exact probability distributions of performance metrics at specific
time instances. Moreover, in systems that require dropping out redundant
packets in the queue to improve information freshness (e.g., single
buffer systems \cite{costa2016age}, small buffer systems \cite{kesidis2023age},
and preemptive systems \cite{moltafet2025aoi}), these approximation
techniques are insufficient for accurately assessing the AoI distribution.

An additional challenge in analyzing the AoI distribution in time-varying
systems stems from the intrinsic definition of AoI, which emphasizes
information freshness from the receiver's perspective rather than
traditional performance metrics like queue length or throughput. As
a result, evaluating AoI from the receiver's perspective requires
tracking a more complex set of system variables, including: 1) the
generation time of the last successfully delivered informative packet;
2) the age of the packet currently being processed at the receiver
(if any); and 3) the current queueing status (e.g., processing progress
or remaining workload). The interdependence of these factors significantly
increases the analytical complexity of deriving the AoI distribution
in systems with time-varying characteristics.

As a starting point to tackle the AoI in time-varying systems, we
consider a relatively basic yet representative $M_{t}/G/1/1$ system
with probabilistic preemption. In this system, packet arrivals follow
a non-stationary Poisson process with a time-varying rate. Upon arrival,
a new packet may preempt the one currently in service with a given
probability. The system can have at most one packet at a time, allowing
it to discard redundant packets and thereby improve information freshness.
The probabilistic preemption mechanism offers greater operational
flexibility. Both non-preemptive and fully preemptive systems can
be seen as special cases of this model. We propose a novel analytical
framework based on multi-dimensional PDEs to capture the joint evolution
of the system status over time. Our approach is originally motivated
by the differential equation methods used to analyze the virtual workload
distribution in stationary queueing systems (e.g., \cite{brill1977level,cox1986virtual,gautam2024analysis}).
However, we significantly extend this framework in two key ways: 1)
We employ multiple state dimensions to model the AoI process, the
age of the in-service packet, and the remaining workload, all of which
are essential for capturing information freshness of the information
receiver. 2) We incorporate an explicit temporal dimension to describe
how these system states evolve under time-varying dynamics. These
extensions result in a high-dimensional, time-dependent PDE system
that is inherently difficult to solve due to the interdependencies
and temporal correlations among the involved processes. To overcome
this complexity, we develop a decomposition technique that enables
us to obtain an exact solution to the AoI distribution at any arbitrary
time. Our main contributions can be summarized as follows: 
\begin{itemize}
\item \textbf{(PDE modeling framework and exact solution for time-varying
system):} We develop a novel analytical framework using multi-dimensional
PDEs to characterize the evolution of system state distributions in
an $M_{t}/G/1/1$ queue with probabilistic preemption. To manage the
complexity of the high-dimensional PDE, we decompose it into a set
of lower-dimensional subsystems that describe the marginal distributions
of the original system. Each subsystem is solved using the coordinate
method associated with the initial and boundary conditions, allowing
us to derive the exact solution to the AoI distribution at any arbitrary
time. We also design efficient numerical approaches for practical
computation of exact solutions. Furthermore, our exact solution reveals
that even when the processing time is negligible, the mean AoI does
not exhibit memoryless behavior, highlighting the intricate temporal
dependencies introduced by the time-varying Poisson arrival process.
\item \textbf{(Closed-form expression for the stationary system):} We extend
our framework to the stationary case with constant sampling rates.
By applying our PDE approach, we derive the LST of the AoI in steady
state. For systems with non-zero preemption probabilities, the AoI
distribution can be computed numerically via inverse LST. For the
$M/G/1/1$ system without preemption, our framework directly yields
a closed-form expression for the AoI distribution. We also derive the closed-form expression for the AoI distribution in the $M/M/1/1$ system with probabilistic preemption. This analytical
framework also has the potential to be extended for AoI analysis in
other queueing systems.
\item \textbf{(Insights into AoI distribution):} Our numerical experiments
reveal that under time-varying sampling, the AoI distribution responds
to changes in the sampling rate with a non-trivial delay, due to historical
dependencies and processing latency. Our results also show that the
AoI violation probability with a mid-level threshold is most sensitive
to the change of sampling rate. We find that a fixed preemption probability
does not consistently reduce the AoI violation probability: it may
be beneficial during certain time periods but detrimental during others
when the sampling rate shifts. In both the time-varying and stationary
settings, we discover that the AoI violation threshold plays a crucial
role: certain distributions and preemption probabilities reduce violation
probability under tight thresholds, but may be less effective when
the constraints are relaxed.
\item \textbf{(Optimal system design under AoI violation constraints):}
We address the parameter optimization problem in time-varying systems,
aiming to determine the sampling rate and preemption probability that
minimize sampling cost while ensuring AoI violation probabilities
are met across different time periods. We propose a heuristic method
that uses our stationary results to initialize the solution in each
time slot efficiently, and adjusts the sampling and preemption rates between time
slots when violation requirements change. Numerical results confirm
that our method achieves a low sampling cost while satisfying the
time-varying AoI constraints. 
\end{itemize}
The rest of this paper is organized as follows. Section \ref{sec:Related-work}
first reviews the related literature. We then present the system model
and the PDE-based modeling framework in Section \ref{sec:System-model-and},
followed by the analytical results and solution methodology in Section
\ref{sec:Solution-derivation-approach}. In Section \ref{sec:Discussion-on-the},
we adapt our framework to the stationary scenario. Section \ref{sec:Numerical-studies}
provides numerical studies and discusses the system design problem.
Finally, Section \ref{sec:Concluding-remarks-and} concludes the paper
and discusses potential directions for future research.

\section{Related work}

\label{sec:Related-work}

As our research focuses on AoI analysis within time-varying queueing
systems, in this section, we review literature pertinent to information
freshness in queueing systems, the distribution of AoI, and the analysis
of time-varying systems.

\subsection{Information freshness in queueing systems}

Since the processes by which data packets are transmitted and processed
in communication and computer networks can be modeled as queueing
systems, most existing studies rely on queueing theory to analyze
and derive the mean AoI in various system configurations. For instance,
Kaul \emph{et al.} \cite{kaul2012real} derived the mean AoI in $M/M/1$,
$M/D/1$, and $D/M/1$ systems. Costa \emph{et al.} \cite{costa2016age}
further investigated mean AoI in single buffer systems including $M/M/1/1$,
$M/M/1/2$, and $M/M/1/2^{*}$ configurations. Najm and Nasser \cite{najm2016age}
considered the mean AoI in systems with the Last-Come-First-Serve
(LCFS) scheme and gamma-distributed processing times. Zou \emph{et
al}. \cite{zou2019waiting} further studied the mean AoI in $M/G/1/1$
and $M/G/1/2^{*}$ systems with waiting procedures. Soysal and Ulukus
\cite{soysal2021age} provided exact expressions for mean AoI in $G/G/1/1$
systems, both with and without preemption.

There are also studies that consider the mean AoI in systems with
multiple data streams. For example, Najm and Telatar \cite{najm2018status}
calculated the mean AoI for a multi-class $M/G/1/1$ system with preemption.
Xu and Gautam \cite{xu2021peak} applied a Markov-renewal analysis
to derive the mean AoI peaks in $M/G/1$ multi-class queueing systems
with priority. Yates and Kaul \cite{yates2019age} applied an SHS
analysis to derive the closed-form expressions for the mean AoI in
multi-class $M/M/1$ systems under various service disciplines, including
First-Come-First-Serve (FCFS), LCFS with preemption in the server,
and LCFS with preemption allowed only in waiting. Chen \emph{et al.}
\cite{chen2022age} derived the mean AoI for multi-class $M/G/1/1$
system without preemption. More recently, Xu \emph{et al.} \cite{xu2023should}
used a renewal-type analysis to determine the mean AoI in an $M/G/1/1$
multi-stream system featuring a sleep-wake server, where the server
operates under various sleeping and wake-up schemes. Fahim \emph{et
al.} \cite{fahim2024analyzing} analyzed the mean AoI in multiclass
$M/G/1/2^*$ systems where packets being served or buffered can be
preempted, unless their recorded number of preemptions exceeds predetermined
thresholds.

These studies primarily focus on analyzing or optimizing the average
AoI under various stationary queueing system settings. However, they
generally do not explore the AoI distribution or the probability of
having extreme AoI values, which are critical for understanding system
reliability and performance in time-sensitive applications.

\subsection{Distribution of AoI}

Recognizing that mean AoI does not capture the tail of the AoI distribution
or AoI violation probability, some recent studies began to characterize
the distribution of AoI. Inoue \emph{et al.} \cite{inoue2019general}
provided the LST of AoI for $M/G/1$ systems under different service
disciplines, including FCFS, preemptive LCFS, and non-preemptive LCFS.
Champati \emph{et al.} \cite{champati2019distribution} used a sample
path analysis to characterize the AoI violation probability for $D/G/1/1$
and $M/G/1/1$ systems, and provided upper bounds for the violation
probability in $G/G/1/1$ and $G/G/1/2^*$ systems. Yates \cite{yates2020age}
adopted an SHS approach to investigate the moment generating function
of the AoI in a line network with preemptive memoryless servers. Kesidis
\emph{et al.} \cite{kesidis2023age} derived the Laplace Transform
of the AoI in small-buffer systems using the Palm and Markov-renewal
theory. Fiems \cite{fiems2023age} considered a two-source updating
system and derived the LST of AoI using an effective service time
approach. Dogan and Akar \cite{dogan2021multi} utilized a sample
path argument and the theory of Markov Fluid Queues (MFQ) to numerically
compute the AoI distributions in a multi-class $M/PH/1/1$ queueing
system with probabilistic preemption and packet errors. Moltafet \emph{et
al}. \cite{moltafet2022moment} provided the moment generating function
(MGF) for the AoI in multi-source $M/G/1/1$ systems with self-preemptive
policies. Inoue and Mandjes \cite{inoue2025characterizing} further
derived the LST of AoI in a system with a background stream, and provided
computational approaches to calculate the mean AoI in the multi-source
model with phase-type inter-arrival times. Moltafet \emph{et al}.
\cite{moltafet2025aoi} also derived the MGF of the $M/G/1/1$ system
with probabilistic preemption.

While these studies significantly advance our understanding
of AoI distributions in stationary systems, they generally do not
address scenarios with time-varying sampling rates, which are prevalent
in practical dynamic network environments. As we will demonstrate
later, our framework not only addresses the challenges of analyzing
AoI distributions in time-varying systems but also extends naturally
to stationary scenarios.

\subsection{Time-varying queueing networks}

The existing queueing frameworks for analyzing the time-varying queue
systems mainly rely on fluid and diffusion approximations. Kurtz \cite{kurtz1978strong}
first proved the convergence of a sequence of scaled Poisson processes,
which serve as a theoretical foundation for analyzing the queueing
systems with fluid and diffusion limit approximations. Mandelbaum
\emph{et al.} \cite{mandelbaum1999waiting} studied a non-stationary
Markov multi-server queueing model where waiting customers may abandon
and subsequently retry, and derived fluid and diffusion approximations
for the associated waiting time process. Massey \cite{massey2002analysis}
investigated the fluid and diffusion approximations of the $M_{t}/G/\infty$,
$M_{t}/G/L/L$, and $M_{t}/M_{t}/L_{t}$ models. Ko and Gautam \cite{ko2013critically}
proposed a computational framework that relies on adjusted fluid and
diffusion limits to better approximate the queue length process in
time-varying multi-server queues with abandonment and retrials, when
the number of servers in the system is not large. Whitt \cite{whitt2018time}
discussed the approximation schemes including pointwise-stationary
approximations, closure approximations, modified-offered-load approximations
to various types of time-varying systems, such as fluid models, time-varying
infinite-server queues, and time-varying single-server queues. Whitt
and You \cite{whitt2019time} proposed a time-varying robust-queueing
algorithm to compute the workload process in single server queues
with time-varying arrivals. Pender and Ko \cite{pender2017approximations}
developed a computational framework based on the phase-type distribution
and strong approximation theory to approximate the queue length distributions
of time-varying $G_{t}/G_{t}/n_{t}$ queues. However, while these
studies provide powerful tools for approximating queue length distributions
in time-varying systems, they do not address the analysis of AoI or
provide an exact analysis for queueing metrics in such dynamic settings.

To the best of our knowledge, there is currently no work that analyzes
the AoI distribution in time-varying systems. Our study advances the
current literature by proposing a novel framework that relies on multi-dimensional
PDEs to obtain the exact AoI distribution in systems with time-varying
sampling rates. Furthermore, we also show that this framework can
be applied to stationary systems to obtain the AoI distribution in
the steady state.

\section{System model and the PDE framework}

\label{sec:System-model-and}

This section presents the PDE-based framework for modeling the evolution
of system state probabilities. We first introduce the system model
in Section \ref{subsec:System-model}, and then formulate the multi-dimensional
PDEs that characterize the probabilistic behavior of the system state
in Section \ref{subsec:The-PDE-framework}.

\subsection{System model}

\label{subsec:System-model}

We consider an $M_{t}/G/1/1$ system with a single server where packets
arrive (i.e., sampled from the physical process) according to a non-homogeneous
Poisson process with a time-varying rate $\lambda(t)$. Each packet
requires a random processing time (e.g., for transmission or computation),
which follows a general distribution with a cumulative
distribution function (CDF) $F(z)$, with LST given by $\tilde{F}(s)=\int_{0}^{\infty}e^{-sz}dF(z)$.
The server can process at most one packet at a time. In this work,
we consider a time-varying probabilistic preemption scheme. Under
this scheme, an arriving packet immediately enters service if the
server is idle. If the server is busy at time $t$, the arriving packet preempts
the packet in service with probability $\theta(t)$, or is discarded
with probability $1-\theta(t)$. As a result, the effective arrival
rate (admission rate) is $\lambda(t)$ when the server is idle, and
it is $\lambda(t)\theta(t)$ when the server is busy. As we shall
show later, in certain scenarios, altering the preemption probability
can yield a smaller AoI violation probability than either non-preemption
or full preemption. This scheme allows the system operator to adjust
the preemption probability in response to the change of packet arrival rate
and AoI constraints. We use $N(t,t+\tau)$ to denote the number of
packets admitted into the system during the time interval $[t,t+\tau)$
(i.e., the total number of arrivals excluding discarded ones),
and refer to packets that successfully complete their service as informative
packets. Specifically, to facilitate our analysis, we make the following
assumption for the system parameters. 

{\begin{assum}\label{assum1}
We assume that the system parameters satisfy:
(1) Arrival rate $\lambda(t)>0$ and preemption probability $\theta(t)\in[0,1]$
are both piecewise continuous, with a finite number of discontinuities
over any finite interval, and are piecewise differentiable. (2) The
processing time of a data packet is a non-singular random variable
such that the CDF $F(z)$ satisfies (i) $F(0)=0$; (ii) has finitely many
jumps over any finite interval; and (iii) the derivative exists except for a finite number of points.
\end{assum}

In particular, Part~(2) of Assumption 1 allows a broad class of piecewise-smooth continuous, discrete, or mixed service distributions, while
 excluding singular
distributions (see Lebesgue decomposition theorem, see Page 143 of
\cite{rosenthal2006first}) and distributions with infinitely many irregularities on a finite interval. One can easily check that
common random variables such as shifted Poisson, Geometric (with positive support), Uniform, Gamma
all satisfy this condition. The condition $F(0)=0$ ensures that the processing
time distribution has no atom at zero. This assumption is consistent
with practical systems, where packets typically require a strictly
positive amount of time for processing or transmission. We will discuss the scenario of negligible processing time separately in Section \ref{subsec:Model-of-no}.
}

\begin{figure}
\centering\includegraphics[scale=0.33]{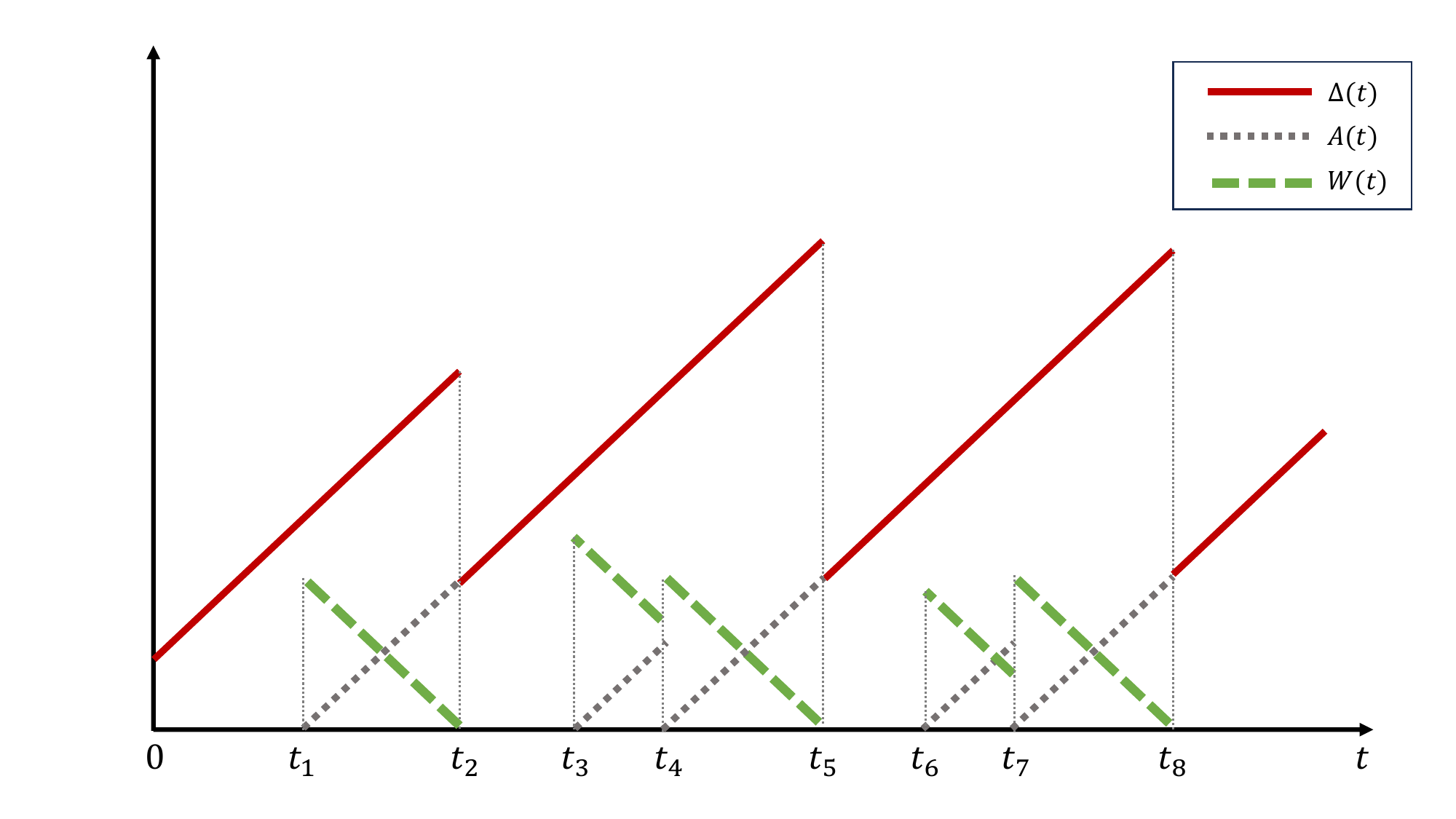}

\caption{A demonstrative figure for a sample path of $(\Delta(t),A(t),W(t))$.
Arrivals occur at times $\{t_{1},t_{3},t_{4},t_{6},t_{7}\}$, preemptions
occur at times $\{t_{4},t_{7}\}$, and packets complete processing
at times $\{t_{2},t_{5},t_{8}\}$.}\label{fig:A-demonstrative-figure}
\end{figure}

The AoI at time $t$ is defined as $\Delta(t)=t-U(t)$, where
$U(t)$ is the sampling time of the most recently processed informative
packet prior to time $t$. We introduce two auxiliary state variables:
$A(t)$, the age of the packet that is currently in process at time
$t$, and $W(t)$, its remaining workload (processing time). If the
system is empty, we set $A(t)=0$ and $W(t)=0$. The remaining workload
of a packet is revealed immediately upon its arrival at the server.
The system status at time $t$ is thus described by a three-dimensional
tuple $(\Delta(t),A(t),W(t))$. A demonstrative graph for a sample
path of $(\Delta(t),A(t),W(t))$ is shown in Fig. \ref{fig:A-demonstrative-figure},
where we can observe that the AoI $\Delta(t)$ will increase with
a constant rate until a packet has finished processing (e.g., at times
$\{t_{2},t_{5},t_{8}\}$). Once a packet has finished processing,
both the remaining workload $W(t)$ and packet age $A(t)$ become
zero, and the AoI $\Delta(t)$ drops to $A(t^{-})$ (the left limit
of $A(t)$). These auxiliary variables are essential because the timing and magnitude of the AoI drop depend on $W(t)$ and $A(t)$. Note that if the system is non-empty,
the packet age will increase and the workload will decrease, both
at a constant rate. This process will end when either the workload
becomes zero or a new arrival preempts the packet in service. For
instance, at time $t_{4}$, a new arrival occurs, resetting the remaining
workload to the processing time of the new packet and resetting the
packet age to zero. To characterize the evolution of system status,
we define the joint CDF of the system state at time $t$ as 
\begin{equation}
H(t,x,y,z)=\boldsymbol{P}(\Delta(t)\leq x,A(t)\leq y,W(t)\leq z),\quad x,y,z\geq0.\label{eq:0-1}
\end{equation}
From this definition, the CDF of the AoI at time $t$ can be directly
obtained as the marginal distribution: 
\begin{equation}
\Phi(t,x)\triangleq\boldsymbol{P}(\Delta(t)\leq x)=H(t,x,\infty,\infty).\label{eq:0-2}
\end{equation}
Our general idea for deriving the probability $H(t,x,y,z)$ is to
analyze its evolution over time $t$, as detailed in the following
subsection. We summarize the main notation used in this paper in Table
\ref{tab:notation_summary}.

\begin{table}[t!]
\centering \caption{Summary of main notations}
\begin{tabular}{cc}
\hline 
\textbf{Symbol}  & \textbf{Description} \tabularnewline
\hline 
$\Delta(t)$  & AoI at time $t$ \tabularnewline
$A(t)$  & Age of the packet in service at time~$t$ (equals~$0$ if the server
is idle) \tabularnewline
$W(t)$  & Remaining processing time of the packet at time~$t$ (equals~$0$
if the server is idle) \tabularnewline
$\lambda(t)$  & Time-varying Poisson arrival rate of packets at time~$t$ \tabularnewline
$\theta(t)$  & Preemption probability at time $t$\tabularnewline
$F(z)$  & CDF of packet processing time \tabularnewline
$dF(z)$  & The Lebesgue-Stieltjes measure on $z$ \tabularnewline
$\tilde{F}(s)$  & LST of processing time: ${\displaystyle \tilde{F}(s)=\int_{0}^{\infty}e^{-sz}\mathrm{d}F(z)}$\tabularnewline
$N(t,t+\tau)$  & Number of packets admitted into the system during interval $[t,t+\tau)$ \tabularnewline
\hline 
\textbf{CDF}  & \textbf{Description} \tabularnewline
\hline 
$H(t,x,y,z)$  & Joint CDF $\boldsymbol{P}(\Delta(t)\leq x,\,A(t)\leq y,\,W(t)\leq z)$ \tabularnewline
$D(t,x,z)$  & $H(t,x,\infty,z)$: Joint distribution of AoI and workload \tabularnewline
$G(t,y,z)$  & $H(t,\infty,y,z)$: Joint distribution of packet age and workload \tabularnewline
$M(t,x)$  & $H(t,x,0,0)$: Probability that $\Delta(t)\leq x$ and the system
is empty at time $t$ \tabularnewline
$B(t,z)$  & $H(t,\infty,\infty,z)$: Distribution of workload \tabularnewline
$\Phi(t,x)$  & $H(t,x,\infty,\infty)$: Distribution of AoI \tabularnewline
\hline 
\end{tabular}\label{tab:notation_summary} 
\end{table}

\subsection{The PDE framework}

\label{subsec:The-PDE-framework}

In this subsection, we introduce the PDE framework that describes
the evolution of the joint cumulative distribution function
$H(t,x,y,z)$ over time. The system dynamics differ depending on whether
the server is idle or busy. An idle state can arise either from the
system remaining idle from the previous time step, or from a busy
state where the remaining workload is small enough to transition to
idle. A busy state, on the other hand, can result from either the
system already being busy, or a transition from idle to busy due to
the arrival of a new packet. This distinction motivates us to construct
two sets of PDEs to capture the system dynamics: one for the empty
(idle) system and another for the non-empty (busy) system. These PDEs
describe the temporal evolution of the system state probabilities
under each condition.

\subsubsection{PDE for the idle system state}

We first consider the conditional probability that the system is idle at time $t+h$, i.e.,
\begin{equation}
\boldsymbol{P}\Big(\Delta(t+h)\leq x,A(t+h)=0,W(t+h)=0|\Delta(t)=a,A(t)=b,W(t)=c\Big),
\end{equation}
where $h>0$ is an infinitesimal time step. We consider two scenarios
that happened during time $[t,t+h]$ as follows: 
\begin{itemize}
\item If the system was empty at time $t$ (i.e., $W(t)=0$), two
types of events can occur during this time period: 1) No arrival occurs
with probability $1-\lambda(t)h+\boldsymbol{o}(h)$ (see Section 5.5
of \cite{ross2019introduction}, where $\boldsymbol{o}(h)$ is
a function such that $\lim_{h\to0^+}\boldsymbol{o}(h)/h=0$). In this case, we need to have $\Delta(t)\leq x-h$. 2)
One or more arrivals occur and complete service before $t+h$.
The probability
that one arrival occurs and completes processing is bounded by $\big(\lambda(t)h+\boldsymbol{o}(h)\big)F(h)=\boldsymbol{o}(h)$
given that $\lim_{h\to0^+}F(h)=F(0)=0$. The probability of having more
than one arrival during $[t,t+h]$ is also given as $\boldsymbol{o}(h)$
(also see Section 5.5 of \cite{ross2019introduction}). Therefore,
the probability that more than one arrival occurs and completes processing
is negligible, i.e., $\boldsymbol{o}(h)$. 
\item If the system was busy at time $t$ (i.e., $W(t)>0$), there
are also two types of events that can occur during this time period: 1)
No new packet is admitted into the system (i.e., $N(t,t+h)=0$).  This occurs if either no arrival happens (with probability $1-\lambda(t)h+\boldsymbol{o}(h)$) or an arrival occurs but does not preempt the
packet in service (with probability $\big(\lambda(t)h+\boldsymbol{o}(h)\big)(1-\theta(t))$).
Therefore, we have $\boldsymbol{P}(N(t,t+h)=0)=1-\lambda(t)\theta(t)h+\boldsymbol{o}(h)$.
Note that this event requires that the remaining workload satisfies
$0<W(t)<h$ to guarantee that $W(t+h)=0$, and also requires that
the packet age $A(t)\leq x-h$ to ensure $\Delta(t+h)\leq x$.
2) At least one arrival is admitted into the system and the service
completes before $t+h$ (i.e., $N(t,t+h)\geq1$). This event requires that at least one packet
is admitted with processing time shorter than $h$, with probability
bounded by $\big(\lambda(t)h+\boldsymbol{o}(h)\big)\theta(t)F(h)=\boldsymbol{o}(h)$. 
\end{itemize}
To summarize the discussion above, we can write the conditional
probability for the system to be idle at $t+h$ with an AoI no greater
than $x$ as: 
\begin{eqnarray}
 &  & \boldsymbol{P}\Big(\Delta(t+h)\leq x,A(t+h)=0,W(t+h)=0|\Delta(t)=a,A(t)=b,W(t)=c\Big)\nonumber \\
 & = & \boldsymbol{P}(N(t,t+h)=0|\Delta(t)=a,A(t)=b,W(t)=c)\Big(1_{\{a\leq x-h,b=0,c=0\}}+1_{\{b\leq x-h,0<c<h\}}\Big)+\boldsymbol{o}(h).\label{eq:1-1}
\end{eqnarray}
The indicator functions in Equation (\ref{eq:1-1}) capture the conditions:
If the system was idle during $(t,t+h)$ (first term), the AoI evolves
from $a$ to $a+h$ and the packet age $A(t+h)$ and workload $W(t+h)$
remain zero; If the system was busy but became idle (second term),
which requires the remaining processing time to satisfy $0<c<h$.
The AoI then resets, and the age of the completed packet $b$ must
satisfy $b\leq x-h$.

By integrating Equation \eqref{eq:1-1} over all possible states at
time $t$ and using first-order approximations for the probabilities
of no arrivals from our discussion above, we obtain the evolution of the idle state CDF: 
\begin{eqnarray}
 &  & H(t+h,x,0,0)=\boldsymbol{P}\Big(\Delta(t+h)\leq x,A(t+h)=0,W(t+h)=0\Big)\nonumber \\
 & = & \int_{a=0}^{\infty}\int_{b=0}^{\infty}\int_{c=0}^{\infty}\boldsymbol{P}\Big(\Delta(t+h)\leq x,A(t+h)=0,W(t+h)=0\nonumber \\
 &  & \Big|\Delta(t)=a,A(t)=b,W(t)=c\Big)dH(t,a,b,c)+\boldsymbol{o}(h)\nonumber \\
 & = & \underbrace{\Big(1-\lambda(t)h\Big)H(t,x-h,0,0)}_{\text{Was idle at time }t}+\underbrace{\Big(1-\lambda(t)\theta(t)h\Big)\Big(H(t,\infty,x-h,h)-H(t,\infty,x-h,0)\Big)}_{\text{Became idle between }t\text{ and }t+h}+\boldsymbol{o}(h),\label{eq:1-2}
\end{eqnarray}
where $dH(t,a,b,c)$ is the Lebesgue--Stieltjes
measure of the system state distribution at time $t$. In the following,
we construct the PDE system by assuming that the derivatives of $H(t,x,y,z)$
exist. This will help us construct the analytical solutions, and we
will show later that under certain assumptions, the derivatives exist
except for a finite number of points. Using Equation \eqref{eq:1-2}, we take the limit as $h\to0^{+}$ and give the
differential equation for the idle-state probability: 
\begin{eqnarray}
 &  & H_{t}(t,x,0,0)=\lim_{h\to0^{+}}\frac{1}{h}\Big(H(t+h,x,0,0)-H(t,x,0,0)\Big)\nonumber \\
 & = & \lim_{h\to0^{+}}\frac{1}{h}\Big(H(t,x-h,0,0)+H(t,\infty,x-h,h)-H(t,\infty,x-h,0)-H(t,x,0,0)\Big)-\lambda(t)H(t,x,0,0)\nonumber \\
 & = & -\lambda(t)H(t,x,0,0)+H_{z}(t,\infty,x,0)-H_{x}(t,x,0,0).\label{eq:1-3}
\end{eqnarray}
This PDE captures the rate of change of the probability that the system
is idle with an AoI less than or equal to $x$.

\subsubsection{PDE for the busy system state}

We next consider the scenario where the system is busy at time $t+h$,
i.e., $W(t+h)>0$. The conditional probability for this state can
be formulated as: 
\begin{eqnarray}
 &  & \boldsymbol{P}(\Delta(t+h)\leq x,A(t+h)\leq y,0<W(t+h)\leq z|\Delta(t)=a,A(t)=b,W(t)=c)\nonumber \\
 & = & \boldsymbol{P}(N(t,t+h)=1|\Delta(t)=a,A(t)=b,W(t)=c)F(z+h)1_{\{a\leq x-h,b=0,c=0\}}\nonumber \\
 &  & +\boldsymbol{P}(N(t,t+h)=1|\Delta(t)=a,A(t)=b,W(t)=c)F(z+h)1_{\{a\leq x-h,c>0\}}\nonumber \\
 &  & +\boldsymbol{P}(N(t,t+h)=0|\Delta(t)=a,A(t)=b,W(t)=c)1_{\{a\leq x-h,b\leq y-h,h<c\leq z+h\}}+\boldsymbol{o}(h).\label{eq:1-4}
\end{eqnarray}
Similar to our analysis for the empty system, we use the term $\boldsymbol{o}(h)$
to characterize the events that can happen during time $[t,t+h]$
but have a probability with $\lim_{h\to0^+}\boldsymbol{o}(h)/h=0$.
The remaining are three mutually exclusive scenarios at time $t$,
which correspond to the three main terms on the right-hand side of
Equation \eqref{eq:1-4}: 
\begin{itemize}
\item The system was idle at time $t$, and one new packet arrived during
$[t,t+h]$; 
\item The system was busy at time $t$, and a newly arriving packet preempted
the one in service; 
\item The system was busy at time $t$, no new packet was admitted, and
the packet in service continued its processing. 
\end{itemize}
It is important to distinguish between the first two scenarios, as
the admission rate of packets differs depending on whether the server
was idle or busy at the time of arrival. If the system is empty, a new
packet will be admitted once it arrives. If the system is busy, the
new packet will be admitted with probability $\theta(t)$. 

We next construct the PDE in this scenario. We first
notice either $A(t)=0$ or $W(t)=0$ implies an empty system. Since
$A(t)$ is defined as the packet age, when $A(t)=0,$ it means that
no packet is in service, which can be observed from Fig. \ref{fig:A-demonstrative-figure}.
It thus follows that $H(t,x,y,0)=H(t,x,0,z)=H(t,x,0,0)$ with $y,z\geq0$
as either $W(t)=0$ or $A(t)=0$ implies the other is also zero.
Using this property, we can marginalize the conditional probability
in Equation \eqref{eq:1-4} to derive the following: 
\begin{eqnarray}
 &  & H(t+h,x,y,z)-H(t+h,x,0,0)\nonumber \\
 & = & \int_{b=0}^{\infty}\int_{a=b}^{\infty}\int_{c=0}^{\infty}\boldsymbol{P}\Big(\Delta(t+h)\leq x,A(t+h)\leq y,\nonumber \\
 &  & 0<W(t+h)\leq z\Big|\Delta(t)=a,A(t)=b,W(t)=c\Big)dH(t,a,b,c)\nonumber \\
 & = & \lambda(t)hH(t,x-h,0,0)F(z+h)+\lambda(t)\theta(t)h\Big(H(t,x-h,\infty,\infty)-H(t,x-h,0,0)\Big)F(z+h)\nonumber \\
 &  & +\Big(1-\lambda(t)\theta(t)h\Big)\Big(H(t,x-h,y-h,z+h)1_{\{y>h\}}-H(t,x-h,y-h,h)1_{\{y>h\}}\Big)+\boldsymbol{o}(h).\label{eq:1-5}
\end{eqnarray}
Here, $H(t,x-h,\infty,\infty)-H(t,x-h,0,0)$ represents the probability
that the system is busy with AoI at most $x-h$ at time $t$. By taking
the limit as $h\to0^{+}$ and assuming that the derivatives exist, we
obtain the differential equation for busy state probability as follows:
\begin{eqnarray}
 &  & H_{t}(t,x,y,z)-H_{t}(t,x,0,0)\nonumber \\
 & = & \lim_{h\to0^{+}}\frac{1}{h}\bigg\{\Big(H(t+h,x,y,z)-H(t,x,y,z)\Big)-\Big(H(t+h,x,y,0)-H(t,x,y,0)\Big)\bigg\}\nonumber \\
 & = & \lambda(t)\theta(t)H(t,x,\infty,\infty)F(z)+\lambda(t)(1-\theta(t))H(t,x-h,0,0)F(z)\nonumber \\
 &  & -\lambda(t)\theta(t)H(t,x,y,z)+\lambda(t)\theta(t)H(t,x,y,0)\nonumber \\
 &  & +\lim_{h\to0^{+}}\frac{1}{h}\bigg\{ H(t,x-h,y-h,z+h)-H(t,x-h,y-h,h)-H(t,x,y,z)+H(t,x,y,0)\bigg\}\nonumber \\
 & = & \lambda(t)\theta(t)H(t,x,\infty,\infty)F(z)+\lambda(t)(1-\theta(t))H(t,x,0,0)F(z)-\lambda(t)\theta(t)H(t,x,y,z)\nonumber \\
 &  & +\lambda(t)\theta(t)H(t,x,y,0)-H_{x}(t,x,y,z)-H_{y}(t,x,y,z)  +H_{z}(t,x,y,z) \nonumber \\
 & &+H_{x}(t,x,y,0)-H_{z}(t,x,y,0).\label{eq:1-6}
\end{eqnarray}

We can now summarize Equations (\ref{eq:1-3}) and (\ref{eq:1-6})
in the following, which we refer to as the \emph{time-varying} system
to be distinguished from the stationary system that we will further
discuss in Section \ref{sec:Discussion-on-the}. 
\begin{itemize}
\item \textbf{Time-varying System: } 
\end{itemize}
\begin{align}
\mbox{\emph{Empty state:} } & H_{t}(t,x,0,0)+\lambda(t)H(t,x,0,0)-H_{z}(t,\infty,x,0)+H_{x}(t,x,0,0)=0,\label{eq:1-7}\\
\mbox{\emph{Busy state:} } & -H_{t}(t,x,y,z)+H_{t}(t,x,0,0)+\lambda(t)\theta(t)H(t,x,\infty,\infty)F(z)\nonumber \\
 & +\lambda(t)\big(1-\theta(t)\big)H(t,x,0,0)F(z)-\lambda(t)\theta(t)H(t,x,y,z)+\lambda(t)\theta(t)H(t,x,y,0)\nonumber \\
 & -H_{x}(t,x,y,z)-H_{y}(t,x,y,z)+H_{z}(t,x,y,z)+H_{x}(t,x,y,0)-H_{z}(t,x,y,0)=0,\label{eq:1-8}\\
\mbox{\emph{Boundary condition: }} & H(t,0,y,z)=0\mbox{ for }y,z\geq0.\label{eq:1-9}
\end{align}
The system of first-order PDEs in Equations (\ref{eq:1-7})-(\ref{eq:1-9})
governs the system evolution for $t>0$ and in the domain $x,y,z\geq0$.
Equation (\ref{eq:1-7}) characterizes the idle state, Equation (\ref{eq:1-8})
captures the busy state, and Equation (\ref{eq:1-9}) serves as a
necessary boundary condition. The boundary condition comes from the
fact that the AoI is always non-negative. This PDE framework, when
supplemented with proper initial conditions that specify the system's
initial state at $t=0$, forms a complete and well-posed problem.
We make the following assumption for the initial condition to make
it valid alongside the boundary condition. 

\begin{assum}\label{assum2}
The initial
condition $H(0,x,y,z)=\boldsymbol{P}(\Delta(0)\leq x,A(0)\leq y,W(0)\leq z)$
satisfies (1) Boundary condition $H(0,0,y,z)=0$ for $y,z\geq0$;
(2) $\boldsymbol{P}(A(0)\geq\Delta(0))=0$; (3) $H(0,x,y,0)=H(0,x,0,z)=H(0,x,0,0)$; (4) $H_{x,y,z}(0,x,y,z)$ is piecewise continuous with a finite
number of jumps over any finite interval, and the busy-state component
$H(0,x,y,z)-H(0,x,0,0)$ is absolutely continuous in $(x,y,z)$ for
$x,y,z>0$; and (5) $H(0,x,0,0)$ has finitely many jumps over any
finite interval and is differentiable except for a finite number of points.
\end{assum}
We make Assumption \ref{assum2} for the following reasons. Part (1) ensures consistency with the boundary condition, namely that $H(t,0,y,z)=0$ for all $t\geq0$. Part (2) guarantees that the age of a packet in service cannot exceed the AoI of the system. Part (3) ensures that either 
$A(0)=0$ or $W(0)=0$ implies an empty system. Finally, Part (4) ensures the existence of the solution for the AoI distribution, as we will demonstrate later. Part (5) provides the required regularity of the initial AoI distribution
when the system is empty, which is not characterized by the mixed derivative
in Part (4).

\section{Solution and analysis}

\label{sec:Solution-derivation-approach}

In Section \ref{subsec:Exact-solution-and}, we first present Theorem
\ref{thm:When-assuming-,}, which summarizes the exact AoI distribution
in the time-varying system defined by Equations (\ref{eq:1-7})-(\ref{eq:1-9}).
We then discuss the computational framework required to obtain the
exact solution in Section \ref{subsec:computation_frame}. In Section
\ref{subsec:Model-of-no}, we discuss a special case where the packet
processing time is negligible.

\subsection{Exact solution}

\label{subsec:Exact-solution-and}

The AoI distribution $\Phi(t,x)$ in the time-varying system can be obtained by
solving Equations (\ref{eq:1-7})-(\ref{eq:1-9}).
Specifically, we let $M(t,x)\triangleq H(t,x,0,0)$ to characterize
the AoI distribution when the system is empty, and $G(t,y,z)\triangleq H(t,\infty,y,z)$
to capture the joint distribution of packet age $A(t)$ and remaining
workload $W(t)$. Notably, $M(t,\infty)$ captures the probability
that the system is empty at time $t$. Let $(t-x)_+=\max\{0,t-x\}$. We summarize the exact solution
to the AoI distribution in the following theorem. 
\begin{thm}
\label{thm:When-assuming-,}When the system parameters
satisfy Assumption \ref{assum1} and the initial condition satisfies Assumption \ref{assum2},
the AoI distribution in the time-varying system is given by 
\begin{eqnarray}
\Phi(t,x) & = & M(t,x)-e^{-\int_{0}^{t}\lambda(u)\theta(u)\mathrm{d}u}\Big(-H(0,x-t,\infty,\infty)+H(0,x-t,\infty,t)\Big)1_{\{x\geq t\}}\nonumber \\
 &  & -\int_{r=(t-x)_+}^{t}\Big(\lambda(r)\theta(r)\Phi(r,x-t+r)\nonumber \\
 &  & +\lambda(r)\big(1-\theta(r)\big)M(r,x-t+r)\Big)\Big(F(t-r)-1\Big)e^{-\int_{r}^{t}\lambda(u)\theta(u)\mathrm{d}u}\mathrm{d}r,\label{eq:2-4-18}
\end{eqnarray}
where 
\begin{eqnarray}
M(t,x) & = & e^{-\int_{0}^{t}\lambda(u)\mathrm{d}u}H(0,x-t,0,0)1_{\{x\geq t\}}+\int_{r=(t-x)_+}^{t}G_{z}(r,x-t+r,0)e^{-\int_{r}^{t}\lambda(u)\mathrm{d}u}\mathrm{d}r,\label{eq:2-4-19}
\end{eqnarray}
\begin{eqnarray}
G_{z}(t,y,0) & = & \int_{r=(t-y)_+}^{t}\Big(\lambda(r)\theta(r)+\lambda(r)\big(1-\theta(r)\big)M(r,\infty)\Big)e^{-\int_{r}^{t}\lambda(u)\theta(u)\mathrm{d}u}\mathrm{d}F(t-r)\nonumber \\
 &  & +e^{-\int_{0}^{t}\lambda(u)\theta(u)\mathrm{d}u}H_{z}(0,\infty,y-t,t)1_{\{y\geq t\}},\label{eq:2-4-20}
\end{eqnarray}
and 
\begin{eqnarray}
M(t,\infty) & = & \int_{r=0}^{t}\Big\{\lambda(r)\theta(r)F(t-r)+\lambda(r)\big(1-\theta(r)\big)M(r,\infty)\big(F(t-r)-1\big)\Big\} e^{-\int_{r}^{t}\lambda(u)\theta(u)\mathrm{d}u}\mathrm{d}r\nonumber \\
 &  & +H(0,\infty,\infty,t)e^{-\int_{0}^{t}\lambda(u)\theta(u)\mathrm{d}u}.\label{eq:2-4-21}
\end{eqnarray}

The solutions to
$\Phi(t,x)$, $M(t,x)$, $G_{z}(t,y,0)$, and $M(t,\infty)$
exist. The functions
$\Phi(t,x)$, $M(t,x)$, and $M(t,\infty)$
are piecewise differentiable in their respective variables or characteristic lines over the domain, except for a finite number of points. The function $G_z(t,y,0)$
is piecewise continuous and piecewise differentiable with respect to $y$.

\end{thm}
\begin{IEEEproof}
The detailed proof is provided in Appendix \ref{subsec:Solution-to-subsystems}. 
\end{IEEEproof}
We can make several observations from Theorem \ref{thm:When-assuming-,}.
First, when time $t$ becomes large, the impact of the initial conditions
will gradually diminish. Specifically, the final term $H(0,\infty,\infty,t)e^{-\int_{0}^{t}\lambda(u)\theta(u)\mathrm{d}u}$
in Equation (\ref{eq:2-4-21}), which reflects the impact of initial
workload distribution, will become negligible as $t\to\infty$ if
$\lim_{t\to\infty}\int_{0}^{t}\lambda(u)\theta(u)\mathrm{d}u=\infty$.
When $\lim_{t\to\infty}\int_{0}^{t}\lambda(u)\theta(u)\mathrm{d}u<\infty$,
we also observe that the term $H(0,\infty,\infty,t)e^{-\int_{0}^{t}\lambda(u)\theta(u)\mathrm{d}u}$
becomes less sensitive to $t$ as $\lim_{t\to\infty}H(0,\infty,\infty,t)=1$.
This means that the distribution $M(t,\infty)$ will lose the dependence
on the initial distribution as time elapses. Similarly, the functions
$G_{z}(t,y,0)$, $M(t,x)$, and $\Phi(t,x)$ increasingly lose their
dependence on the initial distribution. Moreover, we can observe from
Equation (\ref{eq:2-4-21}) that $M(t,\infty)$ will gradually rely
less on its values at earlier times. That is, the integral in Equation
(\ref{eq:2-4-21}) exhibits ``fading memory'' of its history. The
influence of the function value at a much earlier time $r$, $M(r,\infty)$,
is suppressed within the integral because: 1) As the time difference
$t-r$ grows, the term $F(t-r)-1$ approaches zero; 2) if $\lim_{t\to\infty}\int_{0}^{t}\lambda(u)\theta(u)\mathrm{d}u=\infty$,
the exponential weight (i.e., $e^{-\int_{r}^{t}\lambda(u)\theta(u)\mathrm{d}u}$)
also decays to zero as the interval $[r,t]$ widens. Furthermore,
we observe that for large $t$, the AoI distribution $\Phi(t,x)$
primarily depends on the values of $M(t,x)$ over time interval $[t-x,t]$,
as shown by Equation (\ref{eq:2-4-18}). This implies that the effect
of the initial AoI distribution also diminishes over time. This insight
is useful for designing the algorithm for system parameter optimization,
as discussed later in Section \ref{subsec:Optimal-design-for}.

\subsection{Computational framework}

\label{subsec:computation_frame}

Although Theorem \ref{thm:When-assuming-,} provides an exact characterization
of $\Phi(t,x)$, the expression is not available in closed-form. A
key feature of Equation (\ref{eq:2-4-18}) is that for $\theta(t)>0$
for all $t$, the unknown function $\Phi(t,x)$ appears on both sides,
which necessitates solving the equation numerically. Similarly, the
function $M(t,\infty)$ must also be computed numerically when $\theta(t)<1$,
as indicated in Equation (\ref{eq:2-4-21}).

We begin by presenting the computational framework to numerically
evaluate $M(t,\infty)$. Note that when $\theta(t)=1$ for all $t$,
$M(t,\infty)$ can be directly computed via numerical integration
of Equation (\ref{eq:2-4-21}). For the case where $\theta(t)<1$,
Equation (\ref{eq:2-4-21}) is a Volterra integral equation of the
second kind. While various numerical techniques exist for solving
such equations, such as collocation or quadrature methods \cite{linz1985analytical},
we adopt the straightforward method of successive approximations (also
known as Picard iteration) for its simplicity \cite{polyanin2008handbook}.
This approach is detailed in Algorithm \ref{alg:Algorithm-for-calculating}.
The core idea is to discretize the time horizon using a fixed grid
$\mathcal{T}=(t_{0},t_{1},...,t_{n})$, and iteratively solve for
the values of $M(t,\infty)$ at each grid point using Equation (\ref{eq:2-4-21})
as the update rule. Specifically, we use linear interpolation to evaluate
the function at points between the grid nodes, since the primary focus
of this paper is not on the numerical method itself.

\begin{algorithm}
\begin{algorithmic}[1] \Require{$T>0$: The time period for evaluating
$M(t,\infty)$, i.e., $t\in[0,T]$} \State{Set an equally spaced
evaluation grid as $\mathcal{T}=(t_{0},t_{1},...,t_{n})$, with step
size $h=T/n$, such that $t_{i}=i\cdot h$. } \State{Initialize
the value with $M(t_{i},\infty)=w_{i}\leftarrow1$ for $i=0,...,n$}
\State{Set the error tolerance $etol>0$, initialize $error=1.0$}
\While{$error>etol$} \State{Obtain the function $M(t,\infty)$}
by interpolating over the grid $\mathcal{T}$ with the values {$\boldsymbol{w}=(w_{0},w_{1},...,w_{n})$}
\For{$i=0,...,n$} \State Compute the RHS of Equation (\ref{eq:2-4-21})
at $t_{i}$ using the trapezoidal rule by $RHS[i]\leftarrow\int_{r=0}^{t_{i}}\Big\{\lambda(r)\theta(r)F(t_{i}-r)+\lambda(r)(1-\theta(r))M(r,\infty)\big(F(t_{i}-r)-1\big)\Big\} e^{-\int_{r}^{t_{i}}\lambda(u)\theta(u)\mathrm{d}u}\mathrm{d}r+H(0,\infty,\infty,t_{i})e^{-\int_{0}^{t_{i}}\lambda(u)\theta(u)\mathrm{d}u}$
\EndFor \State $error\leftarrow\max_{i}|RHS[i]-w_{i}|$ \State
$w_{i}\leftarrow RHS[i]$ for $i=0,...,n$ \EndWhile \State{\Return
Function $M(t,\infty)$ interpolated at $(\mathcal{T},\boldsymbol{w})$}
\end{algorithmic}\caption{Numerical algorithm for calculating $M(t,\infty)$}
\label{alg:Algorithm-for-calculating} 
\end{algorithm}

Once the function $M(t,\infty)$ has been determined using Algorithm
\ref{alg:Algorithm-for-calculating}, the intermediate function $M(t,x)$
can be calculated directly. This is done by first computing $G_{z}(t,y,0)$
through numerical integration of Equation (\ref{eq:2-4-20}), and
then computing $M(t,x)$ using Equation (\ref{eq:2-4-19}). This provides
all the necessary components to solve for the AoI distribution.

Next, we present the computational framework for calculating the AoI
distribution, $\Phi(t,x)$. As shown in Equation (\ref{eq:2-4-18}),
the recursive term vanishes when $\theta(t)=0$ for all $t$, and
$\Phi(t,x)$ can then be computed directly with standard numerical
integration. The more complex case is when $\theta(t)>0$, where $\Phi(t,x)$
appears on both sides of the equation. A naive numerical solution
would require a two-dimensional approximation in the $(t,x)$ plane,
which is computationally expensive and can accumulate significant
errors. To overcome this, we perform a change of variables to reduce
the problem's dimensionality. By defining $u=t-x$, $\tau=x-t+r$,
we rewrite Equation (\ref{eq:2-4-18}) in terms of a new function:
\begin{eqnarray}
\hat{\Phi}(u,x)=\Phi(u+x,x) & = & M(u+x,x)-e^{-\int_{0}^{u+x}\lambda(s)\theta(s)\mathrm{d}s}\Big(-H(0,-u,\infty,\infty)+H(0,-u,\infty,u+x)\Big)1_{\{u<0\}}\nonumber \\
 &  & -\int_{\tau=\max\{0,-u\}}^{x}\Big(\lambda(\tau+u)\theta(\tau+u)\hat{\Phi}(u,\tau)\nonumber \\
 &  & +\lambda(\tau+u)\big(1-\theta(\tau+u)\big)M(\tau+u,\tau)\Big)\Big(F(x-\tau)-1\Big)e^{-\int_{u+\tau}^{u+x}\lambda(s)\theta(s)\mathrm{d}s}\mathrm{d}\tau.\label{eq:2-4-22}
\end{eqnarray}
The advantage of this transformation is that for any fixed $u$, the
problem becomes a one-dimensional Volterra integral equation in the
variable $x$, which is much simpler to solve numerically. We summarize
the numerical method for computing $\Phi(t,x)$ in Algorithm \ref{alg:Algorithm-for-calculating-1}.
We will verify the effectiveness of this computational framework in
Section \ref{subsec:Numerical-study-for}.

\begin{algorithm}
\begin{algorithmic}[1] \Require $t>0$ and $x\geq0$. Input an
initial distribution $H(0,x,y,z)$ that satisfies Assumption \ref{assum2}. \State{$u\leftarrow t-x$}
\State{Set the grid to evaluate $\hat{\Phi}(u,x)$ as $\mathcal{X}=(x_{0},x_{1},...,x_{m})$,
with $x_{0}=\max\{0,-u\}$ and $x_{m}=x$} \State{Initialize the value of
$w_{i}$ with $\hat{\Phi}(u,x_{i})=w_{i}\leftarrow1$ for $i=0,...,m$}
\State{Set the error tolerance $etol>0$, initialize $error=1.0$}
\While{$error>etol$} \State{Obtain the function $\hat{\Phi}(u,x)$}
by interpolating over the grid $\mathcal{X}$ with the value {$\boldsymbol{w}=(w_{0},...,w_{m})$}
\For{$i=0,...,m$} \If{$x\leq t$} \State Compute the RHS of
Equation (\ref{eq:2-4-22}) at $x_{i}$ using numerical integration
by $RHS[i]\leftarrow M(u+x_{i},x_{i})-\int_{\tau=0}^{x_{i}}\Big(\lambda(u+\tau)\theta(u+\tau)\hat{\Phi}(u,\tau)+\lambda(u+\tau)\big(1-\theta(u+\tau)\big)M(u+\tau,\tau)\Big)\Big(F(x_{i}-\tau)-1\Big)e^{-\int_{u+\tau}^{u+x_{i}}\lambda(s)\theta(s)\mathrm{d}s}\mathrm{d}\tau.$
\Else \State{ Compute the RHS of Equation (\ref{eq:2-4-22}) at
$x_{i}$ using numerical integration by $RHS[i]\leftarrow M(u+x_{i},x_{i})-e^{-\int_{0}^{u+x_{i}}\lambda(s)\theta(s)\mathrm{d}s}\Big(-H(0,-u,\infty,\infty)+H(0,-u,\infty,u+x_{i})\Big)-\int_{\tau=-u}^{x_{i}}\Big(\lambda(u+\tau)\theta(u+\tau)\hat{\Phi}(u,\tau)+\lambda(u+\tau)\big(1-\theta(u+\tau)\big)M(u+\tau,\tau)\Big)\Big(F(x_{i}-\tau)-1\Big)e^{-\int_{u+\tau}^{u+x_{i}}\lambda(s)\theta(s)\mathrm{d}s}\mathrm{d}\tau.$
} \EndIf\EndFor \State $error\leftarrow\max_{i}|RHS[i]-w_{i}|$
\State $w_{i}\leftarrow RHS[i]$ for $i=0,...,m$ \EndWhile \State{\Return
$w_{m}$} \end{algorithmic}\caption{Numerical algorithm for calculating $\Phi(t,x)$}
\label{alg:Algorithm-for-calculating-1} 
\end{algorithm}

\subsection{Model with negligible processing time}

\label{subsec:Model-of-no}

In this subsection, we discuss a special case where packet processing
times are negligible. This model corresponds to practical
scenarios where the processing time is significantly shorter than
the inter-arrival time of packets. To model the AoI in this case,
we denote the temporary AoI distribution as $\Phi(t,x)=\boldsymbol{P}(\Delta(t)\leq x)$,
and construct the conditional probability of the temporary AoI as:
\begin{equation}
\boldsymbol{P}(\Delta(t+h)\leq x|\Delta(t)=a)=\boldsymbol{P}(N(t,t+h)=1)+\boldsymbol{P}(N(t,t+h)=0)1_{\{a+h\leq x\}} + \boldsymbol{o}(h),\label{eq:2-5-1}
\end{equation}
If a packet arrives (with probability $\lambda(t)h$), the AoI resets
to nearly zero and is thus less than $x$. If no packet arrives (with
probability $1-\lambda(t)h$), the AoI increases from $a$ to $a+h$.
From this, we can establish the differential equation for $\Phi(t,x)$
as 
\begin{align}
\Phi_{t}(t,x) & =\lambda(t)+\lim_{h\to0^{+}}\frac{1}{h}\Big[(1-\lambda(t)h)\Phi(t,x-h)-\Phi(t,x)\Big]\nonumber \\
 & =\lambda(t)-\Phi_{x}(t,x)-\lambda(t)\Phi(t,x).\label{eq:2-5-2}
\end{align}
Performing the change of variable by letting $\tau=t$ and $\xi=x-t$,
we simplify the above PDE into the following ODE: 
\begin{equation}
\frac{\partial\Phi(t,x)}{\partial\tau}+\lambda(\tau)\Phi(t,x)=\lambda(\tau).\label{eq:2-5-3}
\end{equation}
The general solution to the ODE is given by: 
\begin{align}
\Phi(t,x) & =e^{-\int_{0}^{\tau}\lambda(u)\mathrm{d}u}\Big(\int_{0}^{\tau}\lambda(r)e^{\int_{0}^{r}\lambda(u)\mathrm{d}u}\mathrm{d}r+g_{6}(\xi)\Big)\nonumber \\
 & =e^{-\int_{0}^{t}\lambda(u)\mathrm{d}u}\Big(e^{\int_{0}^{t}\lambda(u)\mathrm{d}u}-1+g_{6}(x-t)\Big),\label{eq:2-5-4}
\end{align}
where $g_{6}(\cdot)$ is a real-valued function. When $x\geq t$,
considering the initial condition $\Phi(0,x)$, we have $g_{6}(x)=\Phi(0,x)$
for $x\geq0.$ When $x<t$, after applying the boundary condition
$\Phi(t,0)=0$, we obtain $g_{6}(-t)=1-e^{\int_{0}^{t}\lambda(u)\mathrm{d}u}$, which implies that 
\begin{equation}
\Phi(t,x)=1-e^{-\int_{(t-x)_+}^{t}\lambda(u)\mathrm{d}u}+1_{\{x\geq t\}}\Phi(0,x-t)e^{-\int_{0}^{t}\lambda(u)\mathrm{d}u}.\label{eq:2-5-5}
\end{equation}
An interesting observation from Equation (\ref{eq:2-5-5}) is that
as $t$ becomes large, the AoI evaluated at time $t$ will not depend
on the initial condition, but depend on the arrival rate during time
$[t-x,t)$. An intuitive way to understand this result is that if
the age at time $t$ is greater than $x$, it implies that no packet
has arrived in the interval $[t-x,t)$, which occurs with probability
$e^{-\int_{t-x}^{t}\lambda(u)\mathrm{d}u}$. We can also compute
the mean age at time $t$ as 
\begin{align}
\boldsymbol{E}[\Delta(t)] & =\int_{x=0}^{\infty}\Big(1-\Phi(t,x)\Big)\mathrm{d}x\nonumber \\
 & =\int_{x=0}^{\infty}\Big(e^{-\int_{(t-x)_+}^{t}\lambda(u)\mathrm{d}u}-1_{\{x\geq t\}}\Phi(0,x-t)e^{-\int_{0}^{t}\lambda(u)\mathrm{d}u}\Big)\mathrm{d}x.\label{eq:2-5-6}
\end{align}

We also observe from Equation (\ref{eq:2-5-6}) that the expected
AoI at time $t$ is mathematically dependent on the entire history
of the arrival rate $\lambda(u)$ over the interval $[0,t]$. Although
it is true that a high recent arrival rate will dominate the expectation
and minimize the practical influence of the distant past, the entire
history is still required to compute the exact value. This dependence
on a weighted history rather than just the time of the last arrival
demonstrates that the AoI process does not exhibit the memoryless
property in a time-varying setting, as its expectation reflects the
accumulated effects of historical variations in the arrival rate.

\section{Discussion on the stationary scenario}

\label{sec:Discussion-on-the}

In this section, we extend our analysis to the stationary scenario,
where the arrival rate and preemption probability are both constant, i.e.,
$\lambda(t)=\lambda$ and $\theta(t)=\theta$. We focus on the probability
distribution in the steady state (as $t\to\infty$), where the effect
of the initial condition is eliminated. With a slight abuse of notation,
we denote the steady-state joint CDF as $H(x,y,z)\triangleq\lim_{t\to\infty}\boldsymbol{P}(\Delta(t)\leq x,A(t)\leq y,W(t)\leq z)$,
and the steady-state AoI distribution is defined as $\Phi(x)\triangleq H(x,\infty,\infty)$.

We will demonstrate that the analytical framework developed for the
time-varying case in Section \ref{sec:Solution-derivation-approach}
can be adapted to derive the steady-state distribution in the stationary
setting as well. Following a similar structure, we first establish
the PDE framework and provide Theorem \ref{thm:The-LST-of} that summarizes
the solution to the stationary system in Section \ref{subsec:PDE-framework-and}.
In Section \ref{subsec:Special-case-I:}, we use the results in Theorem
\ref{thm:The-LST-of} to find closed-form expressions for AoI in several
special queueing systems.

\subsection{PDE framework and solution}

\label{subsec:PDE-framework-and}

In the steady-state scenario, the system's state probabilities do
not change over time. We can therefore construct the stationary system
of PDEs by setting all time-derivative terms in the time-varying system
to zero. Similar to our discussion of the time-varying system, when
the derivatives of $H(x,y,z)$ exist, the resulting PDEs that describe
the steady-state behavior are given as follows: 
\begin{itemize}
\item Stationary system: 
\end{itemize}
\begin{eqnarray}
\mbox{\emph{Empty state:}} &  & \lambda H(x,0,0)-H_{z}(\infty,x,0)+H_{x}(x,0,0)=0\label{eq:3-0-1}\\
\mbox{\emph{Busy state:}} &  & \lambda\theta H(x,\infty,\infty)F(z)+\lambda(1-\theta)H(x,0,0)F(z)-\lambda\theta H(x,y,z)+\lambda\theta H(x,y,0)\nonumber \\
 &  & -H_{x}(x,y,z)-H_{y}(x,y,z)+H_{z}(x,y,z)+H_{x}(x,y,0)-H_{z}(x,y,0)=0,\label{eq:3-0-2}\\
\mbox{\emph{Boundary condition:}} &  & H(0,y,z)=0\mbox{ for any }y,z\geq0.\label{eq:3-0-3}
\end{eqnarray}

As we shall show later, solving the PDEs above leads to a Volterra
integral equation for the AoI distribution. By applying the LST to
solve this integral equation, we can obtain the final expression for
the LST of AoI, denoted as $\tilde{\Phi}(s)\triangleq\int_{x=0}^{\infty}e^{-sx}\mathrm{d}\Phi(x)$.
The solution is presented in the following theorem, with the detailed
derivation provided in Appendix \ref{subsec:Solution-to-subsystems-1}.
\begin{thm}
\label{thm:The-LST-of}In the stationary scenario where Part (2) of Assumption \ref{assum1} holds, if the preemption
probability is non-zero (i.e., $\theta>0$), the LST of the AoI distribution
is given by 
\begin{equation}
\tilde{\Phi}(s)=s\cdot M^{*}(s)\frac{1+(1-\theta)K^{*}(s)}{1-\theta K^{*}(s)},
\end{equation}
with 
\begin{equation}
M^{*}(s)=\frac{\lambda}{(\lambda+s)s}\cdot\frac{\theta\tilde{F}(\lambda\theta+s)}{1-(1-\theta)\tilde{F}(\lambda\theta)},\quad K^{*}(s)=\frac{\lambda}{\lambda\theta+s}\left(1-\tilde{F}(\lambda\theta+s)\right).
\end{equation}
When there is no preemption (i.e., $\theta=0$), the CDF of the AoI
distribution is directly given as 
\begin{equation}
\Phi(x)=M(x)+\lambda\int_{s=0}^{x}M(s)[1-F(x-s)]\mathrm{d}s,
\end{equation}
where $M(x)=\frac{\lambda}{1-\lambda\tilde{F}^{\prime}(0)}\int_{s=0}^{x}F(s)e^{-\lambda(x-s)}\mathrm{d}s$. 
\end{thm}
Note that for the system without preemption, the theorem provides
the CDF of the AoI distribution directly. When the preemption probability
is non-zero, we can compute the mean and higher-order moments of the
AoI by taking successive derivatives of the LST. Furthermore, the
full distribution of the AoI can be recovered by applying the inverse
LST using the Bromwich integral \cite{abate1995numerical}. Specifically,
the probability density function (PDF) and CDF of the AoI can be obtained as follows: 
\begin{eqnarray}
\Phi_{x}(x) & = & \frac{1}{2\pi i}\lim_{T\to\infty}\int_{\gamma-iT}^{\gamma+iT}e^{sx}\tilde{\Phi}(s)\mathrm{d}s,\label{eq:3-2-12}\\
\Phi(x) & = & \frac{1}{2\pi i}\lim_{T\to\infty}\int_{\gamma-iT}^{\gamma+iT}e^{sx}\frac{\tilde{\Phi}(s)}{s}\mathrm{d}s.\label{eq:3-2-13}
\end{eqnarray}
Here, $\gamma$ is a real constant chosen such that all singularities
of $\tilde{\Phi}(s)$ lie to the left of the line $Re(s)=\gamma$
in the complex plane (see \cite{abate1995numerical}). We will numerically
compute the AoI distributions using Equations (\ref{eq:3-2-12}) and
(\ref{eq:3-2-13}) later in Section \ref{sec:Numerical-studies}.

\subsection{Special cases}

\label{subsec:Special-case-I:}

In this subsection, we utilize the results established in Theorem
\ref{thm:The-LST-of} to derive closed-form expressions of the AoI
distributions in specific queueing systems. The following corollary
first derives a closed-form expression for the AoI distribution in
an M/M/1/1 queueing system with probabilistic preemption. Related
results have appeared in the literature, including the AoI distribution
for the non-preemptive M/M/1/1 system \cite{champati2019distribution},
the LST of the AoI for the preemptive M/G/1 system \cite{inoue2019general},
and the moment-generating function of the AoI for the M/G/1/1 system
with probabilistic preemption \cite{moltafet2025aoi}. However, these
studies do not explicitly provide closed-form expressions for the
AoI distribution under probabilistic preemption. Moreover, while previous
studies like \cite{bedewy2019age} proved that a fully preemptive
scheme achieves a smaller AoI than the non-preemptive scheme when
service times are exponential, we provide an alternative proof in the following. Using
the closed-form expressions of AoI, we show that full preemption results
in a stochastically smaller AoI across all possible age thresholds
compared to any probabilistic preemption $\theta<1$.
\begin{cor}
\label{cor:When-the-processing}For the $M/M/1/1$ system with probabilistic
preemption ($\theta>0$), where the processing time is exponentially distributed
with rate $\mu$ with $\lambda\neq\mu$ and $\mu\neq\lambda(1-\theta)$, the CDF of AoI is given by 
\begin{align}
\Phi(x) & =1+\frac{\mu^{2}(\mu+\lambda\theta)}{(\lambda-\mu)(\lambda+\mu)\bigl(\mu-\lambda(1-\theta)\bigr)}e^{-\lambda x}\nonumber \\
 & -\frac{\lambda(\mu+\lambda\theta)}{\theta(\lambda-\mu)(\lambda+\mu)}e^{-\mu x}-\frac{\lambda\mu(1-\theta)}{\theta(\lambda+\mu)\bigl(\mu-\lambda(1-\theta)\bigr)}e^{-(\mu+\lambda\theta)x}.\label{eq:5-2-1}
\end{align}
Moreover, the AoI in the $M/M/1/1$-preemptive system (with $\theta=1$),
denoted by $\Delta_{1}(t)$, is stochastically smaller than the system
with probabilistic preemption with $\theta \in (0,1)$, denoted by $\Delta_{\theta}(t)$,
in the steady state. That is, for all $x\geq0$: 
\begin{equation}
\lim_{t\to\infty}\boldsymbol{P}(\Delta_{1}(t)\leq x)\geq\lim_{t\to\infty}\boldsymbol{P}(\Delta_{\theta}(t)\leq x) \quad \mbox{for any }0<\theta<1.
\end{equation}
\end{cor}
\begin{proof} The detailed proof and the discussion on the special cases where  $\lambda=\mu$ or $\mu=\lambda(1-\theta)$ are given in Appendix \ref{sec:Proof-of-Corollary}.\end{proof}

In the following, we provide the AoI distribution for the $M/D/1/1$
system with deterministic processing time.
\begin{cor}\label{cor:MD11}
For the $M/D/1/1$ system without preemption, with the deterministic
processing time $1/\mu,$ the CDF of AoI is given by 
\begin{eqnarray*}
\Phi(x)=\begin{cases}
0, & \mbox{if }x\in[0,\frac{1}{\mu});\\
\frac{\lambda\mu}{\lambda+\mu}x-\frac{\lambda}{\lambda+\mu}, & \mbox{if }x\in[\frac{1}{\mu},\frac{2}{\mu});\\
1-\frac{\mu}{\lambda+\mu}e^{-\lambda x+2\frac{\lambda}{\mu}}, & \mbox{if }x\in[\frac{2}{\mu},\infty).
\end{cases}
\end{eqnarray*}
\end{cor}
\begin{IEEEproof}
    The proof is provided in Appendix \ref{sec:proof-of-cor2}.
\end{IEEEproof}

\section{Numerical studies and extensive discussions}

\label{sec:Numerical-studies}

In this section, we present numerical studies for both the time-varying
and stationary systems to validate our analytical results and provide
further insights. We then address a parameter optimization problem
in the time-varying setting in Section \ref{subsec:Optimal-design-for}.

\subsection{Numerical study for the time-varying system}

\label{subsec:Numerical-study-for} We begin by performing numerical
studies on the time-varying system to validate our derivations and
develop insights. Specifically, we consider the following two initial
conditions.

\begin{cond} \label{cond:cond1}We consider
the condition under which the system is empty at time 0, and we let
$H(0,x,0,0)=1-e^{-\lambda(0)x}.$ It is easy to verify that this condition
satisfies Assumption \ref{assum2}. We then have 
\begin{eqnarray}
H(0,x,\infty,\infty)=H(0,x,\infty,z)=H(0,x,0,0) & = & 1-e^{-\lambda(0)x},
\end{eqnarray}
and $H_{z}(0,\infty,y,z)=0$.
\end{cond}

\begin{cond}
\label{cond:cond2} We let the initial
distribution of $H(0,x,y,z)$ satisfy $\boldsymbol{P}(W(0)\leq z|\Delta(0)=x)=\frac{1}{2}+\frac{\min\{z,1\}}{2}$,
$\boldsymbol{P}(A(0)\leq y|\Delta(0)=x,W(0)=z>0)=\frac{\min\{y,x\}}{x}$,
$\boldsymbol{P}(A(0)=0|\Delta(0)=x,W(0)=0)=1$, and $\boldsymbol{P}(\Delta(0)\leq x|W(0)=z)=\frac{\min\{(x-1)_+,2\}}{2}$.
We can also verify that this initial distribution satisfies Assumption
\ref{assum2}. The necessary items to characterize the initial conditions in Theorem
\ref{thm:When-assuming-,} are given as follows.
\begin{eqnarray}
H(0,x,\infty,\infty) & = & \frac{\min\{(x-1)_+,2\}}{2},\\
H(0,x,\infty,z) & = & \frac{\min\{(x-1)_+,2\}}{2}\Big(\frac{1}{2}+\frac{\min\{z,1\}}{2}\Big),\\
H(0,x,0,0) & = & \frac{\min\{(x-1)_+,2\}}{4}
\end{eqnarray}
\begin{eqnarray}
H_{z}(0,\infty,y,z) & = & \int_{x=1}^{3}P(A(0)\leq y|\Delta(0)=x, W(0)>0)\frac{1}{4}dx\cdot1_{\{0<z\leq1\}}\nonumber \\
 & = & \begin{cases}
\frac{y}{4}\ln3 \cdot 1_{\{0<z\leq1\}}, & \mbox{ when }0\leq y<1;\\
(y\ln\frac{3}{y}+y-1)\frac{1}{4}\cdot 1_{\{0<z\leq1\}}, & \mbox{ when }1\leq y<3;\\
\frac{1}{2}\cdot 1_{\{0<z\leq1\}}, & \mbox{ when }y\geq3.
\end{cases}
\end{eqnarray}
\end{cond}

\begin{figure}[t!]
\centering\subfloat[]{\includegraphics[scale=0.33]{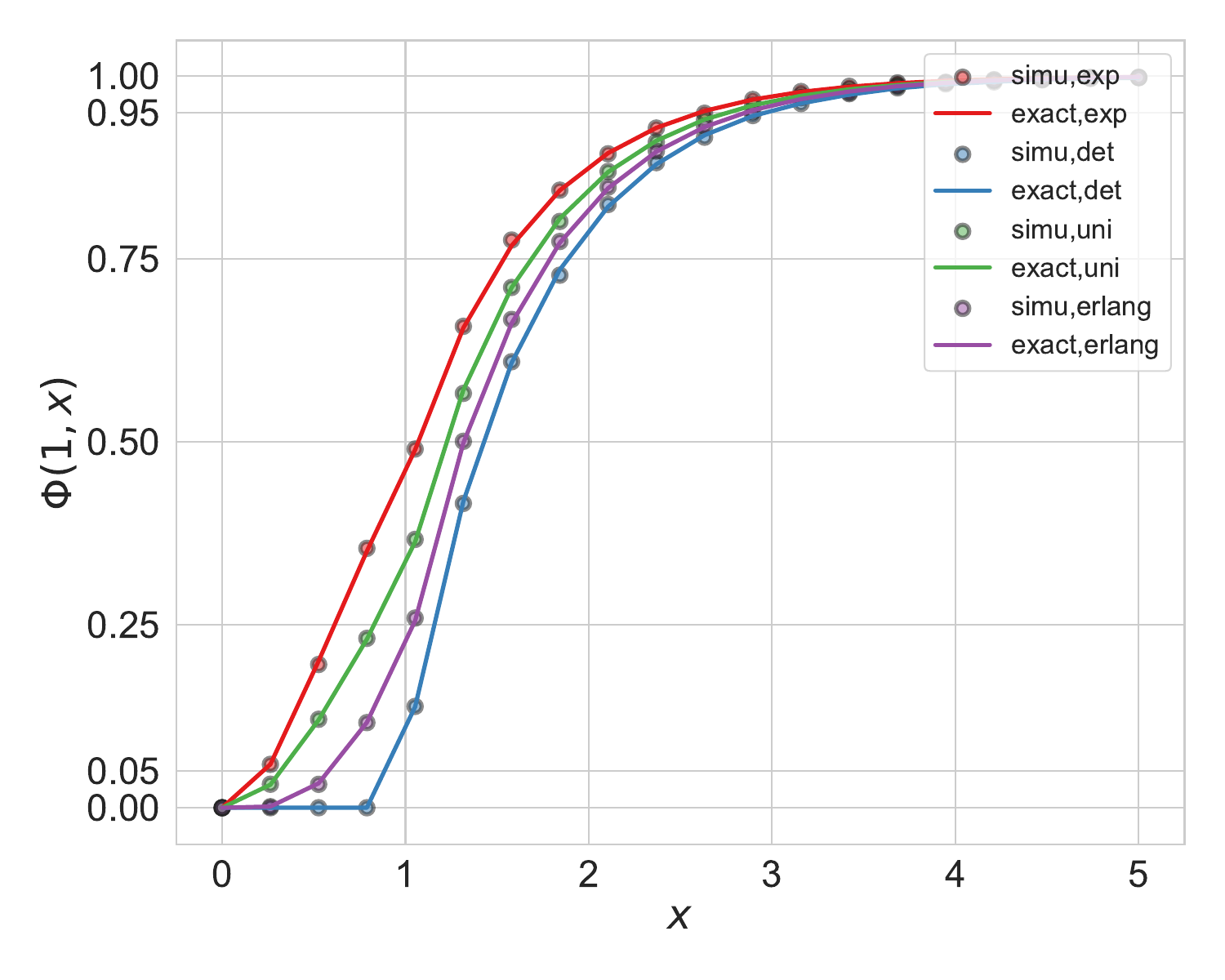}

}\subfloat[]{\includegraphics[scale=0.33]{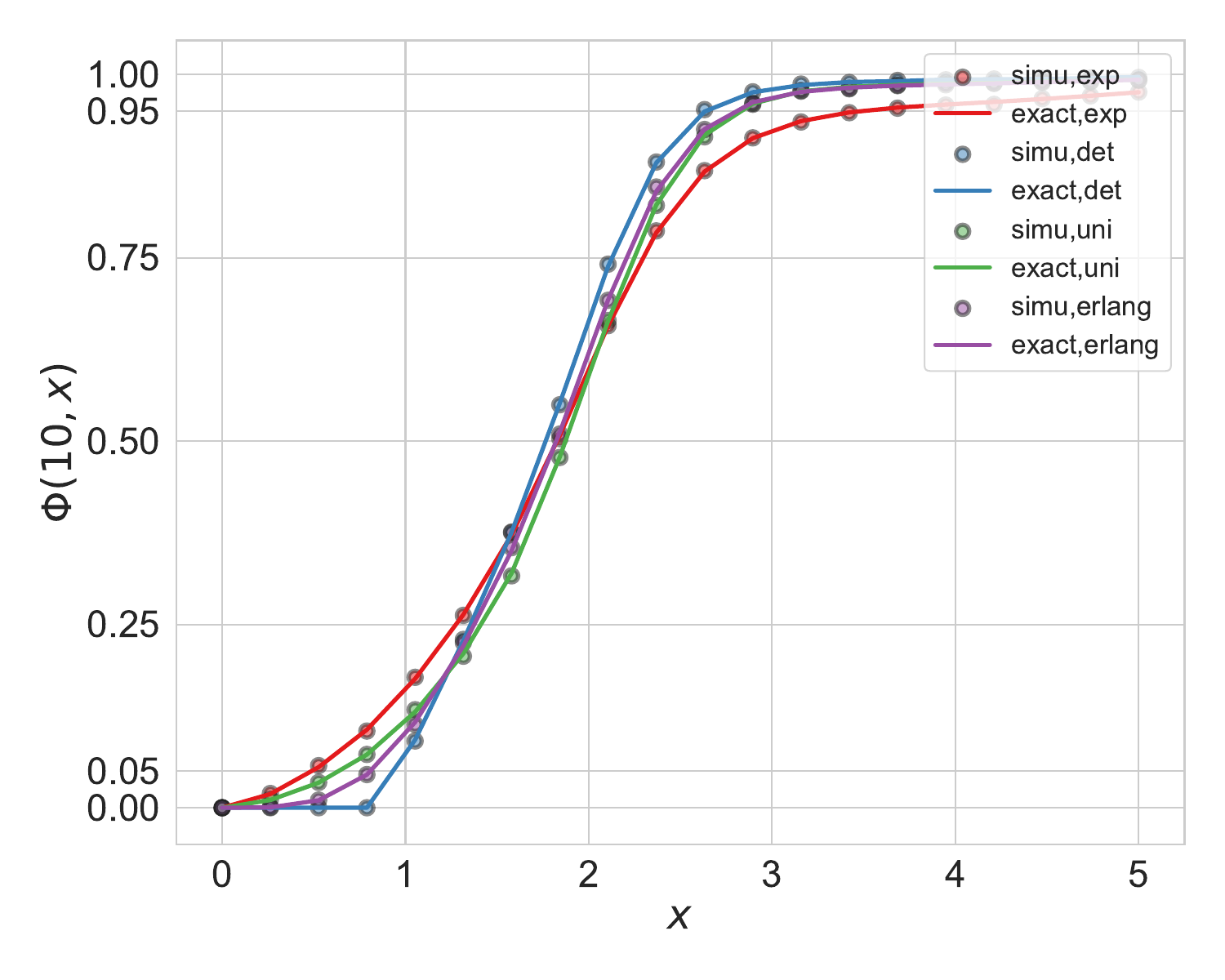}

}

\subfloat[]{\includegraphics[scale=0.33]{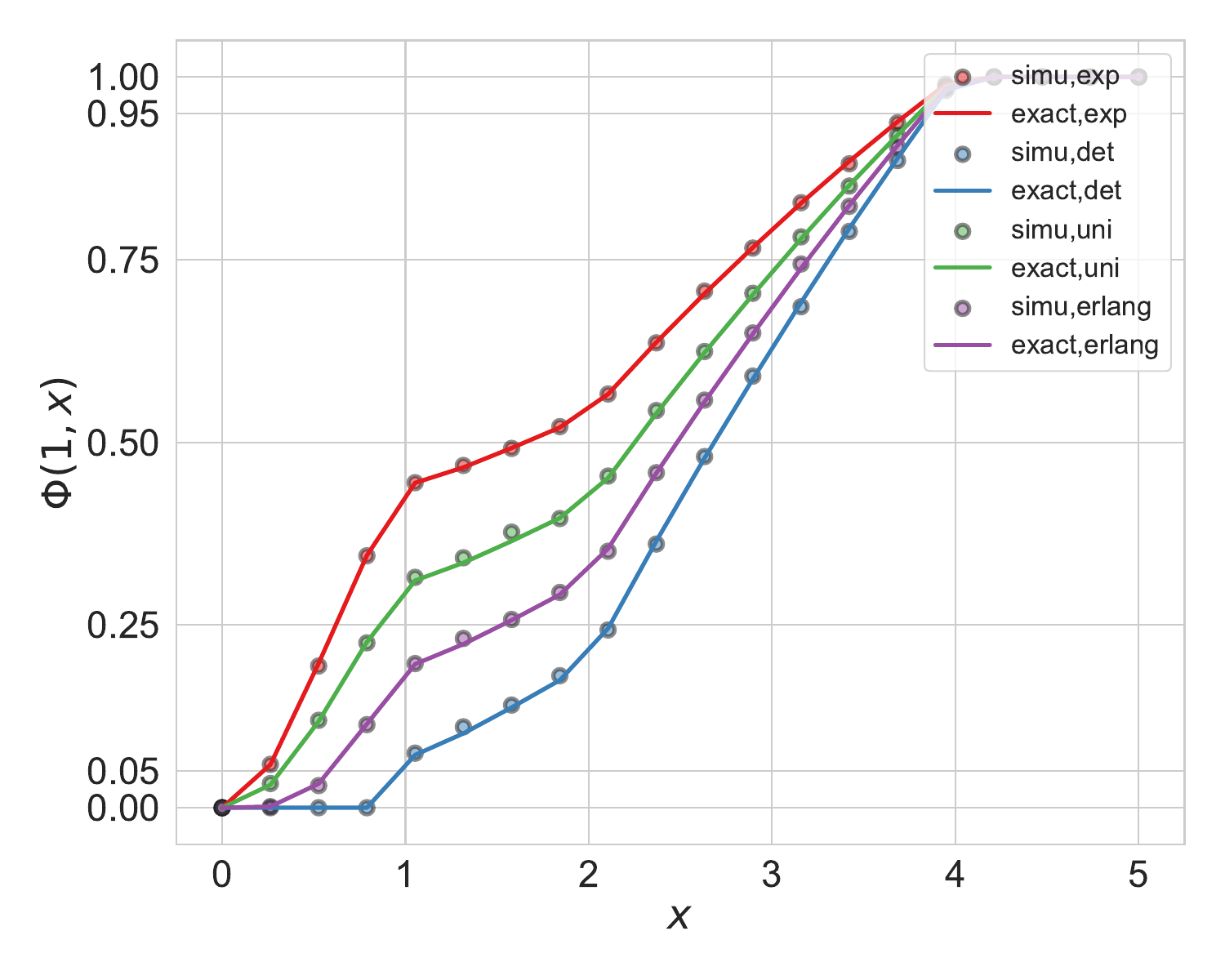}

}\subfloat[]{\includegraphics[scale=0.33]{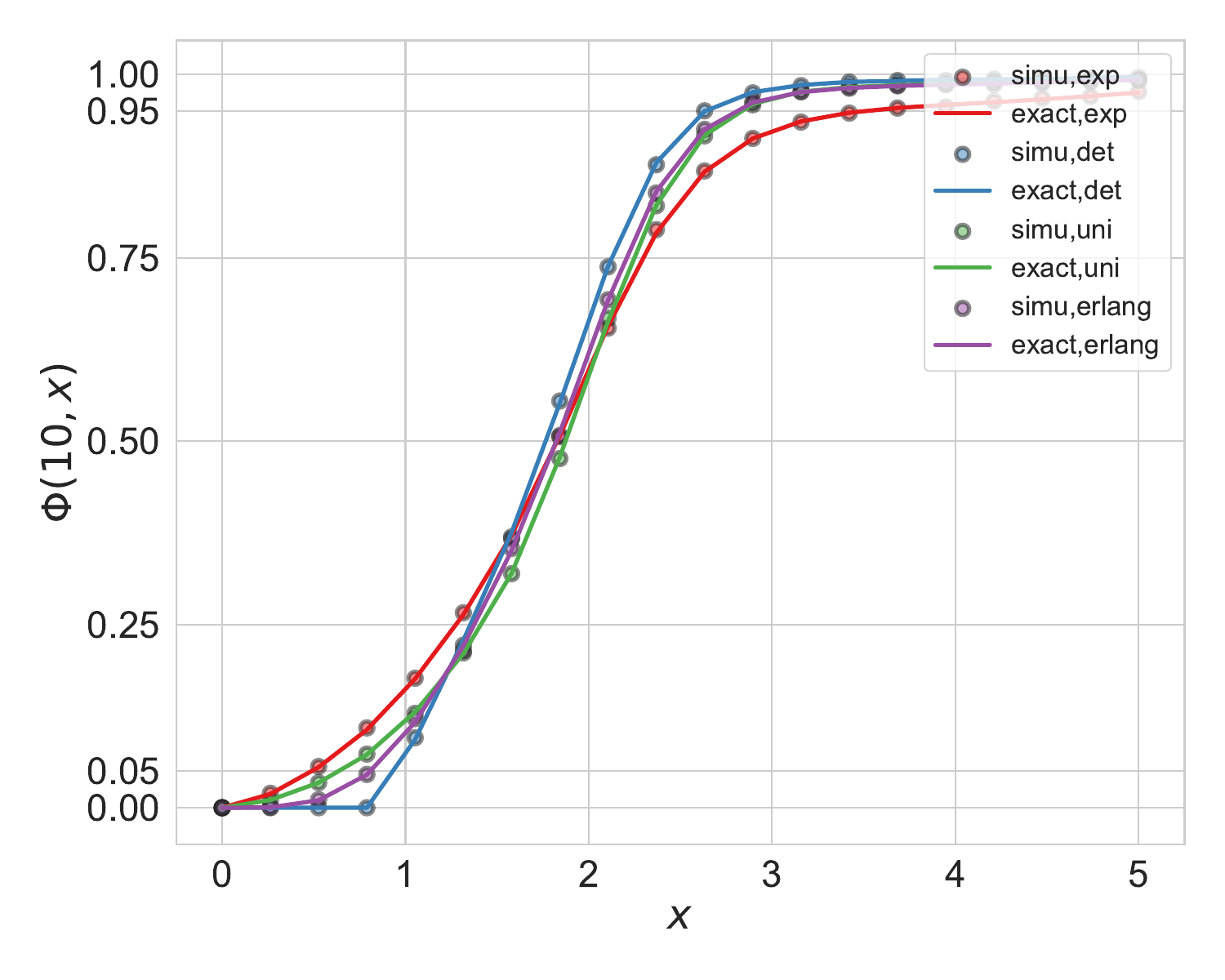}

}

\caption{Validation of the analytical AoI distribution (solid lines) against
simulation results (points) for several processing time distributions.
The initial condition is Condition \ref{cond:cond1}
for subfigures (a) and (b), and is Condition \ref{cond:cond2}
for subfigures (c) and (d). The CDF is shown at (a) $t=3$; (b) $t=10$;
(c) $t=3$; and (d) $t=10$. Common parameters for all subfigures are
arrival rate $\lambda(t)=1.5+\sin(1.8t)$, base processing rate $\mu=1.2$,
and preemption probability $\theta(t)=0.5\cos(1.8t)+0.5$. For the
processing time distribution, the abbreviations correspond to `exp'$\sim\exp(\mu)$,
`uni'$\sim\text{Uniform}(0,2/\mu)$, `det'$\sim(1/\mu)$, `erlang'$\sim\text{Erlang}(5,1/5\mu)$.}
\label{fig:Comparison-of-the-1-1}
\end{figure}

We next run simulations to verify our theoretical derivations. All
the simulation and numerical algorithms were coded in Python and run
on a computer with an Intel(R) Core(TM) Ultra9 185H, 2.30 GHz CPU,
64 GB RAM, and Windows 11 operating system. We validate our derivations
for the time-varying case by comparing the analytical AoI distribution
results from Theorem \ref{thm:When-assuming-,} with simulation results.
Fig. \ref{fig:Comparison-of-the-1-1} illustrates this comparison
for several different processing time distributions (e.g. exponential,
uniform, deterministic, and Erlang). Each simulation data point is
the result of 100,000 instances that measure the AoI of the system
at time $t$. For Figs. \ref{fig:Comparison-of-the-1-1}(a) and (b),
the initial condition at time 0 satisfies Condition \ref{cond:cond1},
and the initial condition satisfies Condition \ref{cond:cond2}
for Figs. \ref{fig:Comparison-of-the-1-1}(c) and (d). As shown in
Fig. \ref{fig:Comparison-of-the-1-1}, the simulation results closely
match the analytical solutions, confirming the accuracy of our framework.
Specifically, by comparing Figs. \ref{fig:Comparison-of-the-1-1}(a)
and (c), we find that the AoI distributions at $t=1$ show distinct
shapes as they depend highly on the initial distribution at time $0$.
The AoI distributions at time $t=10$, as shown in Figs. \ref{fig:Comparison-of-the-1-1}(b)
and (d), have similar shapes, which show that the AoI distribution
has a weaker dependence on the initial distribution as time elapses. 

\begin{figure}[t!]
\centering \subfloat[]{\includegraphics[scale=0.33]{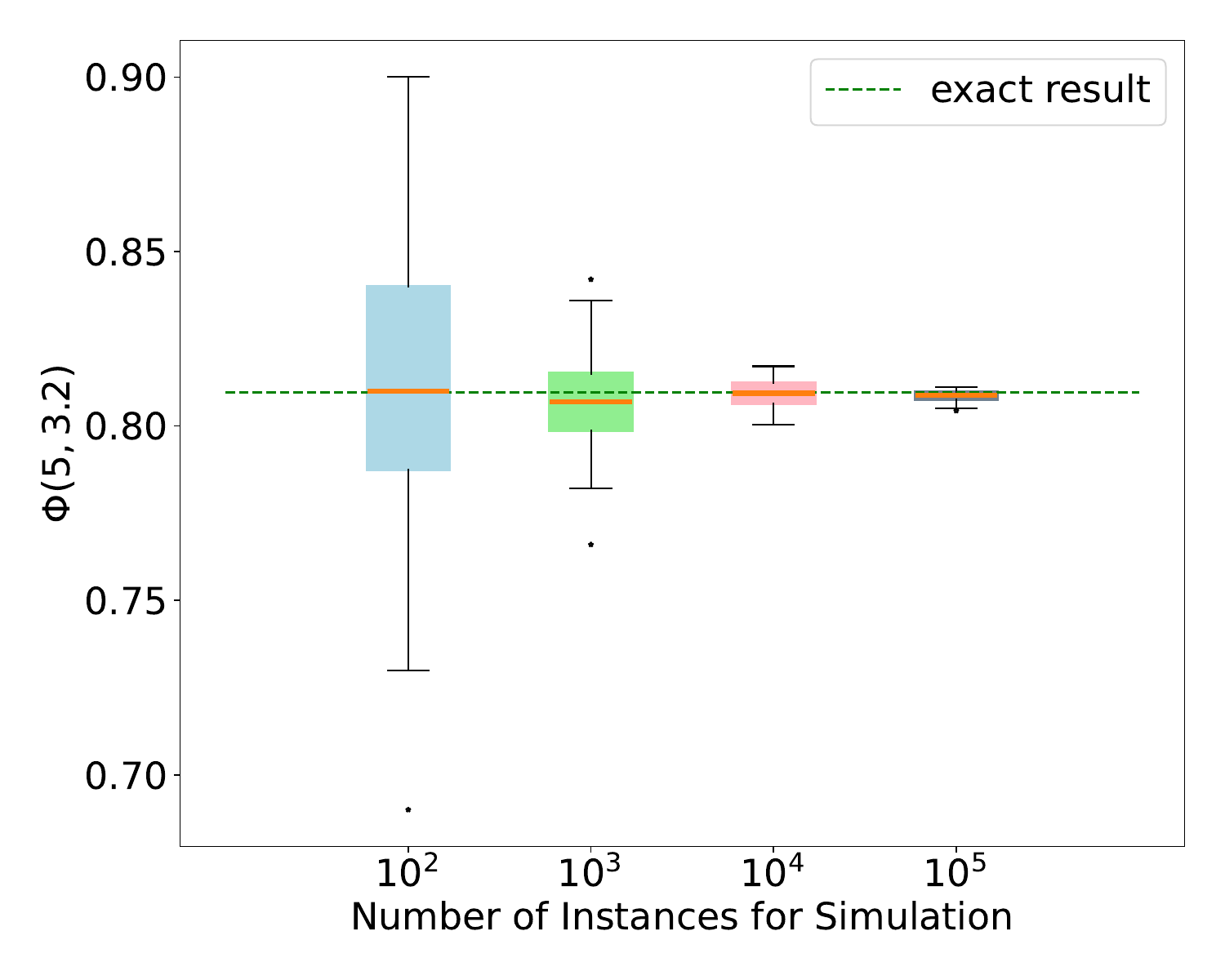} 

}\subfloat[]{\includegraphics[scale=0.33]{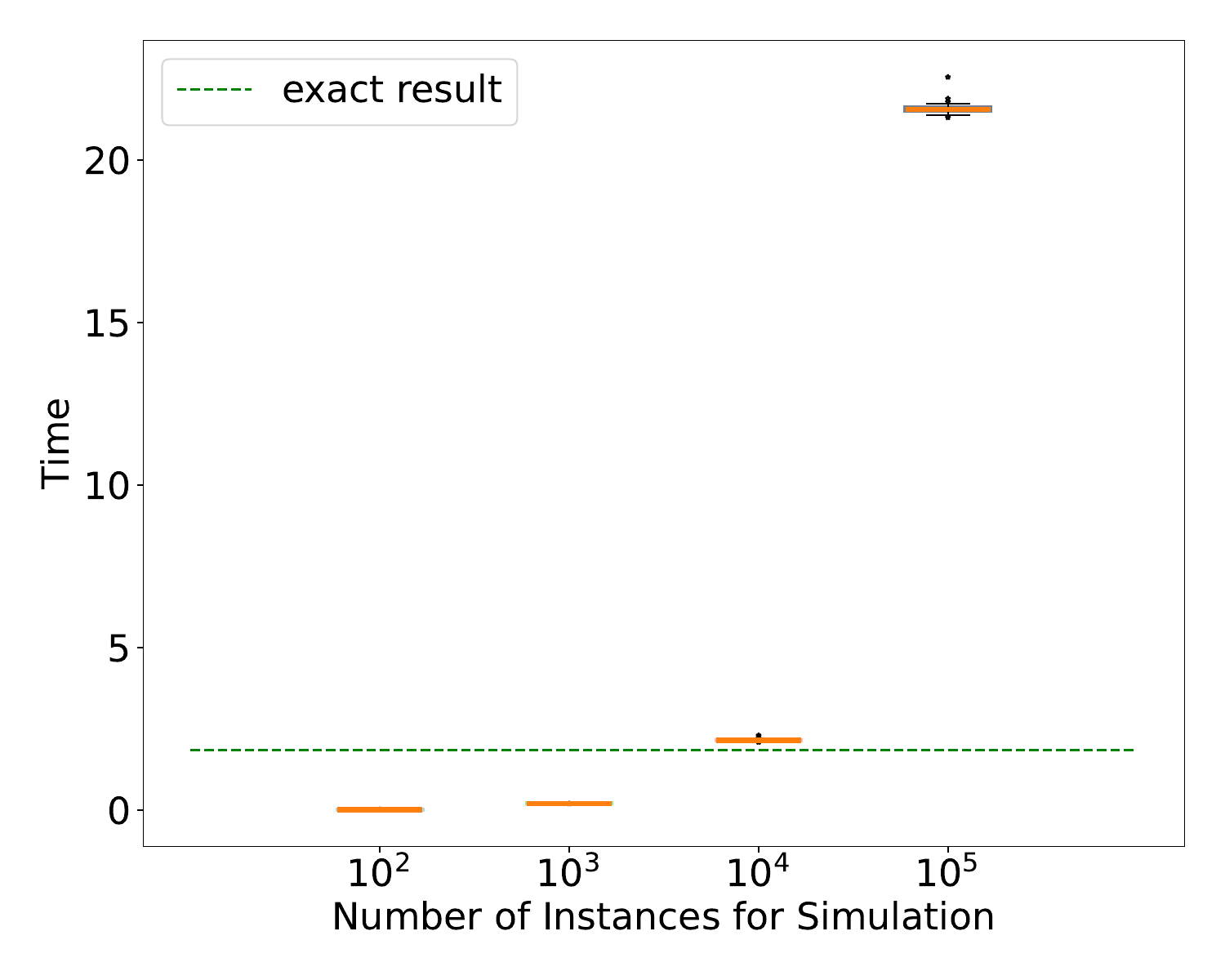}

}\caption{Comparison of simulation with exact result. (a) Comparison of AoI
distribution. (b) Comparison of computation time (in seconds). For
both subfigures, we have $\lambda(t)=2+\sin(1.8t)$, $\theta(t)=0.5\cos(1.8t)+0.5$,
$\mu=1.2$, and the processing time is Erlang distributed with parameters
$(5,1/5\mu)$. The initial condition satisfies Condition \ref{cond:cond1}. The boxplots
for the simulation results are from 100 runs of simulation of different
numbers of instances. The box spans from the first quartile (Q1) to
the third quartile (Q3), with the line inside the box representing
the median. The whiskers extend from the edges of the box to the furthest
data points within 1.5 times the interquartile range (IQR). Data points
beyond the whiskers are considered outliers (also called fliers).}
\label{fig:comparison of simulation with exact result} 
\end{figure}

\begin{figure}[t!]
\centering\subfloat[]{\includegraphics[scale=0.33]{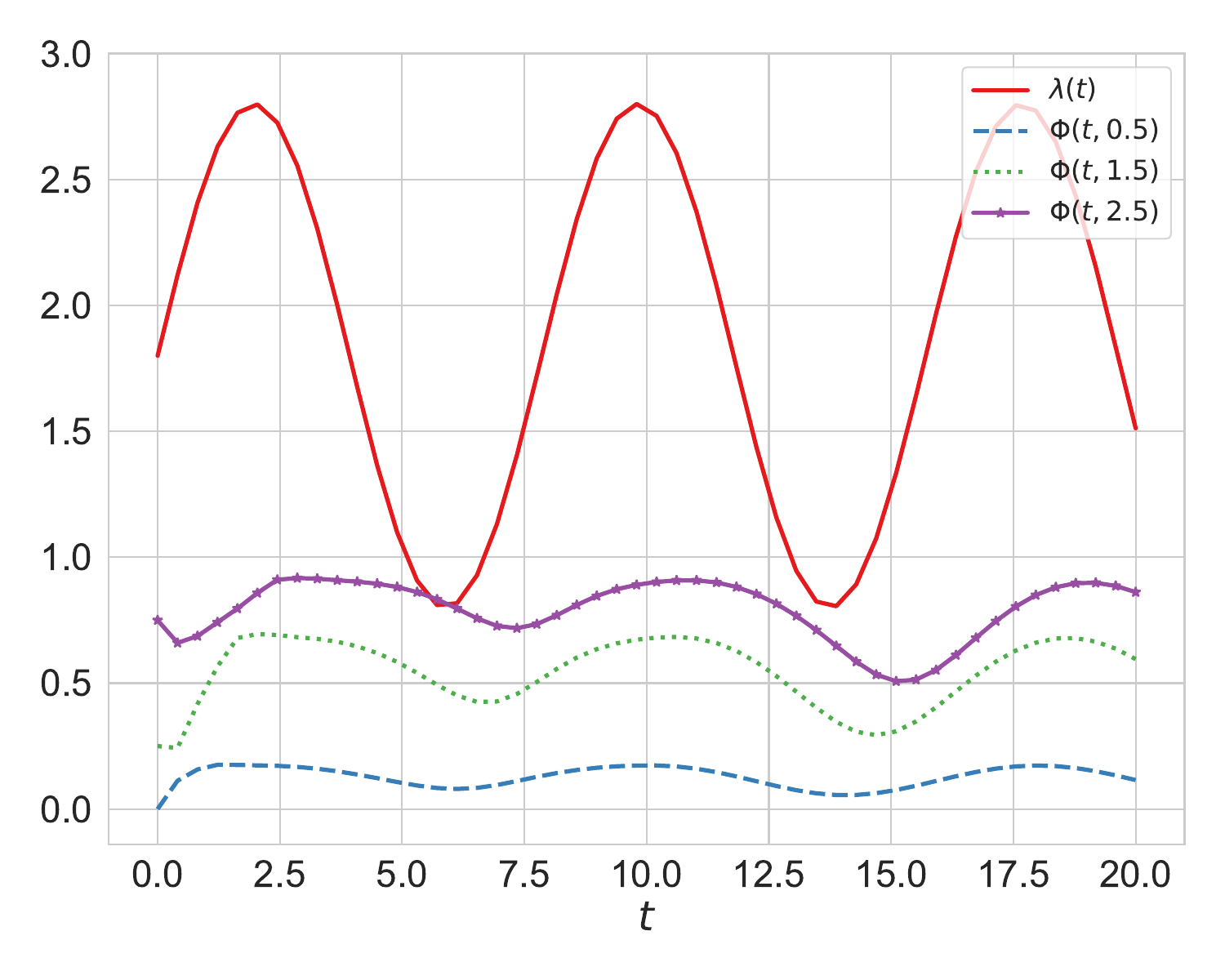}

}\subfloat[]{\includegraphics[scale=0.33]{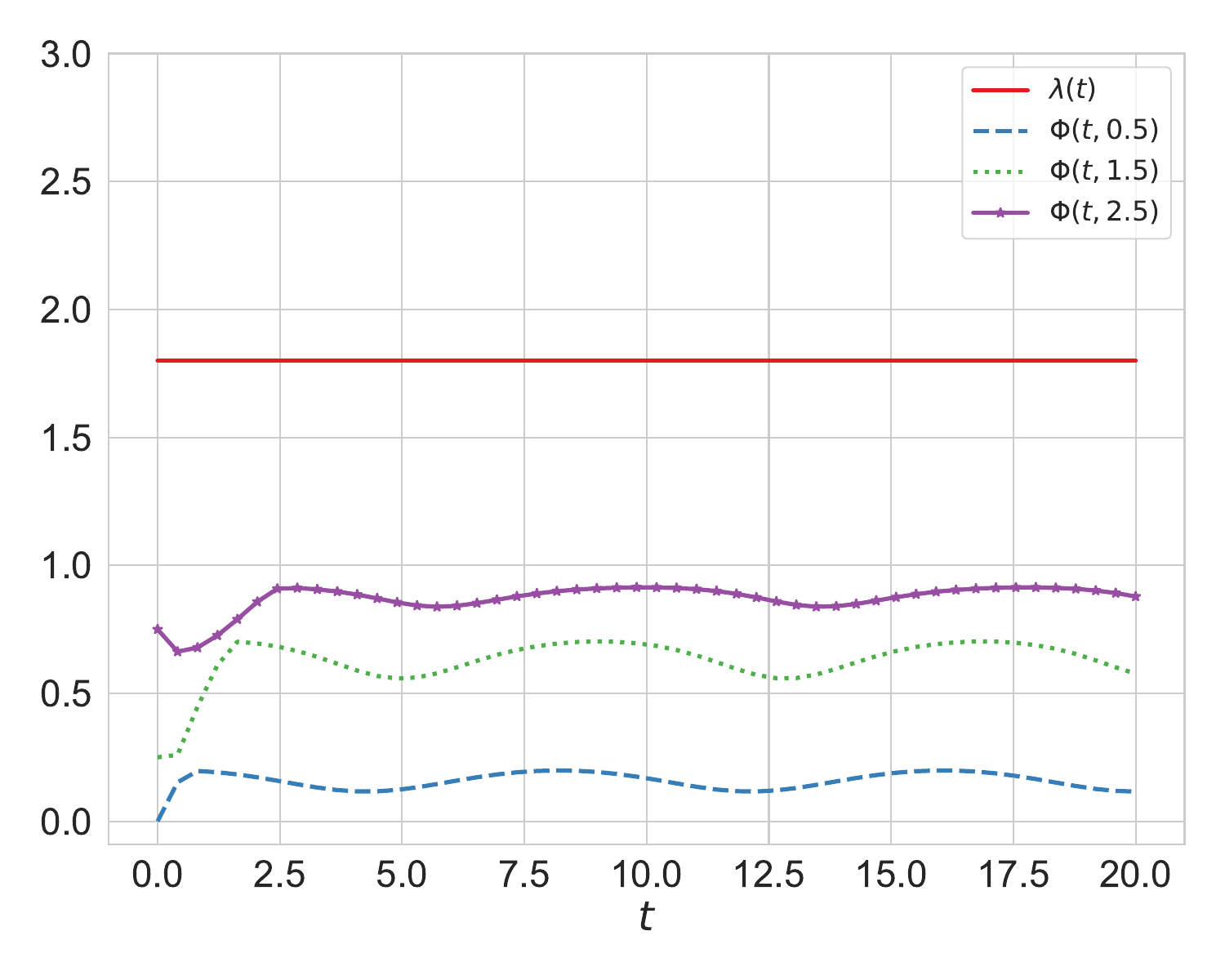}

}

\caption{AoI distribution and the arrival rate over time. (a) $\lambda(t)=1.8+\sin(0.8t)$
and $\theta(t)=0.2$. (b) $\lambda(t)=1.8$ and $\theta(t)=0.5\cos(0.8t)+0.5$.
For both subfigures, the processing time is exponential with rate
$\mu=1.5$ and the initial condition satisfies Condition \ref{cond:cond2}.}\label{fig:AoI-distribution-over}
\end{figure}

Note that we need to apply the numerical algorithms described in Section
\ref{subsec:computation_frame} to compute the exact results from
Theorem \ref{thm:When-assuming-,}. We now compare the accuracy and
computation time between simulation and the numerical approach. In
Fig. \ref{fig:comparison of simulation with exact result}, we evaluate
both methods for calculating the distribution $\Phi(5,3.2)$. As shown
in Fig. \ref{fig:comparison of simulation with exact result}(a),
the variation in simulation results decreases as the number of instances
increases. The range between whiskers (1.5 times the IQR) falls below
10\% of the median only when the number of instances reaches $10^{5}$.
However, Fig. \ref{fig:comparison of simulation with exact result}(b)
shows that running such a large number of simulation instances takes
nearly 10 times longer than the numerical approach, highlighting the
efficiency of our proposed numerical algorithms.

We then plot the AoI distribution $\Phi(t,x)$ over time $t$ in Fig.
\ref{fig:AoI-distribution-over}. Specifically, the sampling rate
is a sine function and the preemption probability is a constant in Fig. \ref{fig:AoI-distribution-over}(a).
In contrast, the sampling rate is a constant and the preemption probability
is sine function in Fig. \ref{fig:AoI-distribution-over}(b). From
Fig. \ref{fig:AoI-distribution-over}(a), we observe that the distribution
$\Phi(t,1.5)$ exhibits the largest variation over time. This occurs
because $x=1.5$ is a middle-range threshold. When the threshold is
very small, the probability that the temporary AoI is below this value
remains low, regardless of $\lambda(t)$, resulting in little variation.
Conversely, when the threshold is large, the AoI is almost always
below this value, again leading to limited variation. Thus, at an
intermediate threshold, the AoI distribution is most sensitive to
changes in the sampling rate $\lambda(t)$. Furthermore, we observe
from Fig. \ref{fig:AoI-distribution-over}(b) that when the sampling
rate is a constant, a time-varying preemption probability can also cause
significant fluctuations in the AoI distribution, particularly for
thresholds at $x=1.5$ and $x=2.5$. This shows that by altering the
preemption probability, the system can achieve significantly different
AoI violation probabilities. 

In addition, we observe from Fig. \ref{fig:AoI-distribution-over}(a)
that the AoI distribution $\Phi(t,x)$ tends to reach its maximum
at a later time than the sampling rate $\lambda(t)$. For instance,
the sampling rate $\lambda(t)$ peaks before $t=10$ while the probability
$\Phi(t,2.5)$ reaches its maximum around $t=11$. Similarly, the
sampling rate hits its minimum near $t=5.5,$ but the probability
$\Phi(t,2.5)$ reaches its lowest point around $t=7.5$. The reason
for this mismatch is that the AoI is influenced by both the sampling
rate and the processing rate. Even when a fresh packet arrives due
to a high sampling rate, it requires a non-negligible processing time
before it can update the system's information. As a result, there
is a delay between changes in the sampling rate and the corresponding
changes in the AoI distribution. Fig. \ref{fig:AoI-distribution-over}(b)
further indicates that a similar delay exists between adjustments
in the preemption probability and the AoI distribution. These observations
suggest that, to increase the AoI probability at a future time, the
sampling rate and preemption probability should be adjusted in advance.
However, as the following numerical studies will show, the optimal
choice of these parameters also depends on the underlying processing
time distribution.

\begin{figure}[t!]
\centering\subfloat[]{\includegraphics[scale=0.33]{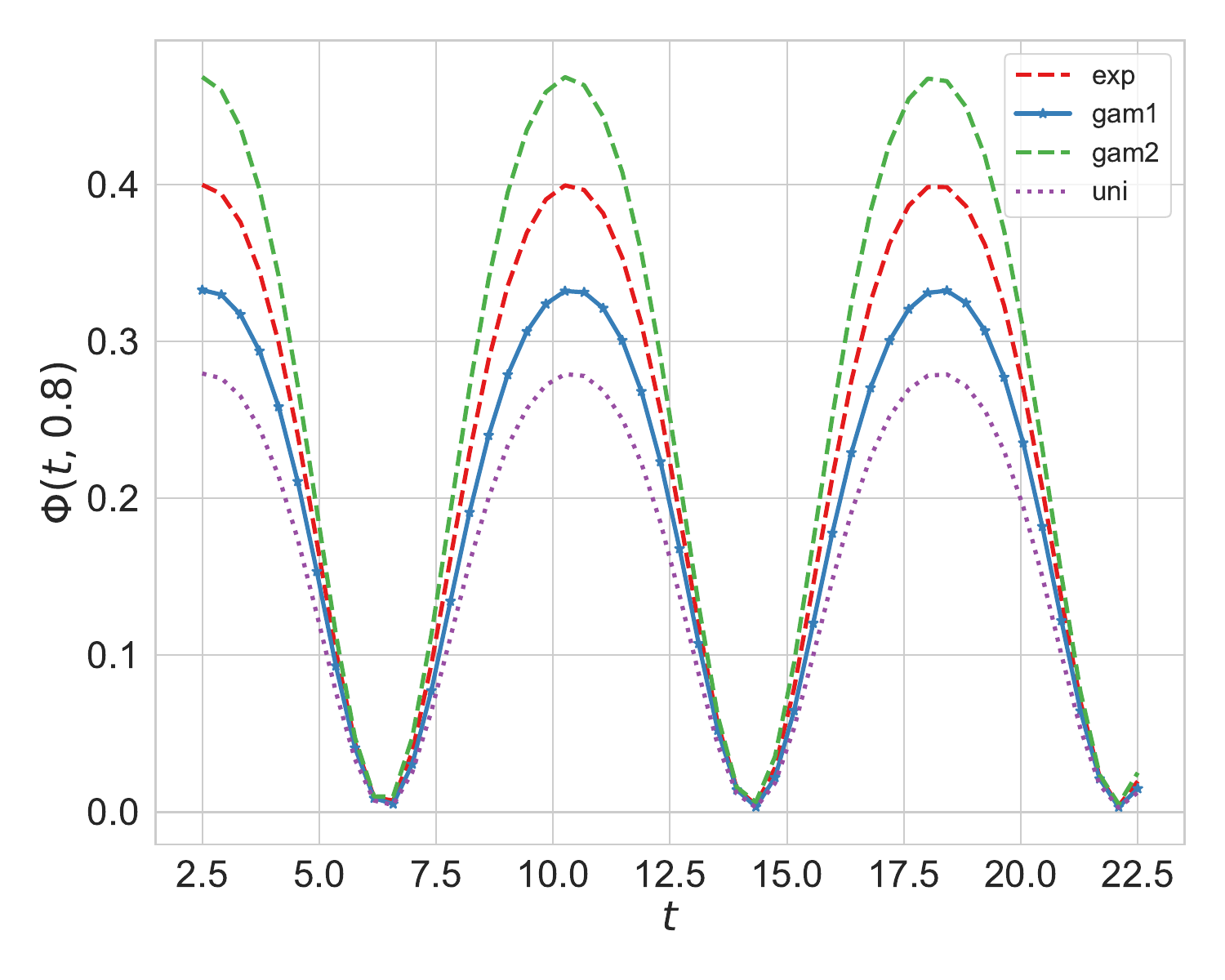}

}\subfloat[]{\includegraphics[scale=0.33]{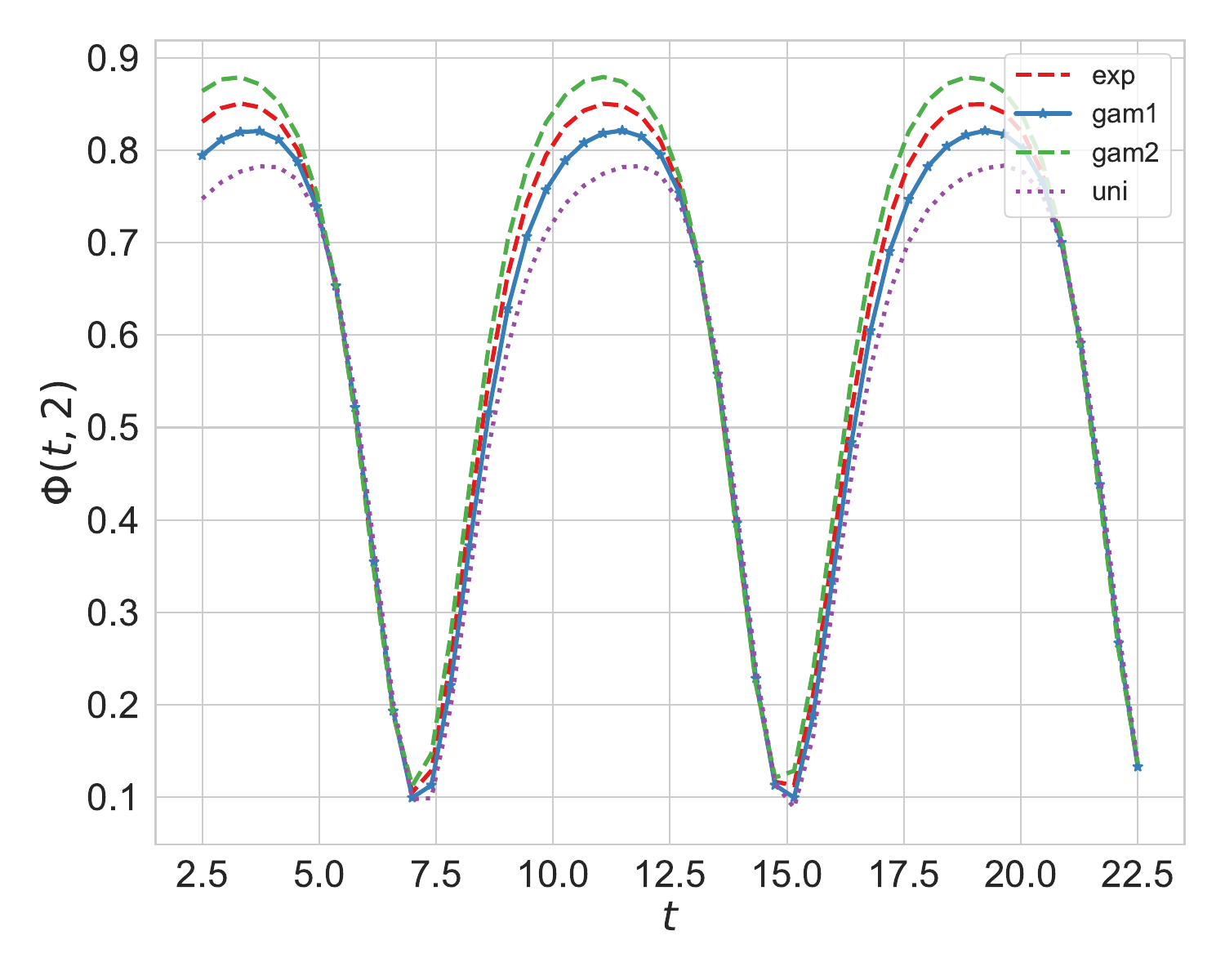}

}

\subfloat[]{\includegraphics[scale=0.33]{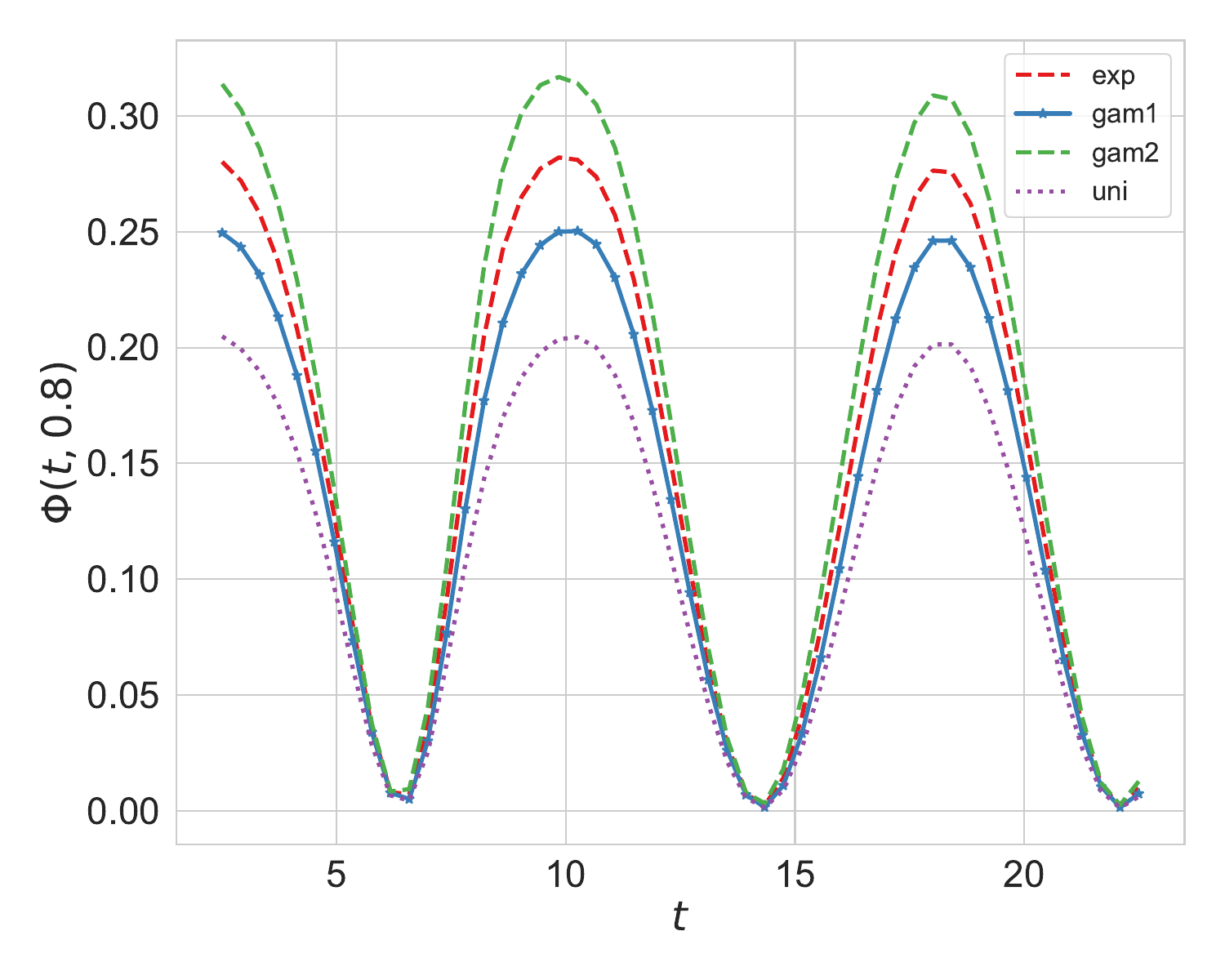}

}\subfloat[]{\includegraphics[scale=0.33]{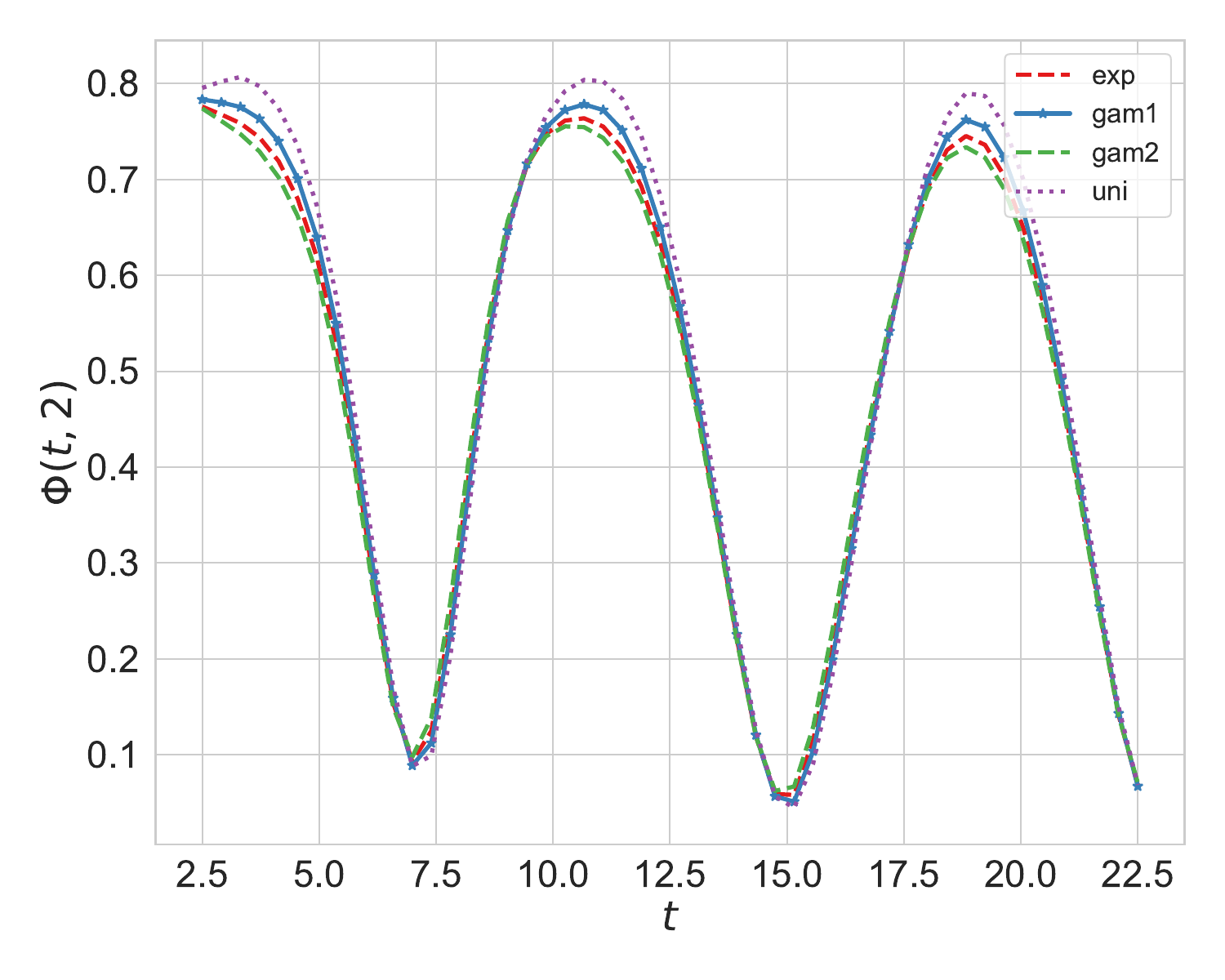}

}

\caption{The impact of preemption probability $\theta(t)$ and processing time distribution
on the AoI distribution $\Phi(t,x)$. The top row (a,b) uses a constant
preemption probability $\theta(t)=1$, and the bottom row (c,d) uses a constant
preemption probability $\theta(t)=0.1$. The left column (a, c) shows results
for a tight AoI threshold of $x=0.8$, while the right column (b,d)
uses a relaxed threshold of $x=2.0$. For all the subfigures we have
a sinusoidal arrival rate $\lambda(t)=1+\sin(0.8t)$, $\mu=1.5$,
`exp'$\sim\exp(\mu)$, `uni'$\sim\text{Uniform}(0,2/\mu)$, `gam1'$\sim\text{Gamma}(\mu,1/\mu^{2})$,
and `gam2'$\sim\text{Gamma}(1/\mu,1)$.}
\label{fig:AoI-distribution-over-1} 
\end{figure}

Fig. \ref{fig:AoI-distribution-over-1} illustrates how the AoI distribution
$\Phi(t,x)=\boldsymbol{P}(\Delta(t)\leq x)$ responds to sampling
profile, the variability of processing times, and the preemption probability
$\theta(t)$. Two AoI thresholds ($x=0.8$ and 2) are considered.
Subplots (a)-(d) in Fig. \ref{fig:AoI-distribution-over-1} show four
processing distributions that share the same mean but differ in variance,
where `gam1' is a log-concave Gamma distribution, and `gam2' is log-convex\footnote{Recall that a Gamma distribution with PDF $f(x;\alpha,\beta)=\frac{x^{\alpha-1}e^{-x/\beta}}{\Gamma(\alpha)\beta^{\alpha}}$
is log-convex when $0<\alpha<1$ and is log-concave when $\alpha>1$.}. In both Figs. \ref{fig:AoI-distribution-over-1}(a) and (b), we
have the preemption probability $\theta(t)=1$, where the `gam2' service
time achieves the highest probability $\Phi(t,x)$. In Figs. \ref{fig:AoI-distribution-over-1}(c)
and (d), we have a lower preemption probability $\theta(t)=0.1$.
We find that when considering $\Phi(t,0.8)$, the `gam2' distribution
still achieves the highest probability. However, for the larger threshold
$x=2$, the `gam2' has the lowest probability $\Phi(t,2)$, while
the uniform distribution achieves the highest. This indicates that
no single service time distribution is uniformly optimal for all AoI
thresholds. Moreover, when comparing Figs. \ref{fig:AoI-distribution-over-1}(a) and (c), we find that
by lowering the preemption probability, the AoI distributions $\Phi(t,0.8)$
for all the four service time distributions decrease. This implies
that when the threshold is low, frequent preemption may help the system
to achieve a small AoI more likely. When comparing Figs. \ref{fig:AoI-distribution-over-1}(b) and (d),
frequent preemption benefits `gam2' for larger thresholds but is detrimental
for the uniform distribution. This is because under uniform service
times, preemption interrupts processing, causing the server to restart
the service with a larger remaining workload.

\begin{figure}
\centering\subfloat[]{\includegraphics[scale=0.33]{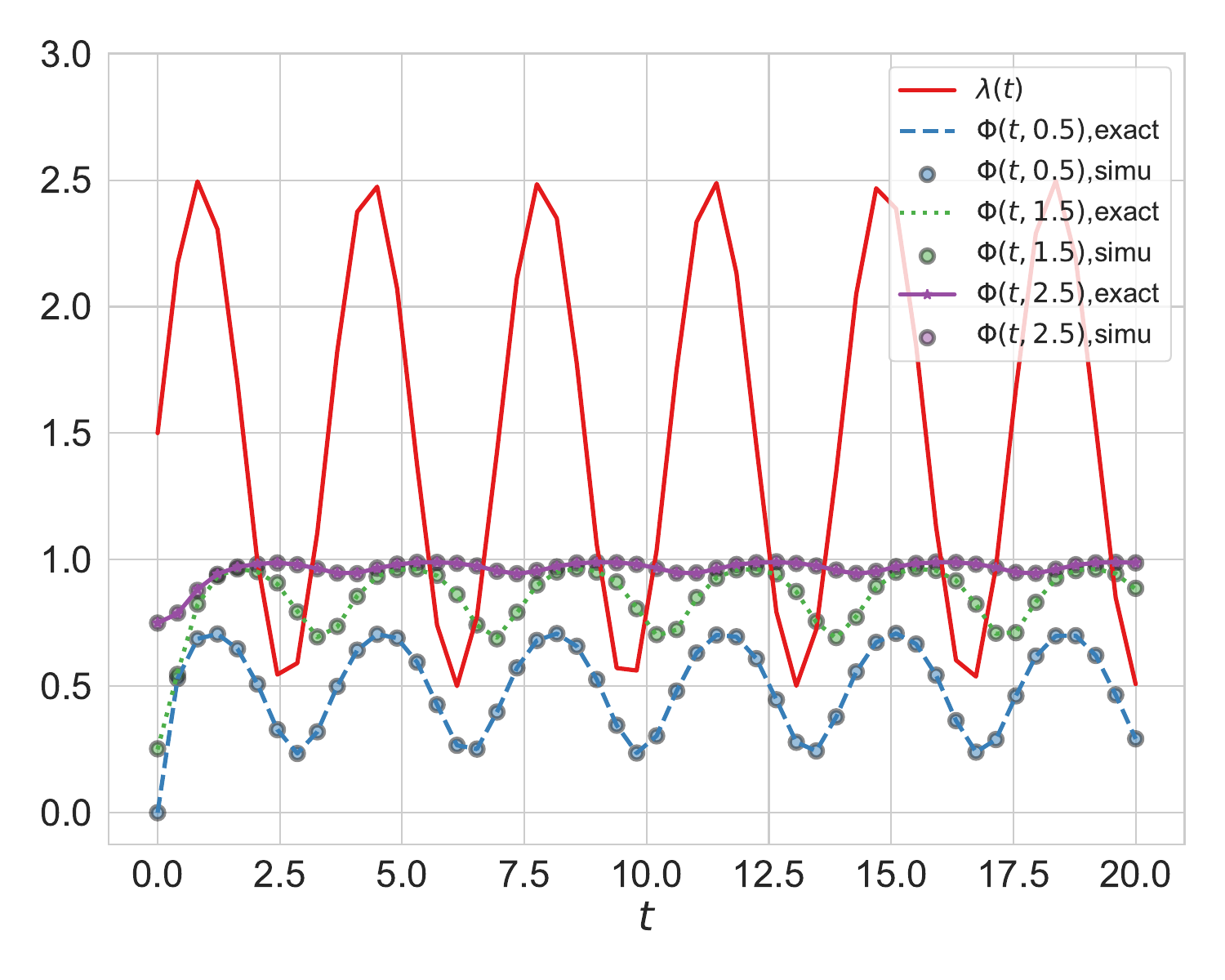}

}\subfloat[]{\includegraphics[scale=0.33]{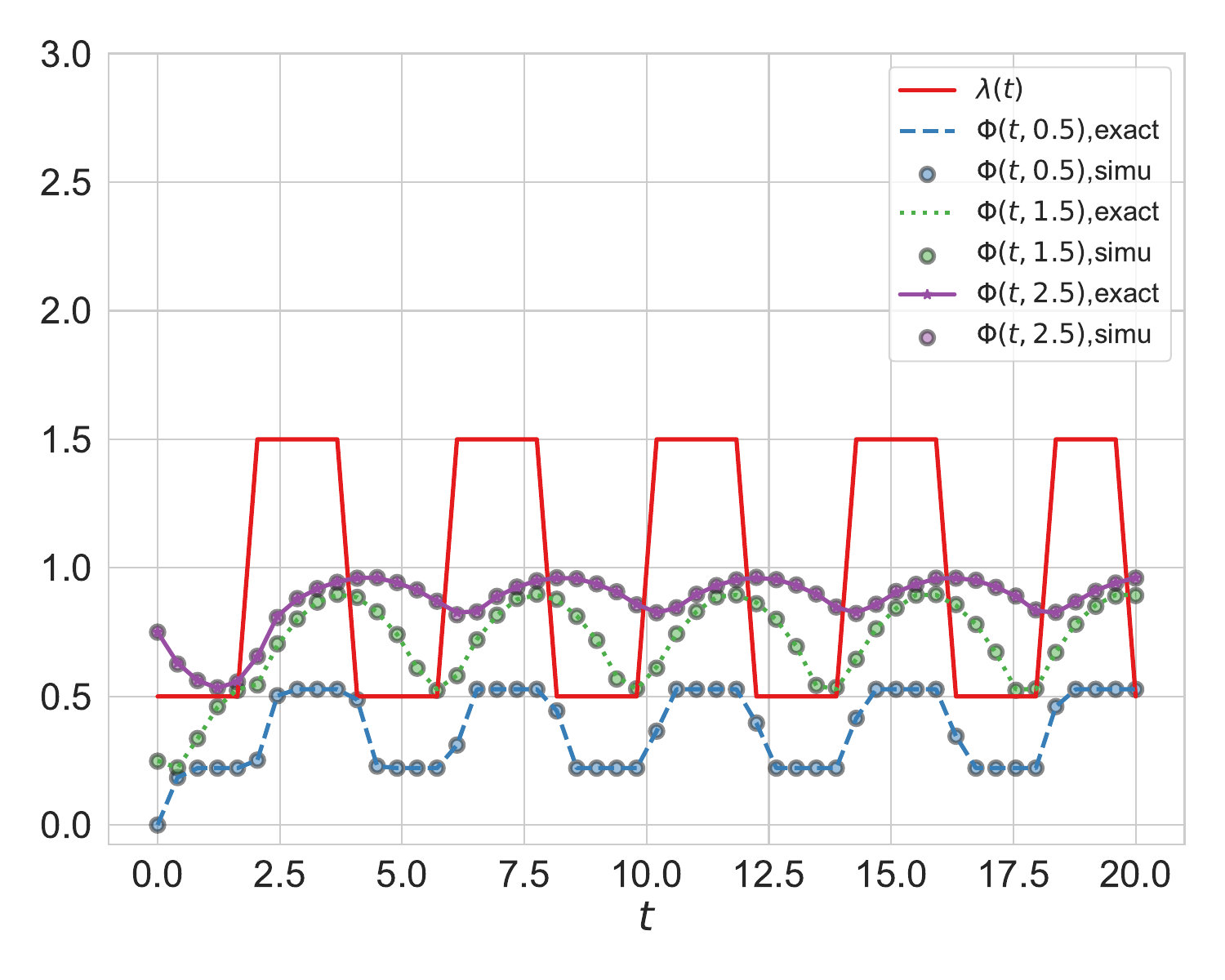}

}\caption{AoI distribution with negligible processing time. Subfigure (a) uses
a sinusoidal arrival rate, $\lambda(t)=1.5+\sin(1.8t)$, while 
Subfigure (b) uses a piecewise constant (square wave) rate that alternates
between 1.5 and 0.5 every three time units.}\label{fig:AoI-distribution-with-1}

\end{figure}

We examine the case of negligible processing time in Fig. \ref{fig:AoI-distribution-with-1},
where the two subfigures have different sampling rates. From both
Figs. \ref{fig:AoI-distribution-with-1}(a) and (b), we find that
the simulation results match with the analytical solution in Section
\ref{subsec:Model-of-no}, validating our derivations. Moreover, we
find that the AoI distribution $\Phi(t,x)$ with a smaller threshold
$x$ is more sensitive to the change of sampling rate. For instance,
when the sampling rate increases, the distribution $\Phi(t,0.5)$
increases first, followed by $\Phi(t,1.5)$ and $\Phi(t,2.5)$. This
occurs because, for larger thresholds, the AoI distribution depends
more on the cumulative sampling over the interval $[t-x,t]$, as observed
from Equation (\ref{eq:2-5-5}), making it less responsive to instantaneous
changes in the sampling rate.

\subsection{Numerical study for the stationary system}

\label{subsec:Numerical-study-for-1}

\begin{figure}[t!]
\centering\subfloat[]{\includegraphics[scale=0.33]{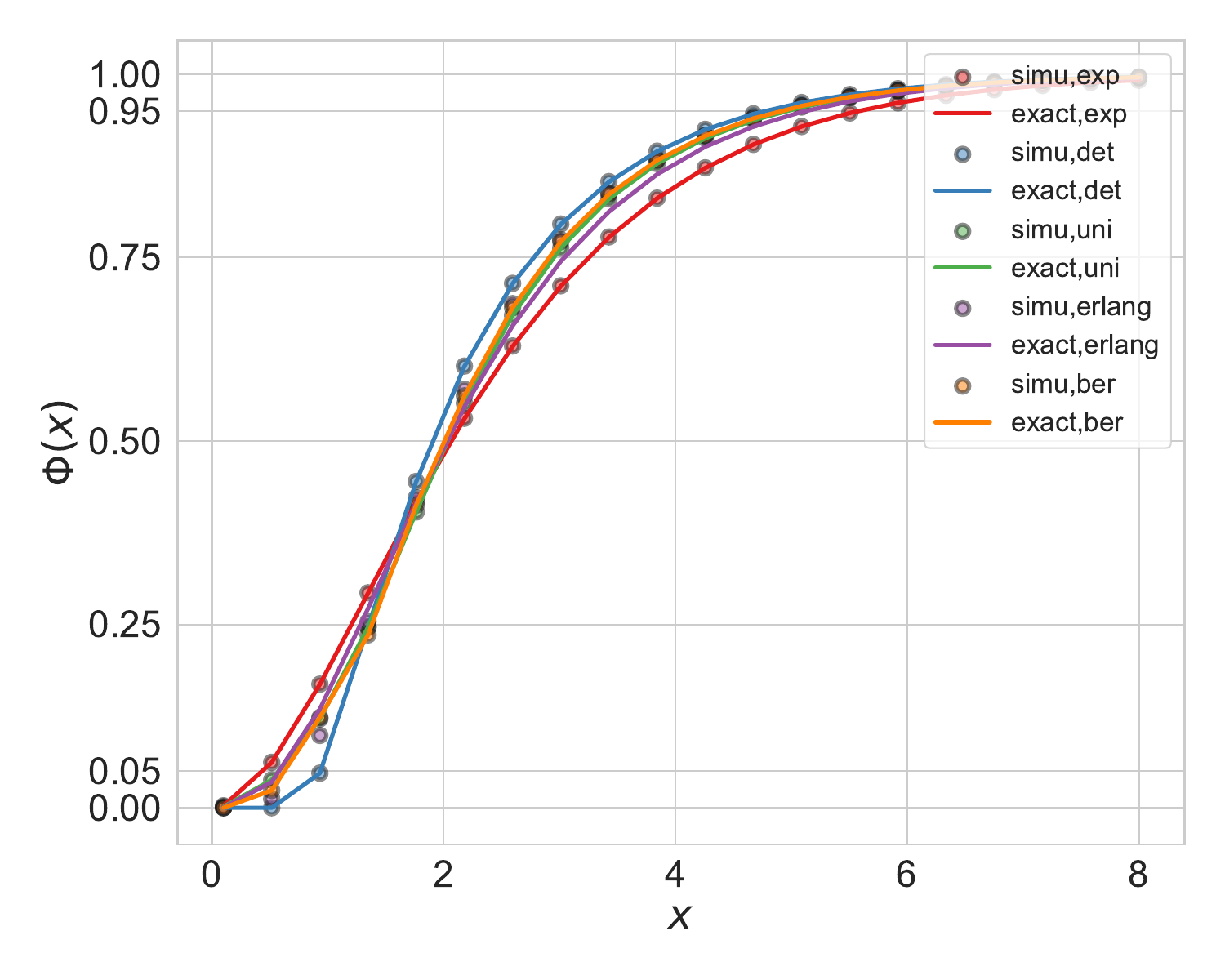}

}\subfloat[]{\includegraphics[scale=0.33]{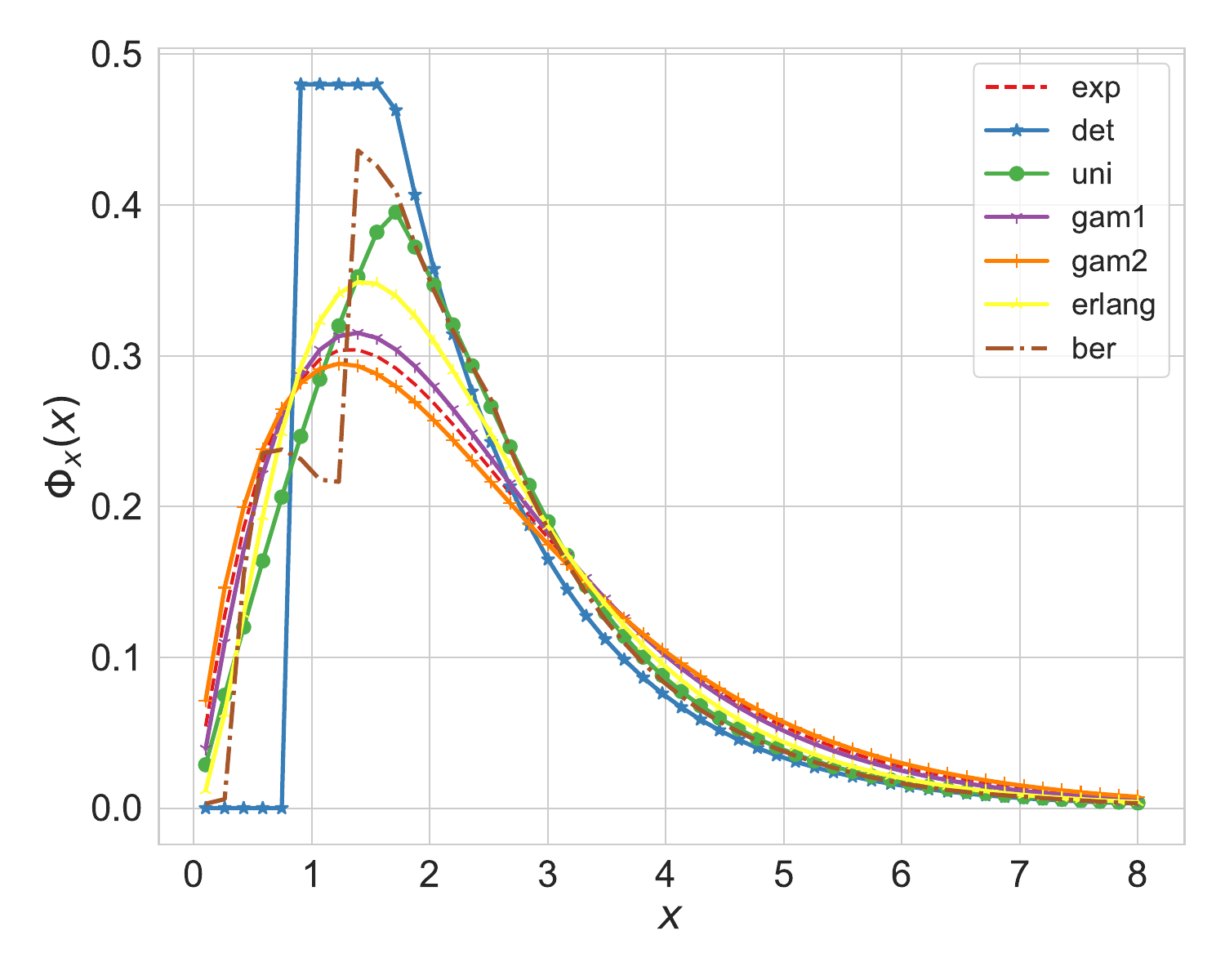}

}

\subfloat[]{\includegraphics[scale=0.33]{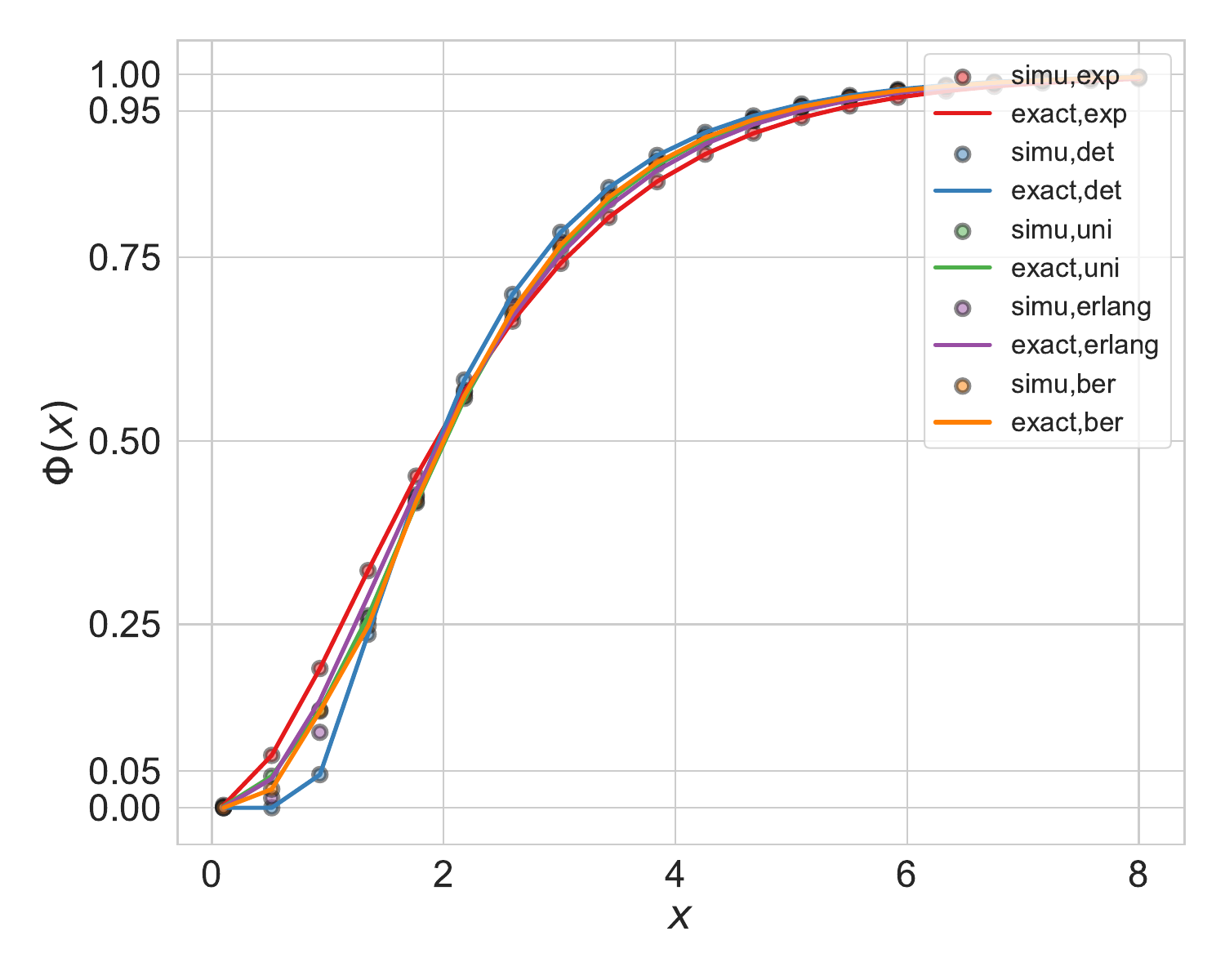}

}\subfloat[]{\includegraphics[scale=0.33]{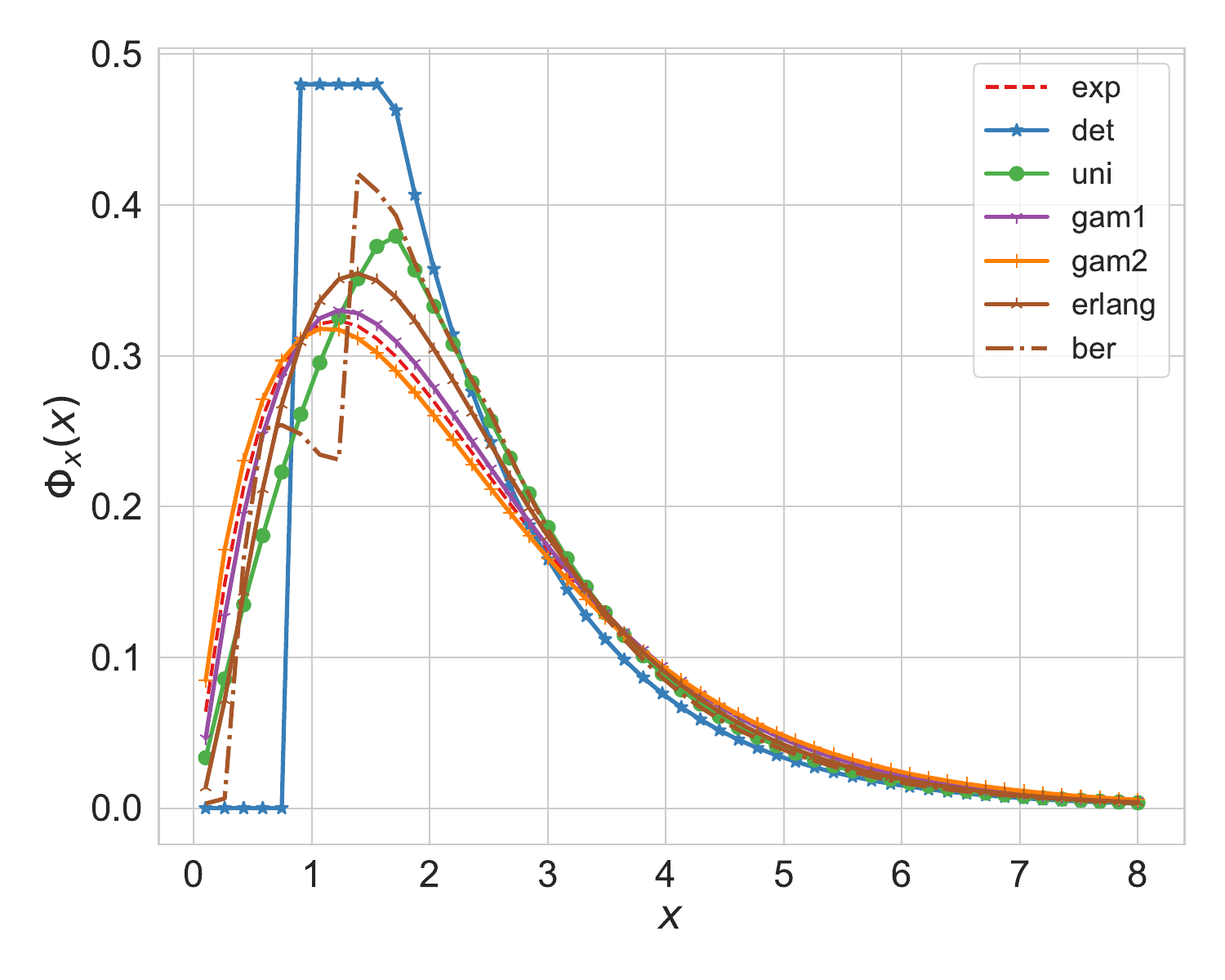}

}

\caption{Validation of the analytical AoI distribution (solid lines) against
simulation results (points) for the stationary $M/G/1/1$ system.
The top row (a,b) shows the system without preemption ($\theta=0$),
while the bottom row (c,d) shows the system with probabilistic preemption
($\theta=0.3$). The left column (a,c) displays the CDFs, and the
right column (b,d) displays the corresponding PDFs. Common parameters
for all plots are $\lambda=0.8$ and $\mu=1.2$. The abbreviations
for the processing time distributions are: `exp'$\sim\exp(\mu)$,
`det'$\sim1/\mu$, `uni'$\sim\text{Uniform}(0,2/\mu)$, `gam1'$\sim\text{Gamma}(\mu,1/\mu^{2})$,
`gam2'$\sim\text{Gamma}(1/\mu,1)$, `erlang'$\sim\text{Erlang}(5,1/5\mu)$, 'ber'$\sim\boldsymbol{P}(W=\frac{1}{2\mu})=\boldsymbol{P}(W=\frac{3}{2\mu})=\frac{1}{2}$.}
\label{fig:Comparison-of-the-1} 
\end{figure}

We validate the AoI distribution derived for the $M/G/1/1$ system
in Theorem \ref{thm:The-LST-of} by comparing it with simulation results.
As shown in Fig. \ref{fig:Comparison-of-the-1}(a), the simulation
outcomes (points) precisely match our analytical expressions (solid
lines), confirming the accuracy of the derivation. The plot also reveals
an interesting trade-off: exponential processing times yield a higher
probability of achieving a very low AoI, but may result in a smaller
probability of staying below larger AoI thresholds compared to other
distributions. In Fig. \ref{fig:Comparison-of-the-1}(b), we present
the PDF of the AoI under various processing time distributions. The
results demonstrate that the AoI distribution in the $M/G/1/1$ system
is generally right-skewed, suggesting that while low AoI values are
achievable, there is a non-negligible probability of higher age values
due to occasional longer processing times.

We further verify the AoI distribution in the $M/G/1/1$ system with
nonzero preemption probabilities ($\theta=0.3$). The analytical results
in Fig. \ref{fig:Comparison-of-the-1}(c) are obtained by numerically
evaluating the Bromwich integral. Again, the comparison in Fig. \ref{fig:Comparison-of-the-1}(c)
shows a close match between the analytical solutions and the simulation
data, confirming the correctness of our LST derivation. Fig. \ref{fig:Comparison-of-the-1}(d)
shows that the AoI distribution remains right-skewed across all processing
time distributions, even when preemption is allowed. This consistent
skewness reveals a structural property of these systems: while low
AoI values are most probable, there is always a non-negligible tail
corresponding to longer age values.

\begin{figure}[t!]
\centering \subfloat[]{\includegraphics[scale=0.33]{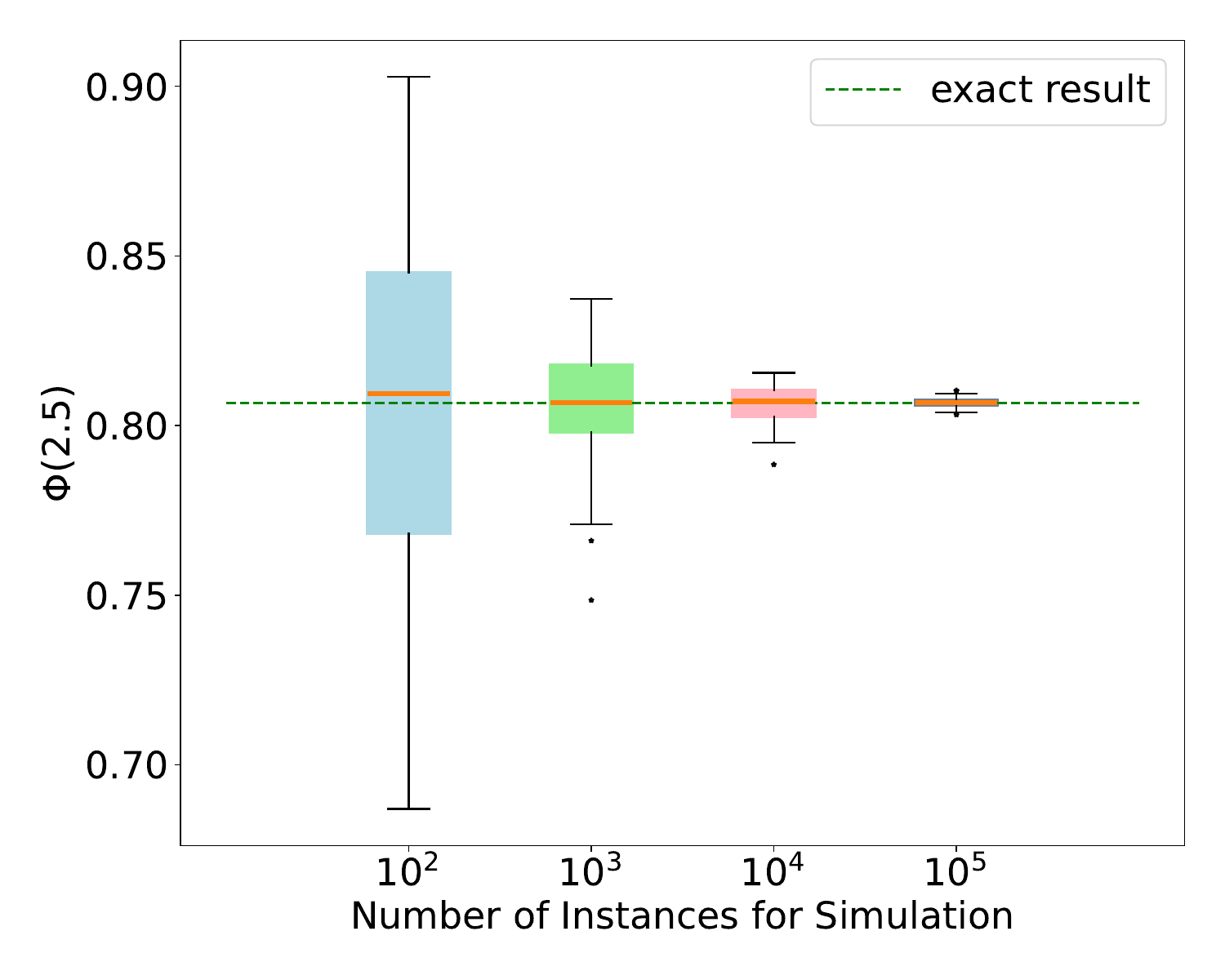} 

}\subfloat[]{\includegraphics[scale=0.33]{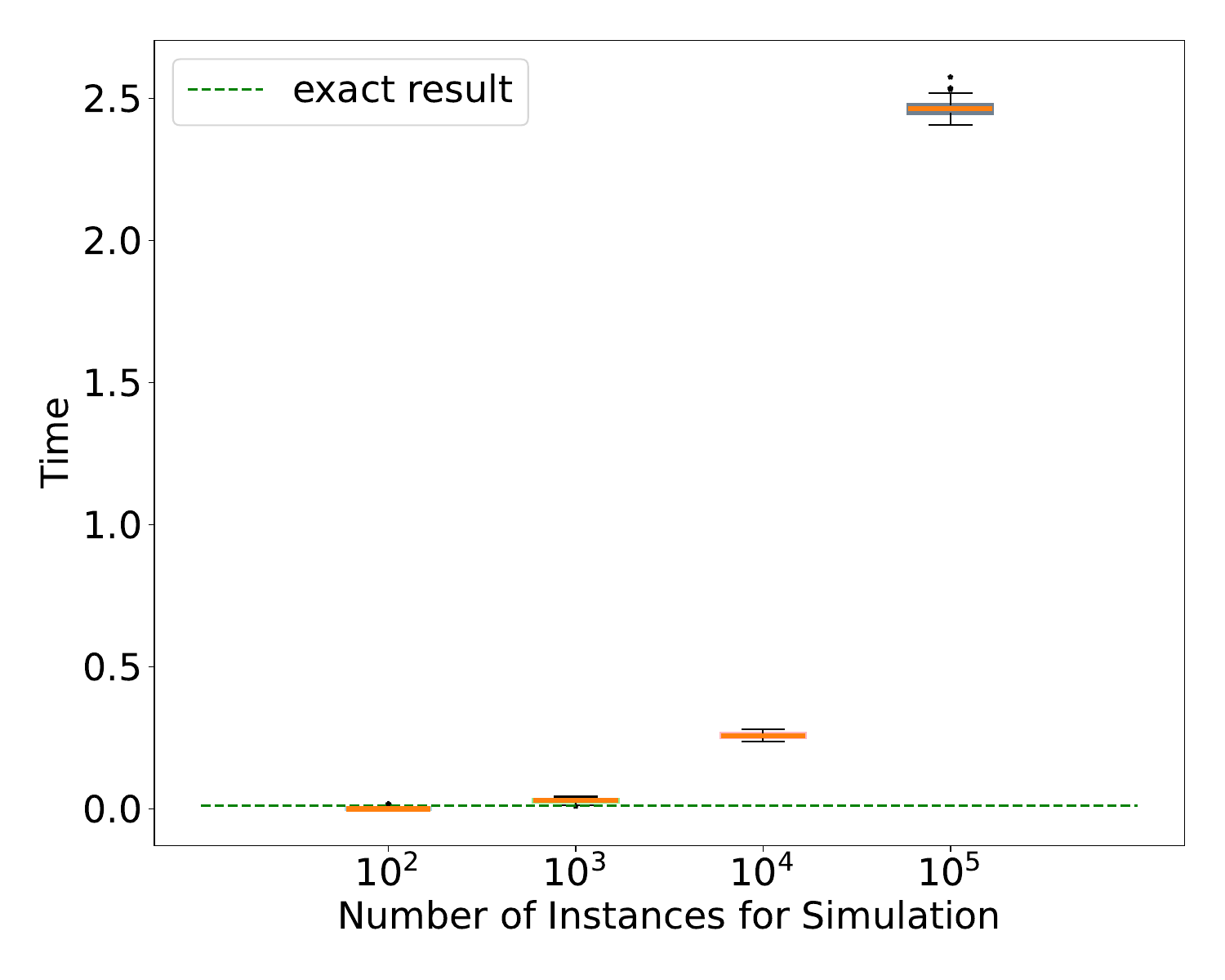}

}\caption{Comparison of simulation with exact result. (a) Comparison of AoI
distribution. (b) Comparison of computation time (in seconds). For
both subfigures, we have $\lambda=2$, $\mu=1.2$, $\theta=0.5$,
and the processing time is Erlang distributed with parameters $(5,1/5\mu)$.
The boxplots for the simulation results are from 100 runs of simulation
of different numbers of instances. The box spans from the first quartile
(Q1) to the third quartile (Q3), with the line inside the box representing
the median. The whiskers extend from the edges of the box to the furthest
data points within 1.5 times the interquartile range (IQR). Data points
beyond the whiskers are considered outliers (also called fliers).}
\label{fig:comparison of simulation with exact result2} 
\end{figure}

In Fig. \ref{fig:comparison of simulation with exact result2}, we
compare the simulation and numerical methods for calculating the distribution
$\Phi(2.5)$. As shown in Fig. \ref{fig:comparison of simulation with exact result2}(a),
the simulation results exhibit large variation when using a small
number of instances. Fig. \ref{fig:comparison of simulation with exact result2}(b)
further shows that the computation time required by the numerical
method is negligible compared to that of the simulation, implying
the high efficiency of the Bromwich integral approach.

\begin{figure}[t!]
\centering \subfloat[]{\includegraphics[scale=0.33]{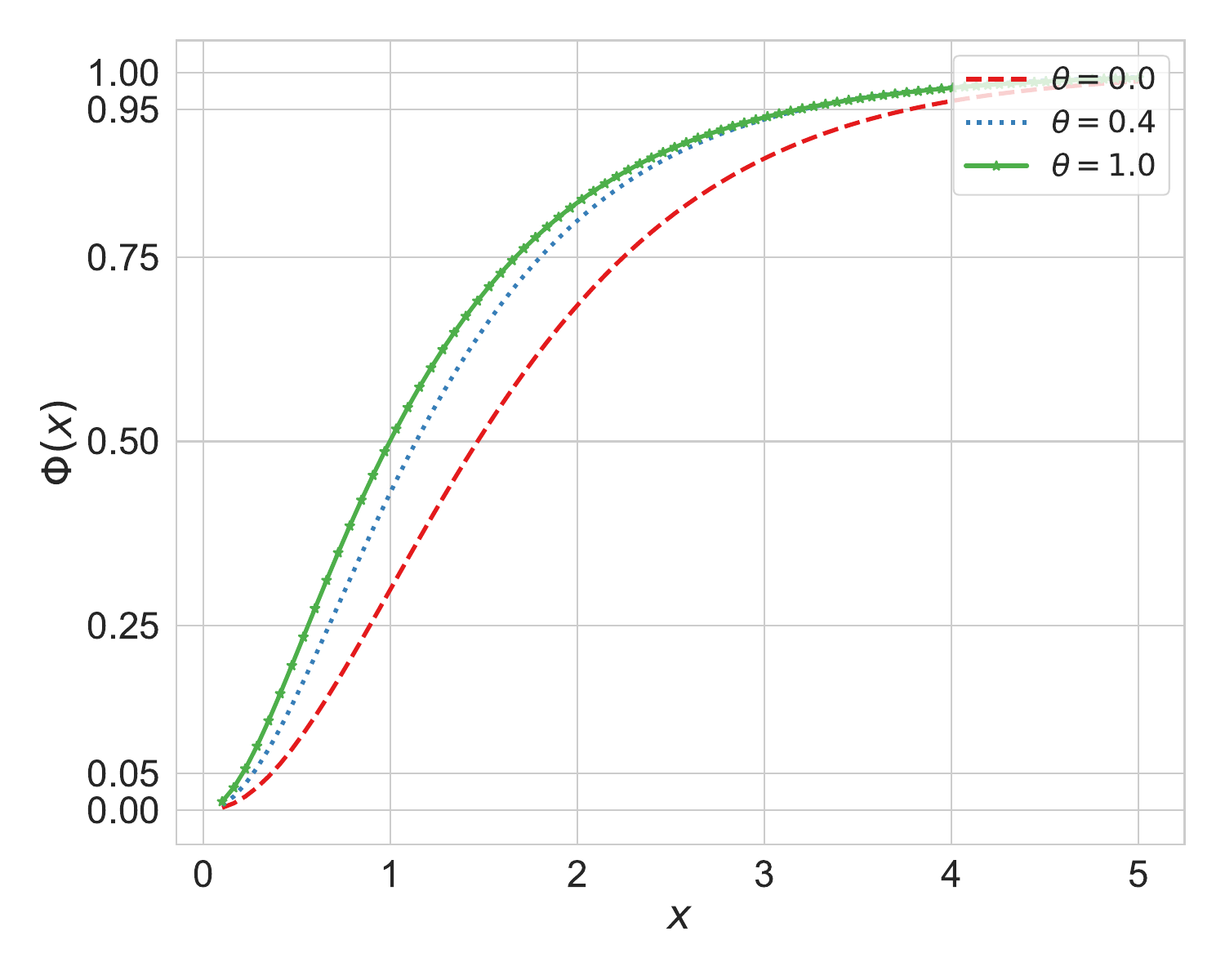} 

}\subfloat[]{\includegraphics[scale=0.33]{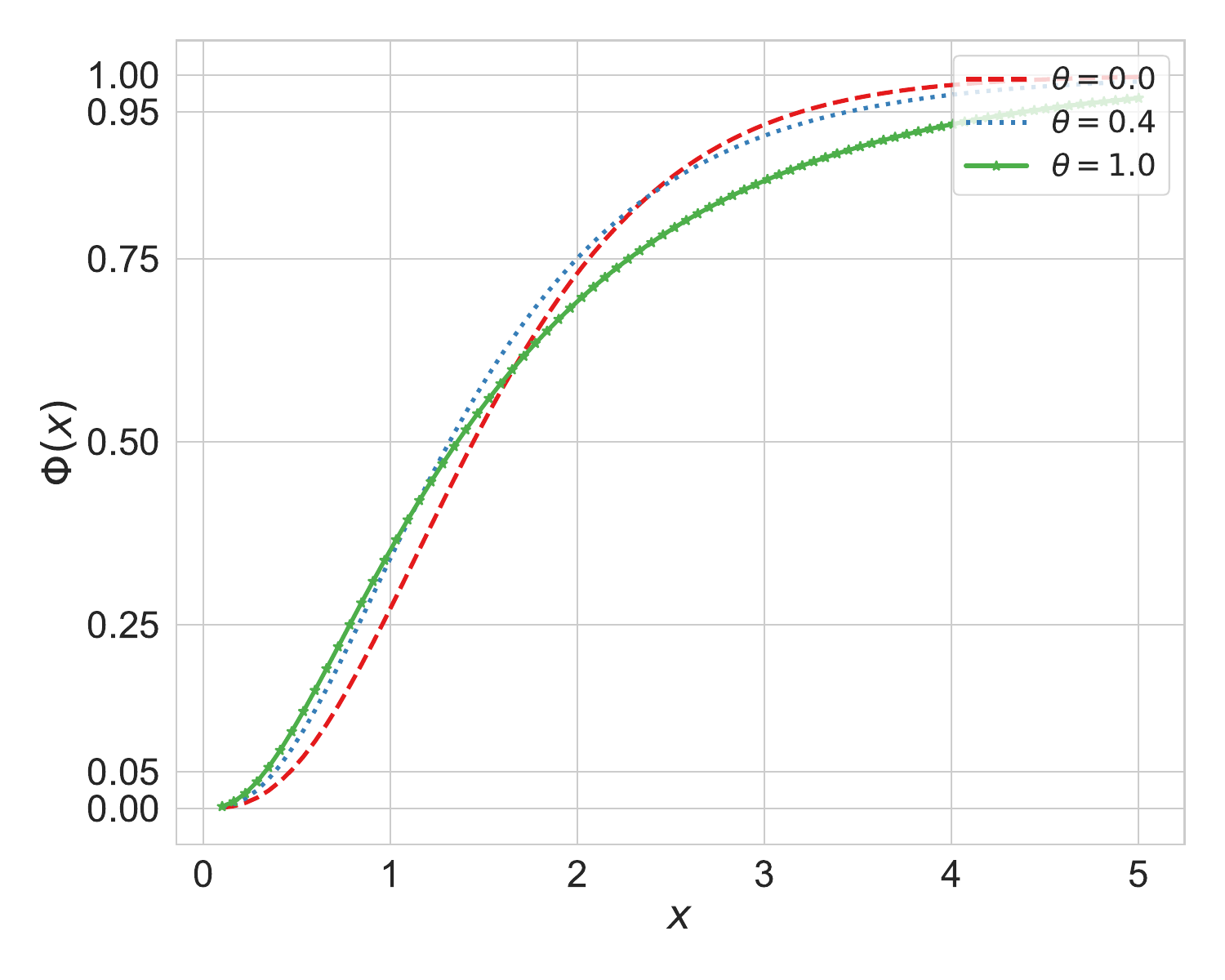}

}\caption{The CDF of AoI for different preemption probabilities $\theta$. (a)
Processing time follows $\text{Gamma}(\mu,1/\mu^{2})$. (b) Processing
time follows $\text{Erlang}(2,1/2\mu)$. Common parameters are $\lambda=3.5$
and $\mu=1.2$. }
\label{fig:AoI-distribution-with} 
\end{figure}

In Fig. \ref{fig:AoI-distribution-with}, we present the AoI distribution
under different preemption probabilities. When the processing time
follows a Gamma distribution $Gamma(\mu,1/\mu^{2})$, a higher preemption
probability uniformly achieves a higher AoI probability for all thresholds
$x$, as we observed from Fig. \ref{fig:AoI-distribution-with}(a).
However, when the processing time is Erlang distributed with parameters
$Erlang(2,1/2\mu)$, Fig. \ref{fig:AoI-distribution-with}(b) shows
that when the threshold $x$ is small, setting $\theta=1$ (full preemption)
yields the highest $\Phi(x)$ (i.e., the lowest AoI violation probability).
For intermediate thresholds (e.g., $1.2\leq x\leq2.3$), a moderate
preemption probability, such as $\theta=0.4$, performs better than
both the non-preemptive ($\theta=0$) and fully preemptive ($\theta=1$)
schemes. When $x\geq2.3$, the non-preemptive case ($\theta=0$) gives
the largest $\Phi(x)$. This observation indicates that the optimal
preemption probability is threshold-dependent, and should be chosen
according to the specific AoI violation requirement.

\begin{figure}[t!]
\centering\subfloat[]{\includegraphics[scale=0.23]{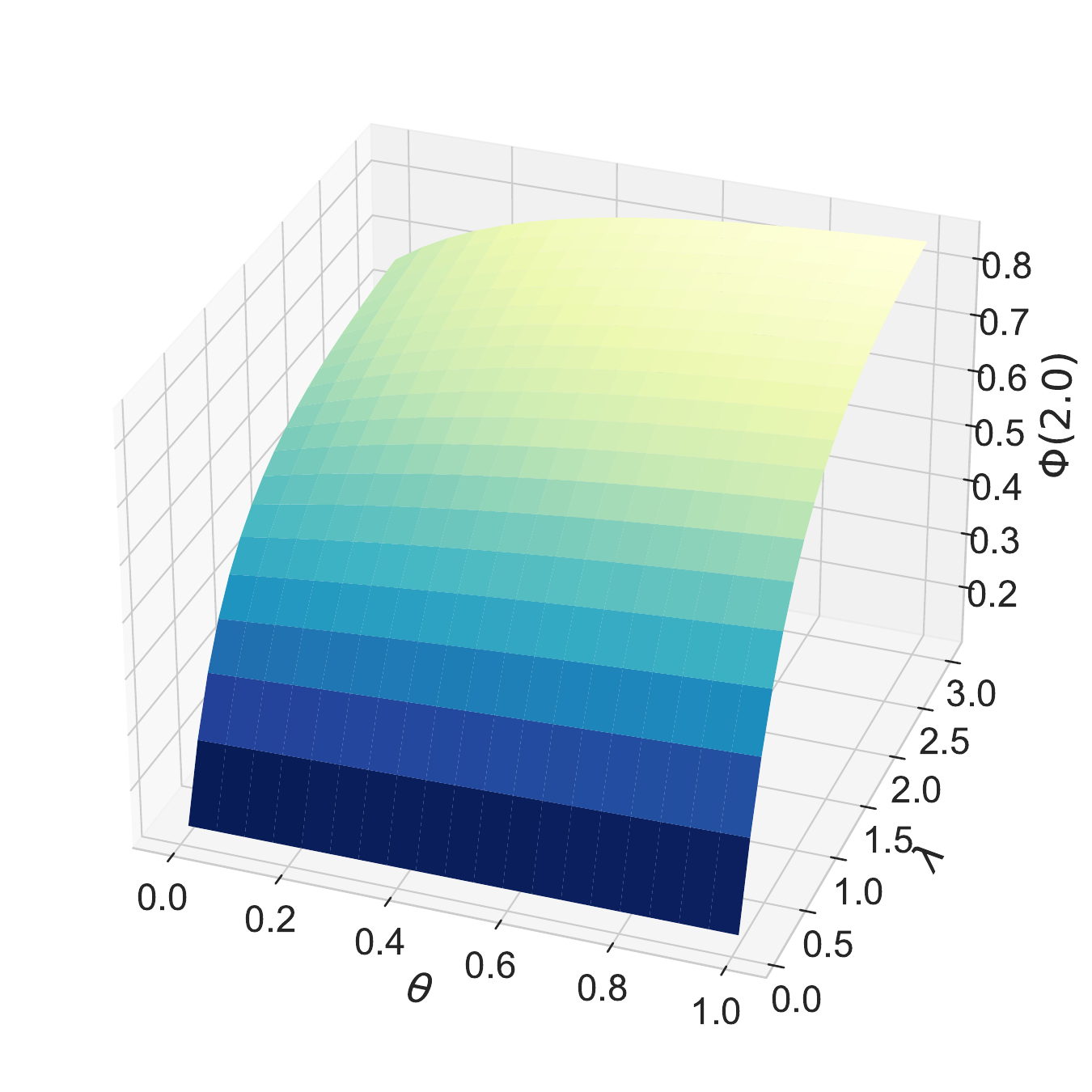}

}\subfloat[]{\includegraphics[scale=0.23]{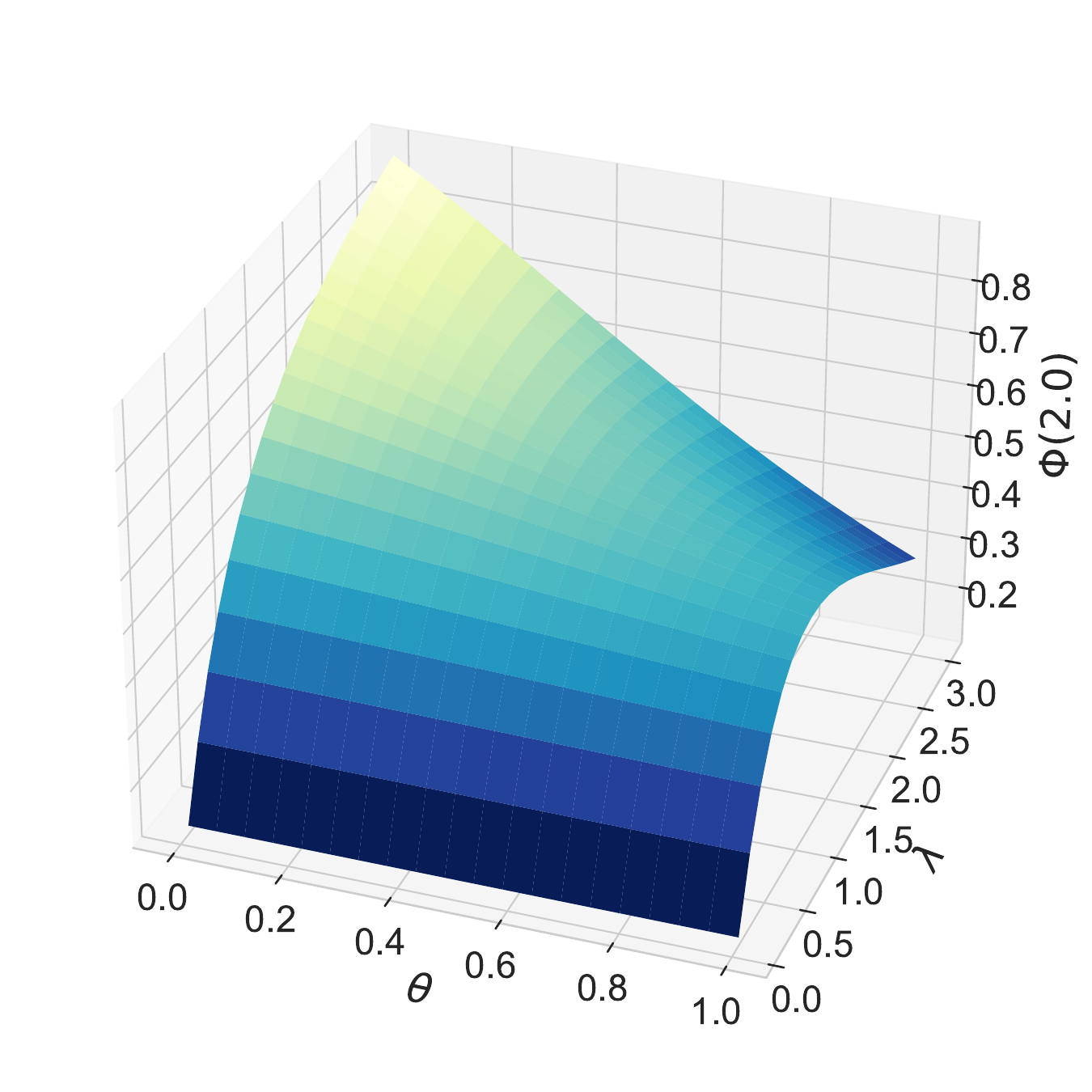}

}\subfloat[]{\includegraphics[scale=0.23]{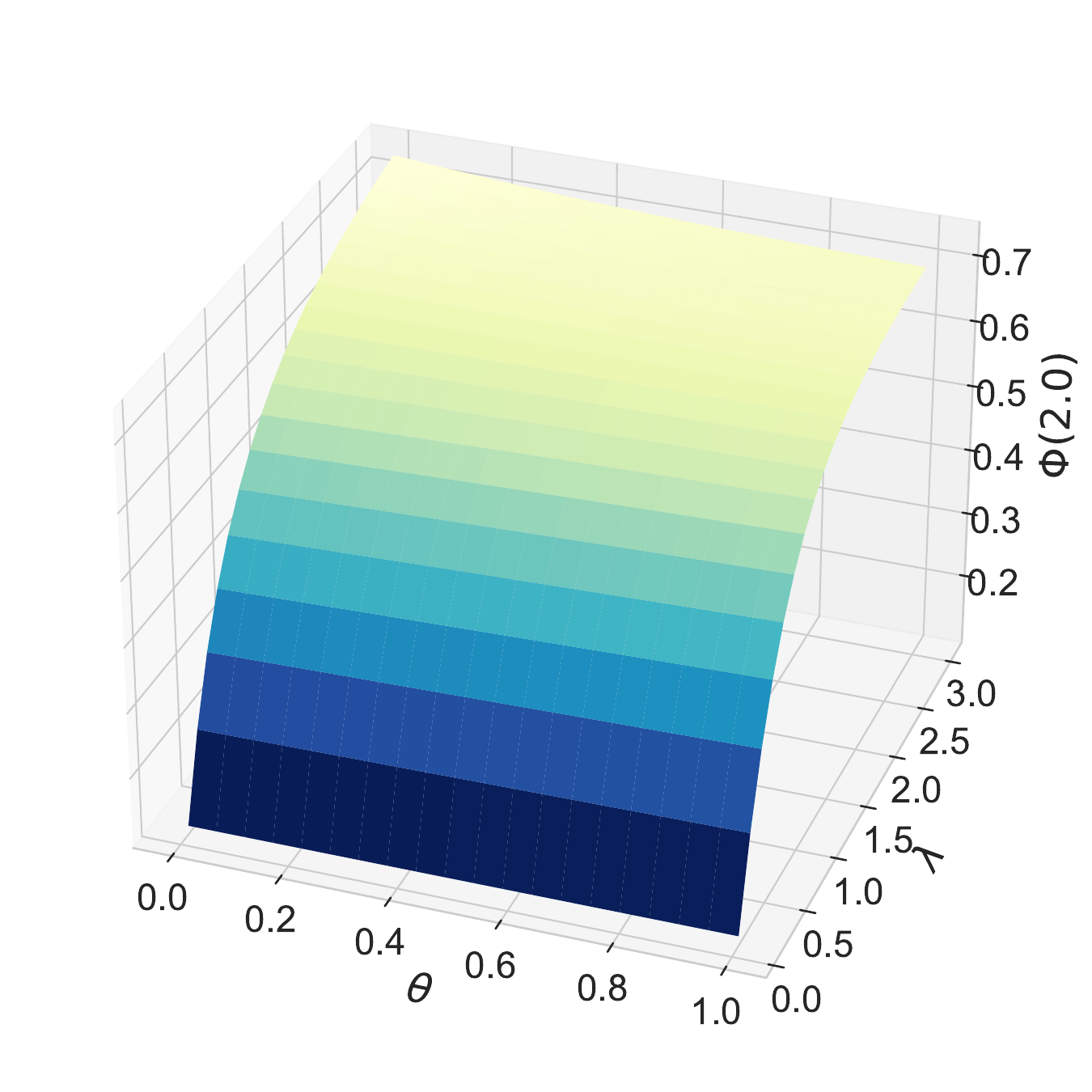}

}

\caption{The AoI distribution $\Phi(x)$ as a function of the sampling rate
$\lambda$ and preemption probability $\theta$ for a fixed AoI threshold
$x=2.0$. Each subplot shows a different processing time distribution:
(a) Exponential, (b) Deterministic, and (c) Uniform. For all the subplots
we have average processing rate $\mu=1.2$.}
\label{fig:Comparison-of-the} 
\end{figure}

Next, we investigate how the preemption probability $\theta$ and
the sampling rate jointly affect the AoI probability $\Phi(x)$ for
a fixed threshold $x$. The numerical results are shown in Fig. \ref{fig:Comparison-of-the}.
The numerical results for a threshold of $x=2$ are shown in the 3D
surface plots in Fig. \ref{fig:Comparison-of-the}. Each subplot corresponds
to a different processing time distribution.

Fig. \ref{fig:Comparison-of-the}(a) shows that under exponential
processing times, increasing the preemption probability $\theta$
always increases the likelihood of meeting the AoI threshold. This
aligns with Corollary \ref{cor:When-the-processing}, which states
that a fully preemptive system ($\theta=1$) is optimal for exponential
processing times. In contrast, for deterministic processing times,
Fig. \ref{fig:Comparison-of-the}(b) shows that this is no longer
true. When the sampling rate is high, a large preemption probability
can be counterproductive, actually decreasing the probability of meeting
the threshold. This happens because frequent preemption of a deterministic-length
task can significantly prolong its effective completion time, increasing
the chance of a large AoI. Finally, Fig. \ref{fig:Comparison-of-the}(c)
illustrates that for uniformly distributed processing times, the AoI
distribution $\Phi(x)$ is not very sensitive to changes in the preemption
probability.

\subsection{Optimal design for sampling and preemption rates}

\label{subsec:Optimal-design-for}

The preceding sections (Sections \ref{sec:Solution-derivation-approach}
and \ref{sec:Discussion-on-the}) have focused on the ``forward problem'':
deriving the AoI distribution $\boldsymbol{P}(\Delta(t)\leq x)$ when
the system parameters are known. In many practical system designs,
however, the interest lies in the ``inverse problem''. The
goal is to find the system parameters, such as sampling rate $\lambda(t)$
and the preemption probability $\theta(t)$ that can achieve a desired AoI
performance level. For example, system operators may require the probability
that the AoI is below a certain threshold $x_{i}$ in the $i^{th}$ time
interval is at least $p_{i}$. Then what are the necessary
conditions on the arrival rate $\lambda(t)$ and preemption probability
$\theta(t)$ to achieve this target? Optimal designs of sampling and
preemption have been studied in prior work like \cite{bedewy2021optimal,zhu2024optimizing}. We now extend the discussion to the time-varying setting. We assume
that the processing time distribution $F(x)$ is fixed and known,
since transmission and processing times are typically determined by
the communication channel and processor capacity.

In the time-varying case, the AoI distribution $\boldsymbol{P}(\Delta(t)\le x|\lambda(\cdot),\theta(\cdot))$
is dependent on time. An objective of natural design is to achieve
the desired performance using minimal resources, such as minimizing
the total number of packets sampled. In many applications, these performance
requirements may change over time, within a defined operational period
$[t_{0},t_{n})$. For instance, a system might demand a low AoI threshold
($x_{i}$) with high probability ($p_{i}$) during critical time intervals,
while allowing for more relaxed constraints during others to conserve
resources. This leads to an optimization problem, which can be stated
as: 
\begin{align}
\min_{\lambda(t),\theta(t)}\quad & \mathcal{J}=\int_{t_{0}}^{t_{n}}\lambda(s)\mathrm{d}s\\
\mbox{s.t. }\quad & \boldsymbol{P}(\Delta(t)\leq x_{i}|\lambda(\cdot),\theta(\cdot))\geq p_{i},\quad\forall t\in[t_{i},t_{i+1})\mbox{ for }i=0,...,n-1.\label{eq:constraints}
\end{align}
This optimization problem is challenging for several reasons. First,
the search space of the sampling rate function $\lambda(t)$ can be
selected from infinitely many piecewise continuous functions over
the time interval $[t_{0},t_{n})$. Second, probability $\boldsymbol{P}(\Delta(t)\leq x_{i}|\lambda(\cdot),\theta(\cdot))$
is a non-monotone function of both the sampling rate and the preemption
probability, especially when the processing time is not exponential.
Third, the probability at time $t$ depends on the entire history
of the sampling rate, making the problem difficult to solve directly.

 To overcome these challenges, we propose a heuristic
approach for the parameter optimization problem. The main idea is to design \emph{piecewise constant} system parameters
by (i) constructing an initial solution using stationary approximations,
and (ii) iteratively refining the sampling rate until all time-varying
AoI constraints are satisfied. The core steps are as follows:

\textbf{Step 1: Initialization and piecewise-constant parameterization.}
Let $\boldsymbol{t}=(t_{0},\ldots,t_{n})$ be the time points in Equation
(\ref{eq:constraints}), and let $\boldsymbol{x}=(x_{0},\ldots,x_{n-1})$
and $\boldsymbol{p}=(p_{0},\ldots,p_{n-1})$ be the corresponding
AoI thresholds and probability targets. To verify feasibility numerically,
we introduce an evaluation grid $\boldsymbol{\eta}=(\eta_{1},\ldots,\eta_{m})\subset[t_{0},t_{n})$
and check whether the AoI distribution satisfies the constraints at
these points, and set a maximum number of refinement iterations $ite_{max}$.
In addition, we specify a grid $\boldsymbol{\theta}_{grid}$ of candidate
preemption probabilities and a rate increment $\delta>0$ used in
the refinement step.

\textbf{Step 2: Split the time windows.} From Theorem \ref{thm:When-assuming-,},
we observe that the AoI distribution $\Phi(t,x)$ depends on the entire
system history over $[0,t)$ through $M(t,\infty)$, which depends
on earlier values $M(r,\infty)$ and the sampling rate $\lambda(r)$
for $r\in[0,t)$. However, this dependence weakens as the time difference
between $r$ and $t$ grows, and the distribution is most influenced
by the system behavior in the recent window $[t-x,t)$. Based on this
fact, we construct the refined sub-interval time points $\tilde{\boldsymbol{t}}=(t_{0},\;t_{1}-x_{1},\;t_{1},\;\ldots,\;t_{n-1}-x_{n-1},\;t_{n-1},\;t_{n})$ with $x_{i+1}\leq t_{i+1} - t_{i}$,
and seek piecewise constant functions $\lambda(t)$ and $\theta(t)$
such that, for each refined sub-interval $[\tilde{t}_{j},\tilde{t}_{j+1})$,
$\lambda(t)=\lambda_{j},\theta(t)=\theta_{j}$.

\textbf{Step 3: Estimate initial parameters via stationary approximation.}
For each refined sub-interval, the algorithm performs an initial grid
search for constant parameters that satisfy the relevant AoI constraints.
To make this search computationally efficient, we use the stationary
mapping $\Phi(x;\lambda,\theta)$ (e.g., obtained from Theorem~\ref{thm:The-LST-of})
to approximate the required sampling rate under constant $(\lambda,\theta)$.
In each sub-interval $[t_{k},t_{k+1}-x_{k+1})$, we perform a grid
search over $\theta\in\boldsymbol{\theta}_{grid}$ and select the
minimum $\lambda$ (together with its corresponding $\theta$) such
that the steady-state constraint for the current threshold is satisfied,
i.e., $\Phi(x_{k};\lambda,\theta)\ge p_{k}$. In each sub-interval
$[t_{k+1}-x_{k+1},t_{k+1})$, we similarly select the minimum $\lambda$
and corresponding $\theta$ such that \emph{both} the current and
the upcoming steady-state constraints hold, namely $\Phi(x_{k};\lambda,\theta)\ge p_{k}$
and $\Phi(x_{k+1};\lambda,\theta)\ge p_{k+1}$. This leads to an initial
piecewise-constant design $\{(\lambda_{j},\theta_{j})\}$ over the
refined partition $\tilde{\boldsymbol{t}}$.

\textbf{Step 4: Iteratively refine the solution under the time-varying
model.} The initial parameters obtained from the stationary approximation
are not guaranteed to satisfy the full time-varying constraints. Therefore,
the heuristic enters a refinement loop. Using the current piecewise-constant
functions $\lambda(t)$ and $\theta(t)$ on $[t_{0},t_{n})$, we evaluate
the time-varying AoI distribution on the grid $\boldsymbol{\eta}$:
for each $\eta_{i}$, let $k$ be such that $\eta_{i}\in[t_{k},t_{k+1})$,
and compute $\Phi(\eta_{i},x_{k})$. If a violation $\Phi(\eta_{i},x_{k})<p_{k}$
is detected, we locate the refined sub-interval index $s$ such that
$\eta_{i}\in[\tilde{t}_{s},\tilde{t}_{s+1})$ and increase the local
sampling rate by $\lambda_{s}\gets\lambda_{s}+\delta$. The algorithm
repeats this evaluation--adjustment procedure until either all constraints
are satisfied on $\boldsymbol{\eta}$ (feasible design found) or the
iteration counter reaches $ite_{max}$ (no feasible solution is returned).

\begin{figure}[t!]
\centering\subfloat[]{\includegraphics[scale=0.33]{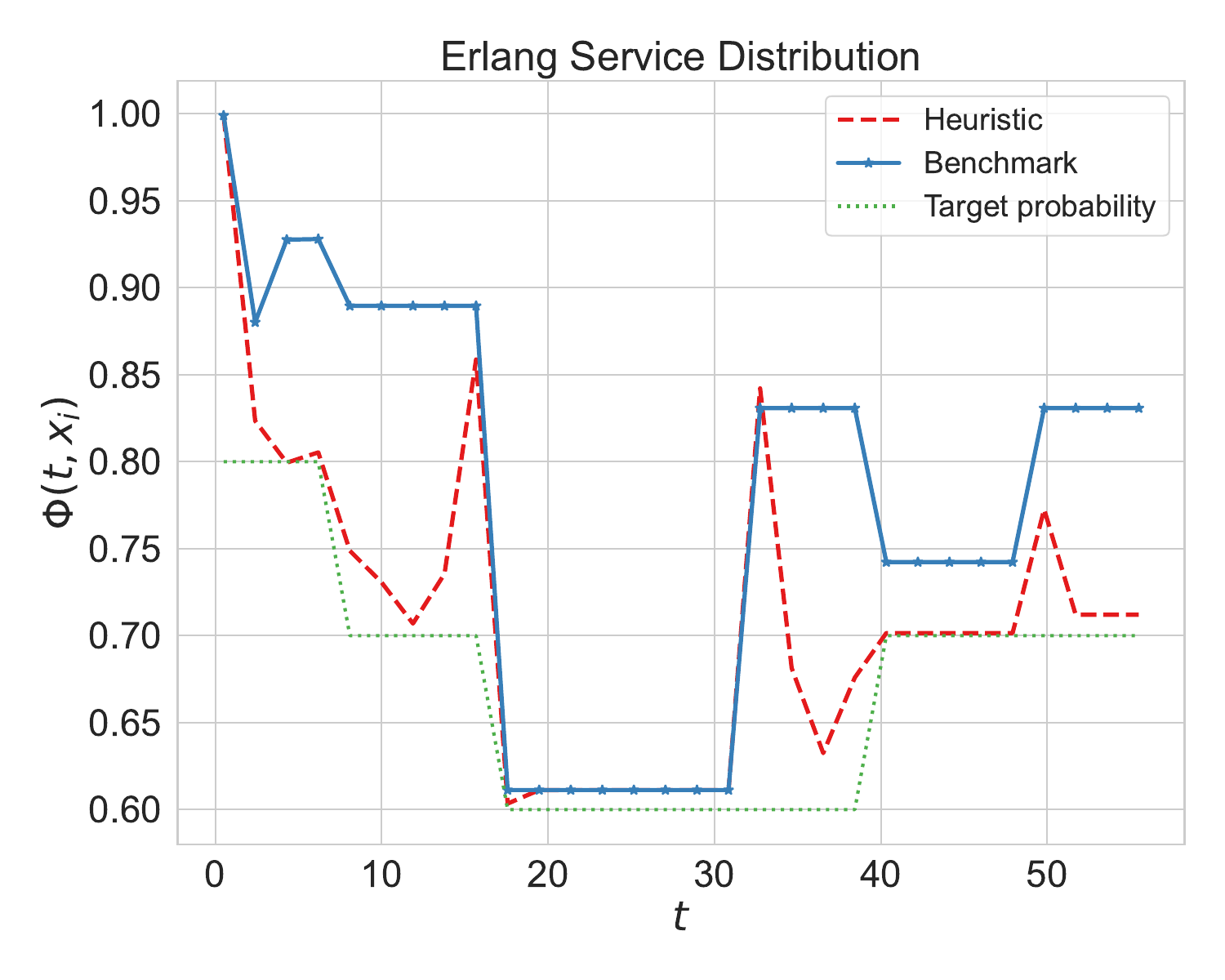}

}\subfloat[]{\includegraphics[scale=0.33]{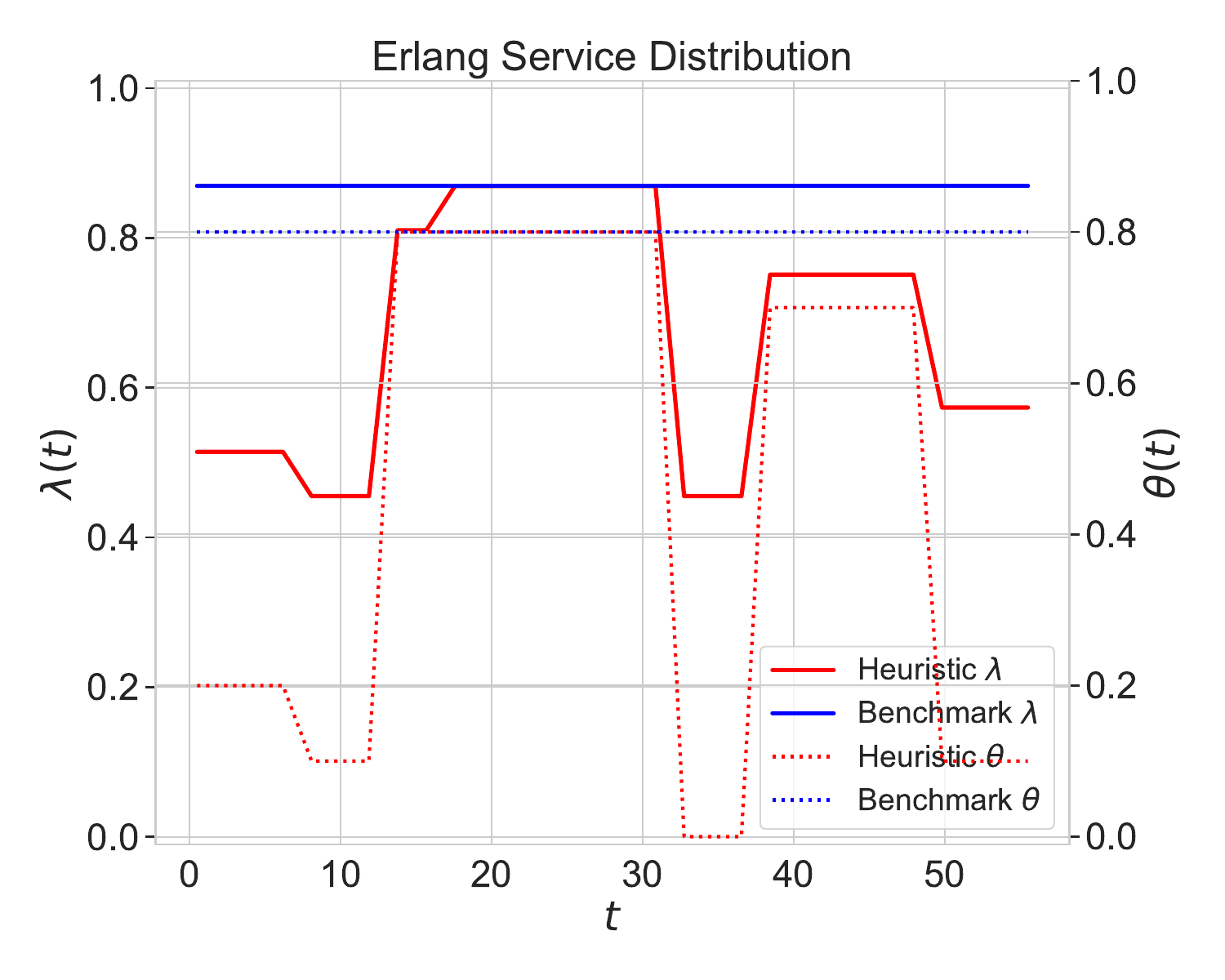}

}

\caption{Performance comparison of the proposed heuristic approach
(red dashed line) and a benchmark method using constant parameters
(blue solid line) under an Erlang service-time distribution. (a) The
AoI distribution $P(\Delta(t)\le x_{i})$ over time (the horizontal
dotted line indicates the target probability over time). (b) The sampling
rate $\lambda(t)$ and preemption probability $\theta(t)$ for each
method. The experiment uses time intervals $(t_{0},\ldots,t_{n})=(0,8,16,24,32,40,48,56)$
with time-varying AoI thresholds $(x_{0},\ldots,x_{n-1})=(4,3.5,2,2,3,2.5,3)$
and probability targets $(p_{0},\ldots,p_{n-1})=(0.8,0.7,0.6,0.6,0.6,0.7,0.7)$.
The processing time follows an Erlang distribution with shape $k=2$
and mean $1/\mu$ where $\mu=1.5$. The initial condition is given
by Condition \ref{cond:cond2}.}
\label{fig:Performance-theta-change}
\end{figure}

The pseudocode of the heuristic approach is provided in  Appendix~\ref{sec:pseudocode}. We evaluate the effectiveness of the proposed heuristic
by comparing it with a benchmark method that uses the minimum constant
sampling rate required to satisfy the tightest AoI constraint across
all time slots. Fig. \ref{fig:Performance-theta-change} compares
the performance of our adaptive heuristic approach against this static
benchmark. In Fig. \ref{fig:Performance-theta-change}(a), we compare
the AoI distribution $\boldsymbol{P}(\Delta(t)\leq x_{i})$ under
both approaches. The results show that while neither approach violates
the AoI constraint, the heuristic method's performance closely tracks
the target threshold as it changes. In contrast, the benchmark method
is overly conservative over multiple time slots, maintaining a much
higher probability than necessary during less demanding periods. Fig.
\ref{fig:Performance-theta-change}(b) plots the corresponding control
profiles. By adjusting the parameters across sub-intervals, the heuristic
approach achieves a clearly smaller overall sampling effort compared
to the constant-parameter benchmark. Moreover, in this Erlang setting
the heuristic method adapts the sampling rate $\lambda(t)$ and the
preemption probability $\theta(t)$, and the resulting $\theta(t)$
can differ substantially across sub-intervals. This behavior is consistent
with the insights from Fig. \ref{fig:AoI-distribution-with}, where
the best choice of the preemption probability depends on the AoI threshold
(e.g., full preemption is preferable for small AoI thresholds, whereas
little or no preemption can be better for larger AoI thresholds).

\section{Concluding remarks and future research}

\label{sec:Concluding-remarks-and}

In this research, we develop an analytical framework based on
multi-dimensional PDEs to characterize the time evolution of AoI and
other key system metrics, including packet age, and remaining workload.
By decomposing the original system into a set of coupled, low-dimensional
subsystems that capture marginal distributions, we derive exact solutions
for the AoI distribution at arbitrary time instances. To accompany
our theoretical results, we also designed efficient numerical procedures
to compute these solutions.

The framework is further extended to the stationary case, where we
derive the LST of the steady-state AoI. In special cases such as the non-preemptive
$M/G/1/1$ system and $M/M/1/1$ system with probabilistic preemption, we obtain closed-form
expressions for the AoI distribution, and for more general cases,
we compute the AoI numerically via inverse LST techniques. The accuracy
of our framework is validated through extensive simulations in both
time-varying and stationary scenarios. Finally, we address a system
parameter optimization problem by proposing an efficient heuristic
algorithm that adaptively selects the sampling rate to meet AoI constraints
while minimizing the sampling cost.

In future work, we plan to extend this PDE-based framework to analyze
AoI in other time-varying queueing systems with diverse service disciplines,
such as $M_{t}/G/1$ queues operating under FCFS and LCFS policies.
We also aim to investigate more complex scenarios, including systems
handling multiple data streams, which introduce additional complexity
in scheduling and interference among streams. Furthermore, we intend
to explore systems with time-varying processing rates and more general,
non-Poisson arrival time distributions. These extensions will provide
a deeper understanding of AoI behavior in more realistic and heterogeneous
network environments.

{
\section*{Acknowledgment}
The authors would like to thank the anonymous reviewers and the editor for their insightful comments and constructive suggestions, which helped to improve the quality of this paper.
}

\bibliographystyle{IEEEtran}
\bibliography{AoI_3}

@article{li2026condition,
  title={Condition-based maintenance for serial multistage manufacturing system considering reliability and quality over a finite horizon},
  author={Li, Jin and Li, Qi and Zhang, Wei and He, Zhen and He, Shuguang},
  journal={ENGINEERING Management},
  volume={13},
  number={1},
  pages={17},
  year={2026}
}

@book{folland1999real,
  title={Real analysis: modern techniques and their applications},
  author={Folland, Gerald B},
  year={1999},
  publisher={John Wiley \& Sons}
}

@article{evans1911volterra,
  title={Volterra\textquotesingle s integral equation of the second kind, with discontinuous kernel. II},
  author={Evans, Griffith C},
  journal={Transactions of the American Mathematical Society},
  volume={12},
  number={4},
  pages={429--472},
  year={1911}
}

@article{zhu2024optimizing,
  title={Optimizing peak age of information in MEC systems: Computing preemption and non-preemption},
  author={Zhu, Jianhang and Gong, Jie},
  journal={IEEE/ACM Transactions on Networking},
  volume={32},
  number={4},
  pages={3285--3300},
  year={2024},
  publisher={IEEE}
}

@book{rosenthal2006first,
  title={A First look at rigorous probability theory},
  author={Rosenthal, Jeffrey S},
  year={2006},
  publisher={World Scientific Publishing Company}
}

@book{ross2019introduction,
  title={Introduction to probability models},
  author={Ross, Sheldon M},
  year={2019},
  publisher={Academic press}
}

@article{jury1959analysis,
  title={The analysis of sampled-data control systems with a periodically time-varying sampling rate},
  author={Jury, EI and Mullin, FJ},
  journal={IRE Transactions on Automatic Control},
  number={1},
  volume={AC-4},
  pages={15--21},
  year={1959},
  publisher={IEEE}
}

@article{li2018stabilization,
  title={Stabilization of switched linear neutral systems: An event-triggered sampling control scheme},
  author={Li, Tai-Fang and Fu, Jun and Deng, Fang and Chai, Tianyou},
  journal={IEEE Transactions on Automatic Control},
  volume={63},
  number={10},
  pages={3537--3544},
  year={2018},
  publisher={IEEE}
}

@book{tenenbaum1985ordinary,
  title={Ordinary differential equations: an elementary textbook for students of mathematics, engineering, and the sciences},
  author={Tenenbaum, Morris and Pollard, Harry},
  year={1985},
  publisher={Courier Corporation}
}

@article{kesidis2023age,
  title={Age of information using Markov-renewal methods},
  author={Kesidis, George and Konstantopoulos, Takis and Zazanis, Michael A},
  journal={Queueing Systems},
  volume={103},
  number={1},
  pages={95--130},
  year={2023},
  publisher={Springer}
}

@article{abate1995numerical,
  title={Numerical inversion of Laplace transforms of probability distributions},
  author={Abate, Joseph and Whitt, Ward},
  journal={ORSA Journal on computing},
  volume={7},
  number={1},
  pages={36--43},
  year={1995},
  publisher={INFORMS}
}

@article{zhang2023sampled,
  title={Sampled-data control systems with non-uniform sampling: A survey of methods and trends},
  author={Zhang, Xian-Ming and Han, Qing-Long and Ge, Xiaohua and Ning, Boda and Zhang, Bao-Lin},
  journal={Annual Reviews in Control},
  volume={55},
  pages={70--91},
  year={2023},
  publisher={Elsevier}
}

@article{zhang2019networked,
  title={Networked control systems: A survey of trends and techniques},
  author={Zhang, Xian-Ming and Han, Qing-Long and Ge, Xiaohua and Ding, Derui and Ding, Lei and Yue, Dong and Peng, Chen},
  journal={IEEE/CAA Journal of Automatica Sinica},
  volume={7},
  number={1},
  pages={1--17},
  year={2019},
  publisher={IEEE}
}

@article{kurtz1978strong,
  title={Strong approximation theorems for density dependent Markov chains},
  author={Kurtz, Thomas G},
  journal={Stochastic Processes and their Applications},
  volume={6},
  number={3},
  pages={223--240},
  year={1978},
  publisher={Elsevier}
}

@article{massey2002analysis,
  title={The analysis of queues with time-varying rates for telecommunication models},
  author={Massey, William A},
  journal={Telecommunication Systems},
  volume={21},
  pages={173--204},
  year={2002},
  publisher={Springer}
}

@inproceedings{mandelbaum1999waiting,
  title={Waiting time asymptotics for time varying multiserver queues with abandonment and retrials},
  author={Mandelbaum, A and Massey, WA and Reiman, MI and Stolyar, AL},
  booktitle={Proceedings of the Annual Allerton Conference on Communication Control and Computing},
  volume={37},
  pages={1095--1104},
  year={1999},
  organization={Citeseer}
}

@article{whitt2018time,
  title={Time-varying queues},
  author={Whitt, Ward},
  journal={Queueing models and service management},
  volume={1},
  number={2},
  year={2018}
}

@article{pender2017approximations,
  title={Approximations for the queue length distributions of time-varying many-server queues},
  author={Pender, Jamol and Ko, Young Myoung},
  journal={INFORMS Journal on Computing},
  volume={29},
  number={4},
  pages={688--704},
  year={2017},
  publisher={INFORMS}
}

@article{fiems2023age,
  title={Age of information analysis with preemptive packet management},
  author={Fiems, Dieter},
  journal={IEEE Communications Letters},
  volume={27},
  number={4},
  pages={1105--1109},
  year={2023},
  publisher={IEEE}
}

@article{xu2023should,
 title={How Should the Server Sleep? -- {Age}-Energy Tradeoff in Sleep-Wake Server Systems},
  author={Xu, Jin and Wu, Xinyuan and Huang, Qisheng and Sun, Peng},
  journal={IEEE Transactions on Green Communications and Networking},
  volume={7},
  number={4},
  pages={1620--1634},
  year={2023},
  publisher={IEEE}
}

@article{zhong2023stochastic,
  title={Stochastic peak age of information guarantee for cooperative sensing in internet of everything},
  author={Zhong, Ailing and Li, Zhidu and Wu, Dapeng and Tang, Tong and Wang, Ruyan},
  journal={IEEE Internet of Things Journal},
  volume={10},
  number={17},
  pages={15186--15196},
  year={2023},
  publisher={IEEE}
}

@article{inoue2025characterizing,
   author={Inoue, Yoshiaki and Mandjes, Michel},
  journal={IEEE Transactions on Information Theory}, 
  title={Characterizing the Age of Information With Multiple Coexisting Data Streams}, 
  year={2025},
  volume={71},
  number={6},
  pages={4732-4753}
}

@article{fahim2024analyzing,
  title={Analyzing the age of information in prioritized status update systems under an interruption-based hybrid discipline},
  author={Fahim, Tamer E and Rabia, Sherif I and Abd El-Malek, Ahmed H and Zahra, Waheed K},
  journal={Performance Evaluation},
  volume={165},
  pages={102415},
  year={2024},
  publisher={Elsevier}
}

@article{dogan2021multi,
  title={The multi-source probabilistically preemptive {M/PH/1/1} queue with packet errors},
  author={Dogan, Ozancan and Akar, Nail},
  journal={IEEE Transactions on Communications},
  volume={69},
  number={11},
  pages={7297--7308},
  year={2021},
  publisher={IEEE}
}

@article{moltafet2022moment,
  title={Moment Generating Function of Age of Information in Multi-Source {M/G/1/1} Queueing Systems},
  author={Moltafet, Mohammad and Leinonen, Markus and Codreanu, Marian},
  journal={IEEE Transactions on Communications},
  volume={70},
  number={10},
  pages={6503-6516},
  year={2022},
  publisher={IEEE}
}

@inproceedings{champati2019distribution,
  title={On the Distribution of {AoI} for the {GI/GI/1/1} and {GI/GI/1/2} Systems: Exact Expressions and Bounds},
  author={Champati, Jaya Prakash and Al-Zubaidy, Hussein and Gross, James},
  booktitle={IEEE INFOCOM 2019-IEEE Conference on Computer Communications},
  pages={37--45},
  year={2019},
  organization={IEEE}
}

@article{chen2022age,
  title={Age of Information: The Multi-Stream {M/G/1/1} Non-Preemptive System},
  author={Chen, Zhengchuan and Deng, Dapeng and She, Changyang and Jia, Yunjian and Liang, Liang and Fang, Shuyang and Wang, Min and Li, Yonghui},
  journal={IEEE Transactions on Communications},
  volume={70},
  number={4},
  pages={2328--2341},
  year={2022},
  publisher={IEEE}
}

@article{soysal2021age,
  title={Age of information in {G/G/1/1} systems: Age expressions, bounds, special cases, and optimization},
  author={Soysal, Alkan and Ulukus, Sennur},
  journal={IEEE Transactions on Information Theory},
  volume={67},
  number={11},
  pages={7477--7489},
  year={2021},
  publisher={IEEE}
}

@article{yom2014erlang,
  title={{Erlang-R}: A time-varying queue with reentrant customers, in support of healthcare staffing},
  author={Yom-Tov, Galit B and Mandelbaum, Avishai},
  journal={Manufacturing \& Service Operations Management},
  volume={16},
  number={2},
  pages={283--299},
  year={2014},
  publisher={INFORMS}
}

@article{whitt2019time,
  title={Time-varying robust queueing},
  author={Whitt, Ward and You, Wei},
  journal={Operations Research},
  volume={67},
  number={6},
  pages={1766--1782},
  year={2019},
  publisher={INFORMS}
}

@article{ko2013critically,
  title={Critically loaded time-varying multiserver queues: computational challenges and approximations},
  author={Ko, Young Myoung and Gautam, Natarajan},
  journal={INFORMS Journal on Computing},
  volume={25},
  number={2},
  pages={285--301},
  year={2013},
  publisher={INFORMS}
}

@article{cassandras2014event,
  title={The event-driven paradigm for control, communication and optimization},
  author={Cassandras, Christos G},
  journal={Journal of Control and Decision},
  volume={1},
  number={1},
  pages={3--17},
  year={2014},
  publisher={Taylor \& Francis}
}

@inproceedings{florescu2021event,
  title={Event-driven modulo sampling},
  author={Florescu, Dorian and Krahmer, Felix and Bhandari, Ayush},
  booktitle={ICASSP 2021-2021 IEEE International Conference on Acoustics, Speech and Signal Processing (ICASSP)},
  pages={5435--5439},
  year={2021},
  organization={IEEE}
}

@article{zhang2014event,
  title={Event-driven observer-based output feedback control for linear systems},
  author={Zhang, Jinhui and Feng, Gang},
  journal={Automatica},
  volume={50},
  number={7},
  pages={1852--1859},
  year={2014},
  publisher={Elsevier}
}

@article{algabroun2025parametric,
  title={Parametric Machine Learning-Based Adaptive Sampling Algorithm for Efficient {IoT} Data Collection in Environmental Monitoring},
  author={Algabroun, Hatem and H{\aa}kansson, Lars},
  journal={Journal of Network and Systems Management},
  volume={33},
  number={1},
  pages={5},
  year={2025},
  publisher={Springer}
}

@inproceedings{malawade2022ecofusion,
  title={EcoFusion: Energy-aware adaptive sensor fusion for efficient autonomous vehicle perception},
  author={Malawade, Arnav Vaibhav and Mortlock, Trier and Faruque, Mohammad Abdullah Al},
  booktitle={Proceedings of the 59th ACM/IEEE Design Automation Conference},
  pages={481--486},
  year={2022}
}

@article{wang2021energy,
  title={Energy-efficient data and energy integrated management strategy for {IoT} devices based on RF energy harvesting},
  author={Wang, Yang and Yang, Kun and Wan, Weixiang and Zhang, Yitian and Liu, Qiang},
  journal={IEEE Internet of Things Journal},
  volume={8},
  number={17},
  pages={13640--13651},
  year={2021},
  publisher={IEEE}
}

@article{ayan2020probability,
  title={Probability analysis of age of information in multi-hop networks},
  author={Ayan, Onur and G{\"u}rsu, H Murat and Papa, Arled and Kellerer, Wolfgang},
  journal={IEEE Networking Letters},
  volume={2},
  number={2},
  pages={76--80},
  year={2020},
  publisher={IEEE}
}

@inproceedings{zhou2020risk,
  title={Risk-aware optimization of age of information in the Internet of Things},
  author={Zhou, Bo and Saad, Walid and Bennis, Mehdi and Popovski, Petar},
  booktitle={ICC 2020-2020 IEEE International Conference on Communications (ICC)},
  pages={1--6},
  year={2020},
  organization={IEEE}
}

@inproceedings{xu2022aoi,
  title={{AoI}-centric task scheduling for autonomous driving systems},
  author={Xu, Chengyuan and Xu, Qian and Wang, Jianping and Wu, Kui and Lu, Kejie and Qiao, Chunming},
  booktitle={IEEE INFOCOM 2022-IEEE Conference on Computer Communications},
  pages={1019--1028},
  year={2022},
  organization={IEEE}
}

@article{shi2024enhancing,
  title={Enhancing the Safety of Autonomous Driving Systems Via {AoI}-Optimized Task Scheduling},
  author={Shi, Tuo and Xu, Qian and Wang, Jianping and Xu, Chenyuan and Wu, Kui and Lu, Kejie and Qiao, Chunming},
  journal={IEEE Transactions on Vehicular Technology},
  year={2025},
  volume={74},
  number={3},
  pages={3804-3819}
}

@article{huang2023aoi,
  title={{AoI}-aware energy control and computation offloading for industrial {IoT}},
  author={Huang, Jiwei and Gao, Han and Wan, Shaohua and Chen, Ying},
  journal={Future Generation Computer Systems},
  volume={139},
  pages={29--37},
  year={2023},
  publisher={Elsevier}
}

@article{wang2025meta,
  title={Meta Computing-Driven Optimization of {AoI} in Industrial {IoT}: A Hybrid Scheme with Self-Organizing Maps and Reinforcement Learning},
  author={Wang, Min and Wang, Yike and Wang, Nuanlai and Pang, Shanchen and Sun, Xiaobing and Li, Ming},
  journal={IEEE Internet of Things Journal},
  year={2025},
  publisher={IEEE},
  volume={12},
  issue={14},
  pages={26244-26254}
}

@article{seuaciuc2010energy,
  title={Energy harvesting under excitations of time-varying frequency},
  author={Seuaciuc-Os{\'o}rio, Thiago and Daqaq, Mohammed F},
  journal={Journal of Sound and Vibration},
  volume={329},
  number={13},
  pages={2497--2515},
  year={2010},
  publisher={Elsevier}
}

@article{hu2021status,
  title={Status update in {IoT} networks: Age-of-information violation probability and optimal update rate},
  author={Hu, Limei and Chen, Zhengchuan and Dong, Yunquan and Jia, Yunjian and Liang, Liang and Wang, Min},
  journal={IEEE Internet of Things Journal},
  volume={8},
  number={14},
  pages={11329--11344},
  year={2021},
  publisher={IEEE}
}

@article{brill1977level,
  title={Level crossings in point processes applied to queues: single-server case},
  author={Brill, PH and Posner, Morton JM},
  journal={Operations Research},
  volume={25},
  number={4},
  pages={662--674},
  year={1977},
  publisher={INFORMS}
}

@article{cox1986virtual,
  title={The virtual waiting-time and related processes},
  author={Cox, DR and Isham, Valerie},
  journal={Advances in Applied Probability},
  volume={18},
  number={2},
  pages={558--573},
  year={1986},
  publisher={Cambridge University Press}
}

@article{gautam2024analysis,
  title={Analysis of Real-Time Order Fulfillment Policies: When to Dispatch a Batch?},
  author={Gautam, Natarajan and Geunes, Joseph},
  journal={Service Science},
  volume={16},
  number={2},
  pages={85--106},
  year={2024},
  publisher={INFORMS}
}

@article{yates2020age,
  title={The age of information in networks: Moments, distributions, and sampling},
  author={Yates, Roy D},
  journal={IEEE Transactions on Information Theory},
  volume={66},
  number={9},
  pages={5712--5728},
  year={2020},
  publisher={IEEE}
}

@article{xu2021age,
  title={Age of information for single buffer systems with vacation server},
  author={Xu, Jin and Hou, I-Hong and Gautam, Natarajan},
  journal={IEEE Transactions on Network Science and Engineering},
  volume={9},
  number={3},
  pages={1198--1214},
  year={2021},
  publisher={IEEE}
}

@inproceedings{moltafet2025aoi,
  title={{AoI} in {M/G/1/1} Queues with Probabilistic Preemption},
  author={Moltafet, Mohammad and Sadjadpour, Hamid R and Rezki, Zouheir and Codreanu, Marian and Yates, Roy D},
  booktitle={2025 IEEE International Symposium on Information Theory (ISIT)},
  pages={1--5},
  year={2025},
  organization={IEEE}
}

@book{polyanin2008handbook,
  title={Handbook of integral equations},
  author={Polyanin, Polyanin and Manzhirov, Alexander V},
  year={2008},
  publisher={Chapman and Hall/CRC}
}

@book{strauss2007partial,
  title={Partial differential equations: An introduction},
  author={Strauss, Walter A},
  year={2007},
  publisher={John Wiley \& Sons}
}

@book{royden2010real,
  title={Real analysis},
  author={Royden, Halsey and Fitzpatrick, Patrick Michael},
  year={2010},
  publisher={China Machine Press}
}

@article{inoue2019general,
  title={A general formula for the stationary distribution of the age of information and its application to single-server queues},
  author={Inoue, Yoshiaki and Masuyama, Hiroyuki and Takine, Tetsuya and Tanaka, Toshiyuki},
  journal={IEEE Transactions on Information Theory},
  volume={65},
  number={12},
  pages={8305--8324},
  year={2019},
  publisher={IEEE}
}

@article{bedewy2019age,
  title={The age of information in multihop networks},
  author={Bedewy, Ahmed M and Sun, Yin and Shroff, Ness B},
  journal={IEEE/ACM Transactions on Networking},
  volume={27},
  number={3},
  pages={1248--1257},
  year={2019},
  publisher={IEEE}
}

@article{zou2019waiting,
  title={Waiting before serving: A companion to packet management in status update systems},
  author={Zou, Peng and Ozel, Omur and Subramaniam, Suresh},
  journal={IEEE Transactions on Information Theory},
  volume={66},
  number={6},
  pages={3864--3877},
  year={2019},
  publisher={IEEE}
}

@article{xu2021peak,
  author={Xu, Jin and Gautam, Natarajan},
  journal={IEEE Transactions on Information Theory}, 
  title={Peak Age of Information in Priority Queuing Systems}, 
  year={2021},
  volume={67},
  number={1},
  pages={373-390},
  doi={10.1109/TIT.2020.3033501}
}

@article{bedewy2021optimal,
  title={Optimal sampling and scheduling for timely status updates in multi-source networks},
  author={Bedewy, Ahmed M and Sun, Yin and Kompella, Sastry and Shroff, Ness B},
  journal={IEEE Transactions on Information Theory},
  volume={67},
  number={6},
  pages={4019--4034},
  year={2021},
  publisher={IEEE}
}

@inproceedings{najm2016age,
  title={Age of information: The gamma awakening},
  author={Najm, Elie and Nasser, Rajai},
  booktitle={2016 IEEE International Symposium on Information Theory (ISIT)},
  pages={2574--2578},
  year={2016},
  organization={IEEE}
}

@article{yates2019age,
  title={The age of information: Real-time status updating by multiple sources},
  author={Yates, Roy D and Kaul, Sanjit K},
  journal={IEEE Transactions on Information Theory},
  volume={65},
  number={3},
  pages={1807--1827},
  year={2019},
  publisher={IEEE}
}

@inproceedings{najm2018status,
  title={Status updates in a multi-stream {M/G/1/1} preemptive queue},
  author={Najm, Elie and Telatar, Emre},
  booktitle={IEEE Infocom 2018-IEEE Conference On Computer Communications Workshops (Infocom Wkshps)},
  pages={124--129},
  year={2018},
  organization={IEEE}
}

@inproceedings{kaul2012real,
  title={Real-time status: How often should one update?},
  author={Kaul, Sanjit and Yates, Roy and Gruteser, Marco},
  booktitle={INFOCOM, 2012 Proceedings IEEE},
  pages={2731--2735},
  year={2012},
  organization={IEEE}
}

@article{costa2016age,
  title={On the age of information in status update systems with packet management},
  author={Costa, Maice and Codreanu, Marian and Ephremides, Anthony},
  journal={IEEE Transactions on Information Theory},
  volume={62},
  number={4},
  pages={1897--1910},
  year={2016},
  publisher={IEEE}
}

@article{16,
author={Colledani, M. and Gershwin, S. B.},
year={2013},
journal={Annals of Operations Research},
volume={209},
number={1},
title={A decomposition method for approximate evaluation of continuous flow multi-stage lines with general Markovian machines},
publisher={Springer US},
pages={5-40},
language={English}
}

@article{48,
   author = {Cui, L. and Kuo, W. and Loh, H. T. and Xie, M.},
   title = {Optimal allocation of minimal and perfect repairs under resource constraints},
   journal = {IEEE Transactions on Reliability},
   volume = {53},
   number = {2},
   pages = {193-199},
   year = {2004}
}

@book{linz1985analytical,
  title={Analytical and Numerical Methods for Volterra Equations},
  author={Linz, Peter},
  year={1985},
  publisher={Society for Industrial and Applied Mathematics (SIAM)}
}

@book{gripenberg1990volterra,
  title={Volterra Integral and Functional Equations},
  author={Gripenberg, Gustaf and Londen, Stig-Olof and Staffans, Olof},
  volume={34},
  publisher={Cambridge University Press},
  year={1990}
}

\appendices{}

\section{Proof of Theorem \ref{thm:When-assuming-,}}

\label{subsec:Solution-to-subsystems} We here describe the proof of Theorem \ref{thm:When-assuming-,}. The core idea of our
decomposition approach is that we do not need to solve the entire
PDE system (\ref{eq:1-7})-(\ref{eq:1-9}) to obtain the AoI distribution
$\Phi(t,x)$, which corresponds to the marginal distribution after
$A(t)$ and $W(t)$ are marginalized out. Instead, we identify all
components that are necessary to compute $\Phi(t,x)$ and derive each
component by analyzing the subsystem that governs it, where each subsystem
captures a specific marginal distribution of the original system.
As we will show, these components can be obtained through four interconnected
subsystems. Subsystem I marginalizes out the packet age, Subsystem
II characterizes the AoI distribution when the system is empty, Subsystem
III marginalizes out the AoI, and Subsystem IV marginalizes out both
the AoI and the packet age. Among them, Subsystem IV is the most fundamental
one as it only characterizes the workload process. We therefore start
by solving Subsystem IV and then proceed sequentially to the remaining
subsystems. The decomposition idea will be introduced in the next
subsection, followed by a detailed solution of each subsystem in the
subsequent subsections.

\subsection{Subsystem decompositions}

\label{subsec:Subsystem-decompositions}

In the following, we shall explain the decomposition approach step by step.
Specifically, the primary objective of our analysis is the AoI distribution
$\Phi(t,x)=H(t,x,\infty,\infty)$, which is already embedded within
Equation (\ref{eq:1-8}). A key structural observation is that $H(t,x,\infty,\infty)$
does not rely on the variables $y$ or $z$. Therefore, altering the
values of $y$ and $z$ leaves the term $H(t,x,\infty,\infty)$ unchanged
in Equation (\ref{eq:1-8}). Notably, if we set both $y=\infty$ and
$z=\infty$ at the same time in Equation (\ref{eq:1-8}), the term
$\lambda(t)\theta(t)H(t,x,\infty,\infty)F(z)$ cancels with the term
$\lambda(t)\theta(t)H(t,x,y,z)$ due to $F(\infty)=1$. A practical
alternative is to only let $y=\infty$ to eliminate the impact of
the packet age variable from the model. We can thus construct a reduced
system by letting $y=\infty$ in Equation (\ref{eq:1-8}), denoted
as Subsystem I, that involves only the variables $(t,x,z)$. For simplicity
of notation, we define $D(t,x,z)\triangleq H(t,x,\infty,z)$, with
the AoI distribution being expressed as $\Phi(t,x)=D(t,x,\infty)$.
Subsystem I is then written as 
\begin{align}
\mbox{Subsystem I: } & -D_{t}(t,x,z)+M_{t}(t,x)+\lambda(t)\theta(t)\Phi(t,x)F(z)\nonumber \\
 & +\lambda(t)(1-\theta(t))M(t,x)F(z)-\lambda(t)\theta(t)D(t,x,z)+\lambda(t)\theta(t)M(t,x)\nonumber \\
 & -D_{x}(t,x,z)+D_{z}(t,x,z)+M_{x}(t,x)-D_{z}(t,x,0)=0.\label{eq:2-1}
\end{align}

We find from Equation (\ref{eq:2-1}) that all the necessary components
to derive the AoI distribution are the joint distribution $D(t,x,z)$
and the distribution $M(t,x)$. Specifically, if we have the knowledge
of $M(t,x)$, Equation (\ref{eq:2-1}) is then a self-contained PDE
$D(t,x,z)$ given that $\Phi(t,x)=D(t,x,\infty)$. This motivates
us to construct a subsystem to derive $M(t,x)$ then. We find that
$M(t,x)$ is governed by Equation (\ref{eq:1-7}). For clarity, we
isolate this part and summarize it as Subsystem II, which captures
the AoI dynamics when the system is idle: 
\begin{equation}
\mbox{Subsystem II: }M_{t}(t,x)+M_{x}(t,x)+\lambda(t)M(t,x)-H_{z}(t,\infty,x,0)=0.\label{eq:2-2}
\end{equation}

We further observe from Equation (\ref{eq:2-2}) that computing $M(t,x)$
requires knowledge of the term $H_{z}(t,\infty,x,0)$. This observation
motivates the introduction of a third subsystem by setting $x=\infty$
in Equation (\ref{eq:1-8}), and defining $G(t,y,z)\triangleq H(t,\infty,y,z)$.
This yields Subsystem III, which characterizes the evolutions of the
packet age and remaining workload in terms of the variables $(t,y,z)$,
and is formulated as follows: 
\begin{align}
\mbox{Subsystem III: } & -G_{t}(t,y,z)+M_{t}(t,\infty)+\lambda(t)\theta(t)F(z)\nonumber \\
 & +\lambda(t)(1-\theta(t))M(t,\infty)F(z)-\lambda(t)\theta(t)G(t,y,z)+\lambda(t)\theta(t)M(t,\infty)\nonumber \\
 & -G_{y}(t,y,z)+G_{z}(t,y,z)-G_{z}(t,y,0)=0.\label{eq:2-3}
\end{align}

Our objective in analyzing Subsystem III is to compute the function
$G_{z}(t,y,0)=H_{z}(t,\infty,y,0)$, which is required for solving
Subsystem II. In particular, Subsystem III depends on $M(t,\infty)=H(t,\infty,0,0)$,
the probability that the system is empty at time $t$. To determine
$M(t,\infty)$, we must investigate the system's workload dynamics.
We set both $x=\infty$ and $y=\infty$ in Equations (\ref{eq:1-7})
and (\ref{eq:1-8}), define $B(t,z)=H(t,\infty,\infty,z)$, and then
construct the fourth subsystem described as follows: 
\begin{equation}
\mbox{Subsystem IV: }-B_{t}(t,z)+\lambda(t)\theta(t)F(z)+\lambda(t)(1-\theta(t))B(t,0)\big(F(z)-1\big)-\lambda(t)\theta(t)B(t,z)+B_{z}(t,z)=0.\label{eq:2-4}
\end{equation}

\begin{figure}
\centering\includegraphics[scale=0.4]{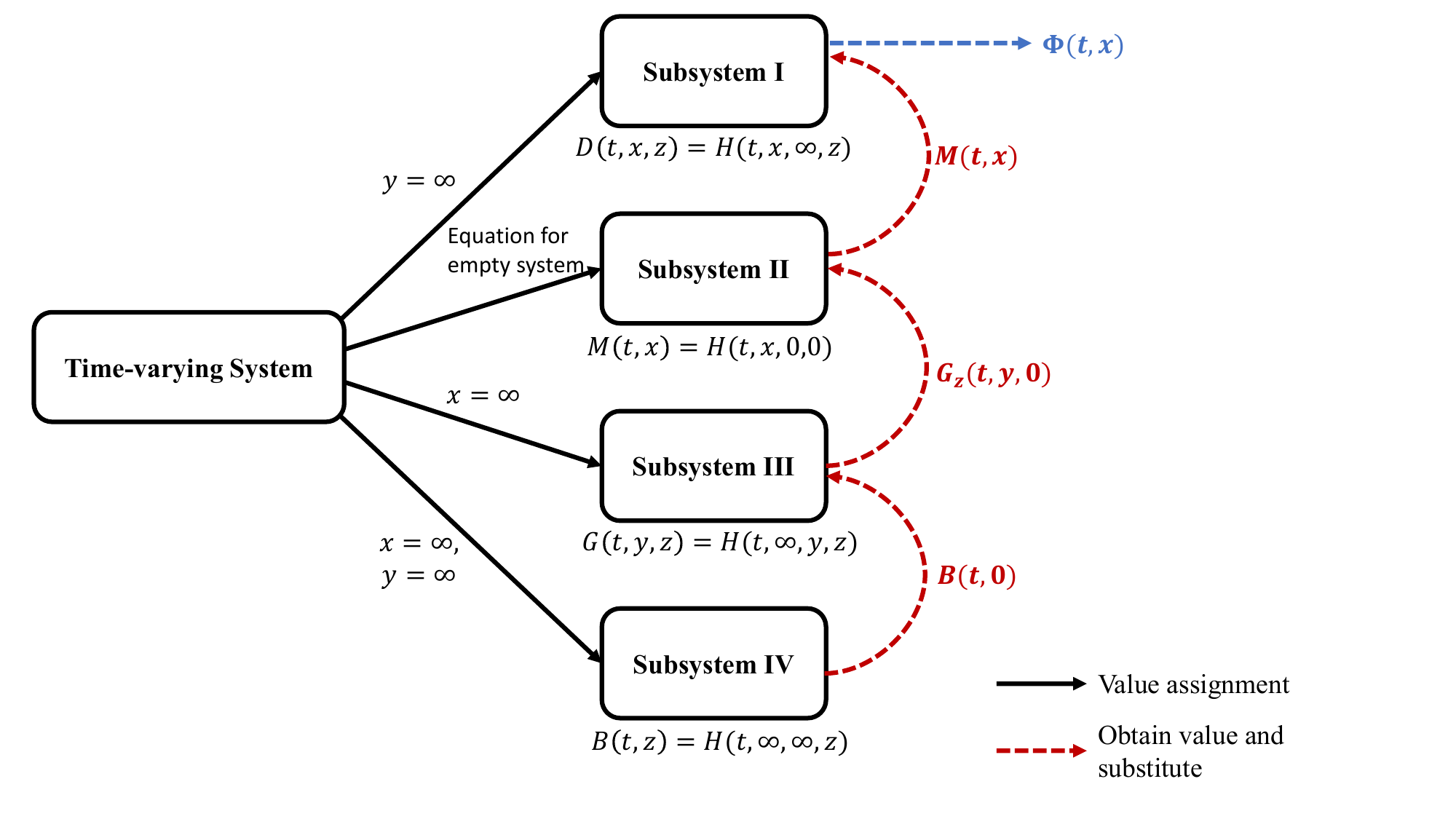}

\caption{Illustration of the decomposition technique used to solve the PDE
system. The original high-dimensional PDE system is decomposed into
four interrelated subsystems (Subsystems I-IV), each characterized
by fewer state variables. Arrows indicate data dependencies between
subsystems.}
\label{fig:Demonstration-of-the}
\end{figure}

Subsystem IV is a self-contained model that solely
characterizes the workload dynamics of the system. This is the most
fundamental subsystem that we will solve first. By solving Subsystem
IV, we obtain $B(t,0)=M(t,\infty)$, which represents the probability
that the system is idle at time $t$. This result is substituted into
Subsystem III to compute $G_{z}(t,y,0)=H_{z}(t,\infty,y,0)$. With
this function determined, we solve Subsystem II to derive $M(t,x)$,
which characterizes the age process when the system is idle. Finally,
we substitute $M(t,x)$ into Subsystem I to obtain the AoI distribution
$\Phi(t,x)=D(t,x,\infty)$. The notations for the marginal distributions
are summarized in Table \ref{tab:notation_summary}, and the sequence
of solving the subsystems is illustrated in Fig. \ref{fig:Demonstration-of-the}.
The remainder of this section presents the detailed solution for each
subsystem.

\subsection{Solving Subsystem IV}

Subsystem IV characterizes the workload process over time as a first-order
PDE with respect to $t$ and $z$. We aim to obtain $B(t,0)=M(t,\infty)$,
which is the probability that the system is idle at time $t$. For
notation simplicity, we rewrite this subsystem as 
\begin{equation}
B_{t}(t,z)-B_{z}(t,z)+\lambda(t)\theta(t)B(t,z)=v_{1}(t,z),\label{eq:2-1-1}
\end{equation}
with $v_{1}(t,z)=\lambda(t)\theta(t)F(z)+\lambda(t)(1-\theta(t))B(t,0)\big(F(z)-1\big)$.
We solve this PDE using the coordinate method (see Section 1.2 of
\cite{strauss2007partial}) as follows. We change the coordinates
by letting $\tau=t$ and $\eta=t+z$, and the derivatives of $B$
can then be transformed as: 
\begin{align}
\frac{\partial B}{\partial t} & =\frac{\partial B}{\partial\tau}\frac{\partial\tau}{\partial t}+\frac{\partial B}{\partial\eta}\frac{\partial\eta}{\partial t}=\frac{\partial B}{\partial\tau}+\frac{\partial B}{\partial\eta},\nonumber \\
\frac{\partial B}{\partial z} & =\frac{\partial B}{\partial\tau}\frac{\partial\tau}{\partial z}+\frac{\partial B}{\partial\eta}\frac{\partial\eta}{\partial z}=\frac{\partial B}{\partial\eta}.\label{eq:2-1-2}
\end{align}
Substituting these into Equation (\ref{eq:2-1-1}), we obtain a simpler
ordinary differential equation (ODE) of the first order: 
\begin{equation}
\frac{\partial B(\tau,\eta-\tau)}{\partial\tau}+\lambda(\tau)\theta(\tau)B(\tau,\eta-\tau)=v_{1}(\tau,\eta-\tau).\label{eq:2-1-3}
\end{equation}
We solve this ODE by multiplying the integrating factor $e^{\int_{0}^{\tau}\lambda(u)\theta(u)\mathrm{d}u}$
on both sides of Equation (\ref{eq:2-1-3}) and have 
\begin{equation}
\frac{\partial}{\partial\tau}\Big(B(\tau,\eta-\tau)e^{\int_{0}^{\tau}\lambda(u)\theta(u)\mathrm{d}u}\Big)=v_{1}(\tau,\eta-\tau)e^{\int_{0}^{\tau}\lambda(u)\theta(u)\mathrm{d}u}.\label{eq:2-1-4}
\end{equation}
The general solution for $B(t,z)$ is given by: 
\begin{eqnarray}
 &  & B(t,z)=B(\tau,\eta-\tau)=e^{-\int_{0}^{\tau}\lambda(u)\theta(u)\mathrm{d}u}\Big\{\int_{r=0}^{\tau}v_{1}(r,\eta-r)e^{\int_{0}^{r}\lambda(u)\theta(u)\mathrm{d}u}\mathrm{d}r+g_{1}(\eta)\Big\}\nonumber \\
 & = & \int_{r=0}^{t}\Big(\lambda(r)\theta(r)F(t+z-r)+\lambda(r)(1-\theta(r))B(r,0)\big(F(t+z-r)-1\big)\Big)e^{-\int_{r}^{t}\lambda(u)\theta(u)\mathrm{d}u}\mathrm{d}r\nonumber \\
 &  & +e^{-\int_{0}^{t}\lambda(u)\theta(u)\mathrm{d}u}g_{1}(t+z),\label{eq:2-1-5}
\end{eqnarray}
where $g_{1}(\cdot)$ is an arbitrary function (see Page 93 of \cite{tenenbaum1985ordinary}).
We now determine the value of $g_{1}(\cdot)$ by the initial condition
of the system. By setting $t=0$ in the solution, we obtain $g_{1}(z)=B(0,z)=H(0,\infty,\infty,z)$,
which represents the CDF of the initial workload at time $0$. This
shows that the initial workload distribution affects the workload
process $B(t,z)$ only through the multiplicative factor $e^{-\int_{0}^{t}\lambda(u)\theta(u)\mathrm{d}u}$. 

Since our goal is to compute the probability that the system is idle
at time $t$, namely $B(t,0)$. We set $z=0$ in Equation (\ref{eq:2-1-5})
and substitute $g_{1}(t)=B(0,t)$ to obtain the final integral equation:
\begin{align}
B(t,0) & =\int_{r=0}^{t}\Big(\lambda(r)\theta(r)F(t-r)+\lambda(r)\big(1-\theta(r)\big)B(r,0)\big(F(t-r)-1\big)\Big)e^{-\int_{r}^{t}\lambda(u)\theta(u)\mathrm{d}u}\mathrm{d}r\nonumber \\
 & +B(0,t)e^{-\int_{0}^{t}\lambda(u)\theta(u)\mathrm{d}u}.\label{eq:2-1-6}
\end{align}
Using the notation $M(t,\infty)=B(t,0)$, we have hence derived Equation
(\ref{eq:2-4-21}) in Theorem \ref{thm:When-assuming-,}.

\subsection{Solving Subsystem III}

Having obtained $M(t,\infty)$ by solving Subsystem IV, we now proceed
to solve Subsystem III. The goal of this step is to obtain the function
$G_{z}(t,y,0)=\lim_{h\to0^{+}}H_{z}(t,\infty,y,h)$, which is a necessary
component to solve Subsystem II. Similar to Subsystem IV, the PDE
for Subsystem III can also be transformed into a first-order ODE using
the coordinate method. By changing variables to $\tau=t$, $\xi=y-t$
and $\eta=z+t$, we can rewrite Subsystem III, given in Equation (\ref{eq:2-3}),
into the following form: 
\begin{equation}
\frac{\partial G(\tau,\xi+\tau,\eta-\tau)}{\partial\tau}+\lambda(\tau)\theta(\tau)G(\tau,\xi+\tau,\eta-\tau)=v_{2}(\tau,\xi+\tau,\eta-\tau),\label{eq:2-2-1}
\end{equation}
where 
\begin{equation}
v_{2}(t,y,z)=\lambda(t)\theta(t)F(z)+\lambda(t)\big(1-\theta(t)\big)M(t,\infty)F(z)+\lambda(t)\theta(t)M(t,\infty)-G_{z}(t,y,0)+M_{t}(t,\infty).\label{eq:2-2-2}
\end{equation}
Note that Equation (\ref{eq:2-2-1}) is also a first-order ODE
(see Page 93 of \cite{tenenbaum1985ordinary}), whose solution is
given by: 
\begin{eqnarray}
 &  & G(t,y,z)=G(\tau,\xi+\tau,\eta-\tau)\nonumber \\
 & = & e^{-\int_{0}^{\tau}\lambda(u)\theta(u)\mathrm{d}u}\Big(\int_{r=0}^{\tau}v_{2}(r,\xi+r,\eta-r)e^{\int_{0}^{r}\lambda(u)\theta(u)\mathrm{d}u}\mathrm{d}r+g_{2}(\xi,\eta)\Big)\nonumber \\
 & = & e^{-\int_{0}^{t}\lambda(u)\theta(u)\mathrm{d}u}\Big(\int_{r=0}^{t}v_{2}(r,y-t+r,z+t-r)e^{\int_{0}^{r}\lambda(u)\theta(u)\mathrm{d}u}\mathrm{d}r+g_{2}(y-t,z+t)\Big),\label{eq:2-2-3}
\end{eqnarray}
and $g_{2}(\cdot,\cdot)$ is an arbitrary function. We now determine
the function $g_{2}(\cdot,\cdot)$ using both the initial and boundary
conditions. We first consider the initial condition by letting $t=0$
in Equation (\ref{eq:2-2-3}). We then have 
\begin{eqnarray}
g_{2}(y,z) & = & G(0,y,z).\label{eq:2-2-4}
\end{eqnarray}
Note that Equation (\ref{eq:2-2-4}) is valid when $y,z\geq0$. Therefore,
when $y\geq t$, we have 
\begin{align}
G(t,y,z) & =e^{-\int_{0}^{t}\lambda(u)\theta(u)\mathrm{d}u}\bigg\{\int_{r=0}^{t}\big(\lambda(r)\theta(r)F(z+t-r)+\lambda(r)(1-\theta(r))M(r,\infty)F(z+t-r)\nonumber \\
 & +\lambda(r)\theta(r)M(r,\infty)-G_{z}(r,y-t+r,0)+M_{t}(r,\infty)\big)e^{\int_{0}^{r}\lambda(u)\theta(u)\mathrm{d}u}\mathrm{d}r+G(0,y-t,z+t)\bigg\}.\label{eq:2-2-5}
\end{align}
Since we are interested in the function $G_{z}(t,y,0)$, we take the
derivative of Equation (\ref{eq:2-2-5}) with respect to $z$ and obtain 
\begin{eqnarray}
G_{z}(t,y,z) & = & e^{-\int_{0}^{t}\lambda(u)\theta(u)\mathrm{d}u}\bigg\{\int_{r=0}^{t}\big(\lambda(r)\theta(r) \nonumber \\
& & +\lambda(r)(1-\theta(r))M(r,\infty)\big)e^{\int_{0}^{r}\lambda(u)\theta(u)\mathrm{d}u}\mathrm{d}F(z+t-r)+G_{z}(0,y-t,z+t)\bigg\}.\label{eq:2-2-6}
\end{eqnarray}
The formula $\mathrm{d}F(z+t-r)$ is the Lebesgue--Stieltjes measure
at time $z+t-r$. Letting $z=0$ in Equation (\ref{eq:2-2-6}), we
have 
\begin{eqnarray}
G_{z}(t,y,0) & = & \int_{r=0}^{t}\big(\lambda(r)\theta(r)+\lambda(r)(1-\theta(r))M(r,\infty)\big)e^{-\int_{r}^{t}\lambda(u)\theta(u)\mathrm{d}u}\mathrm{d}F(t-r)\nonumber \\
& &+e^{-\int_{0}^{t}\lambda(u)\theta(u)\mathrm{d}u}G_{z}(0,y-t,t)\mbox{ with }y\geq t.\label{eq:2-2-7}
\end{eqnarray}

Note that Equation (\ref{eq:2-2-7}) holds only when $y\geq t$. To
derive $g_{2}(\cdot,\cdot)$ when $y<t$, we now use the boundary
condition $G(t,y,0)=M(t,\infty)$, from which we obtain: 
\begin{equation}
M(t,\infty)=G(t,y,0)=e^{-\int_{0}^{t}\lambda(u)\theta(u)\mathrm{d}u}\Big(\int_{r=0}^{t}v_{2}(r,y-t+r,t-r)e^{\int_{0}^{r}\lambda(u)\theta(u)\mathrm{d}u}\mathrm{d}r+g_{2}(y-t,t)\Big).\label{eq:2-2-8}
\end{equation}
We can hence determine the function $g_{2}$ from Equation (\ref{eq:2-2-8}).
For notation simplicity, we let $y'=y-t$ and $t'=t$, and rewrite
Equation (\ref{eq:2-2-8}) as 
\begin{equation}
g_{2}(y-t,t)=g_{2}(y',t')=M(t',\infty)e^{\int_{0}^{t'}\lambda(u)\theta(u)\mathrm{d}u}-\int_{r=0}^{t'}v_{2}(r,y'+r,t-r)e^{\int_{0}^{r}\lambda(u)\theta(u)\mathrm{d}u}\mathrm{d}r.\label{eq:2-2-9}
\end{equation}
We now substitute Equation (\ref{eq:2-2-9}) back into the general
solution Equation (\ref{eq:2-2-3}), and rewrite Equation (\ref{eq:2-2-3})
as 
\begin{eqnarray}
G(t,y,z)& = & e^{-\int_{0}^{t}\lambda(u)\theta(u)\mathrm{d}u}\Big(\int_{r=0}^{t}v_{2}(r,y-t+r,z+t-r)e^{\int_{0}^{r}\lambda(u)\theta(u)\mathrm{d}u}\mathrm{d}r\nonumber \\
& & +M(z+t,\infty)e^{\int_{0}^{z+t}\lambda(u)\theta(u)\mathrm{d}u}-\int_{r=0}^{z+t}v_{2}(r,y-t+r,z+t-r)e^{\int_{0}^{r}\lambda(u)\theta(u)\mathrm{d}u}\mathrm{d}r\Big)\nonumber \\
& = & e^{-\int_{0}^{t}\lambda(u)\theta(u)\mathrm{d}u}\Big(-\int_{r=t}^{z+t}v_{2}(r,y-t+r,z+t-r)e^{\int_{0}^{r}\lambda(u)\theta(u)\mathrm{d}u}\mathrm{d}r\nonumber\\
& &+M(z+t,\infty)e^{\int_{0}^{z+t}\lambda(u)\theta(u)\mathrm{d}u}\Big).\label{eq:2-2-10}
\end{eqnarray}

To find the target function $G_{z}(t,y,0)$, we now apply a different,
but equally valid, boundary condition: $G(t,0,z)=M(t,\infty)$. This
states that the probability of having a packet with age zero is simply
the probability of the system being idle. Applying this to Equation
(\ref{eq:2-2-10}) by setting $y=0$ yields: 
\begin{eqnarray}
 & & M(t,\infty)=G(t,0,z)\nonumber \\
& = & e^{-\int_{0}^{t}\lambda(u)\theta(u)\mathrm{d}u}\left(-\int_{r=t}^{z+t}v_{2}(r,-t+r,z+t-r)e^{\int_{0}^{r}\lambda(u)\theta(u)\mathrm{d}u}\mathrm{d}r+M(z+t,\infty)e^{\int_{0}^{z+t}\lambda(u)\theta(u)\mathrm{d}u}\right)\nonumber \\
& = & e^{-\int_{0}^{t}\lambda(u)\theta(u)\mathrm{d}u}\bigg\{-\int_{r=t}^{z+t}\Big(\lambda(r)\theta(r)F(z+t-r)\nonumber\\
& & +\lambda(r)(1-\theta(r))M(r,\infty)F(z+t-r)+\lambda(r)\theta(r)M(r,\infty)\nonumber \\
& & -G_{z}(r,-t+r,0)+M_{t}(r,\infty)\Big)e^{\int_{0}^{r}\lambda(u)\theta(u)\mathrm{d}u}\mathrm{d}r+M(z+t,\infty)e^{\int_{0}^{z+t}\lambda(u)\theta(u)\mathrm{d}u}\bigg\}.\label{eq:2-2-11}    
\end{eqnarray}
Rearranging this expression, we obtain: 
\begin{eqnarray}
 &  & \int_{r=t}^{z+t}G_{z}(r,-t+r,0)e^{\int_{0}^{r}\lambda(u)\theta(u)\mathrm{d}u}\mathrm{d}r\nonumber \\
 & = & e^{\int_{0}^{t}\lambda(u)\theta(u)\mathrm{d}u}M(t,\infty)+\int_{r=t}^{z+t}\big(\lambda(r)\theta(r)F(z+t-r)+\lambda(r)(1-\theta(r))M(r,\infty)F(z+t-r)\nonumber \\
 &  & +\lambda(r)\theta(r)M(r,\infty)+M_{t}(r,\infty)\big)e^{\int_{0}^{r}\lambda(u)\theta(u)\mathrm{d}u}\mathrm{d}r-M(z+t,\infty)e^{\int_{0}^{z+t}\lambda(u)\theta(u)\mathrm{d}u}.\label{eq:2-2-12}
\end{eqnarray}
Taking the derivative of Equation (\ref{eq:2-2-12}) with respect
to $z$, we have 
\begin{eqnarray}
 &  & G_{z}(z+t,z,0)e^{\int_{0}^{z+t}\lambda(u)\theta(u)\mathrm{d}u}\nonumber \\
 & = & \Big(\lambda(z+t)\theta(z+t)M(t+z,\infty)+M_{t}(t+z,\infty)\Big)e^{\int_{0}^{z+t}\lambda(u)\theta(u)\mathrm{d}u}\nonumber \\
 &  & +\int_{r=t}^{z+t}\Big(\lambda(r)\theta(r)+\lambda(r)(1-\theta(r))M(r,\infty)\Big)e^{\int_{0}^{r}\lambda(u)\theta(u)\mathrm{d}u}\mathrm{d}F(z+t-r)\nonumber \\
 &  & -M_{t}(z+t,\infty)e^{\int_{0}^{z+t}\lambda(u)\theta(u)\mathrm{d}u}-\lambda(z+t)\theta(z+t)M(z+t,\infty)e^{\int_{0}^{z+t}\lambda(u)\theta(u)\mathrm{d}u}\nonumber \\
 & = & \int_{r=t}^{z+t}\Big(\lambda(r)\theta(r)+\lambda(r)(1-\theta(r))M(r,\infty)\Big)e^{\int_{0}^{r}\lambda(u)\theta(u)\mathrm{d}u}\mathrm{d}F(z+t-r).\label{eq:2-2-13}
\end{eqnarray}
Finally, by rearranging and relabeling the variables of Equation (\ref{eq:2-2-13}),
we obtain the desired function: 
\begin{equation}
G_{z}(t,y,0)=\int_{r=t-y}^{t}\Big(\lambda(r)\theta(r)+\lambda(r)(1-\theta(r))M(r,\infty)\Big)e^{-\int_{r}^{t}\lambda(u)\theta(u)\mathrm{d}u}\mathrm{d}F(t-r)\mbox{ with }t>y.\label{eq:2-2-14}
\end{equation}
Summarizing Equations (\ref{eq:2-2-7}) and (\ref{eq:2-2-14}), we
have 
\begin{eqnarray}
G_{z}(t,y,0) & = & \int_{r=\max\{0,t-y\}}^{t}\Big(\lambda(r)\theta(r)+\lambda(r)\big(1-\theta(r)\big)M(r,\infty)\Big)e^{-\int_{r}^{t}\lambda(u)\theta(u)\mathrm{d}u}\mathrm{d}F(t-r)\nonumber \\
 &  & +e^{-\int_{0}^{t}\lambda(u)\theta(u)\mathrm{d}u}H_{z}(0,\infty,y-t,t)1_{\{y\geq t\}}.\label{eq:2-2-15}
\end{eqnarray}
This completes the solution for Subsystem III and provides the result
cited in Theorem \ref{thm:When-assuming-,}.
\begin{rem}
Note that Equation (\ref{eq:2-2-15}) contains the
formula $\mathrm{d}F(t-r)$, which is the Lebesgue--Stieltjes measure
at time $t-r$. When performing the integration by parts, we can rewrite
Equation (\ref{eq:2-2-15}) in the following form:
\begin{eqnarray}
G_{z}(t,y,0) & = & F(\min\{t,y\})\Omega(t,\max\{0,t-y\})+\int_{r=\max\{0,t-y\}}^{t}F(t-r)\frac{\partial\Omega(t,r)}{\partial r}dr\nonumber \\
 &  & +e^{-\int_{0}^{t}\lambda(u)\theta(u)\mathrm{d}u}H_{z}(0,\infty,y-t,t)1_{\{y\geq t\}},\label{eq:2-2-16}
\end{eqnarray}
with $\Omega(t,r)=\Big(\lambda(r)\theta(r)+\lambda(r)\big(1-\theta(r)\big)M(r,\infty)\Big)e^{-\int_{r}^{t}\lambda(u)\theta(u)\mathrm{d}u}$.
This allows us to compute the AoI distribution under general service
time distributions. For instance, when having a deterministic processing
time $1/\mu$, we have for $t>y$ that 
\begin{eqnarray}
G_{z}(t,y,0) & = & F(y)\Omega(t,t-y)+\int_{r=t-y}^{t}1_{\{t-r>\frac{1}{\mu}\}}\frac{\partial\Omega(t,r)}{\partial r}dr\nonumber \\
 & = & 1_{\{y>\frac{1}{\mu}\}}\Omega(t,t-y)+\Omega(t,t-\frac{1}{\mu})1_{\{y>\frac{1}{\mu}\}}-\Omega(t,t-y)1_{\{y>\frac{1}{\mu}\}}\nonumber \\
 & = & \Omega(t,t-\frac{1}{\mu})1_{\{y>\frac{1}{\mu}\}}.\label{eq:2-2-17}
\end{eqnarray}
We have for $t\leq y$ that 
\begin{eqnarray}
G_{z}(t,y,0) & = & F(t)\Omega(t,0)+\int_{r=0}^{t}1_{\{t-r>\frac{1}{\mu}\}}\frac{\partial\Omega(t,r)}{\partial r}dr\nonumber \\
 &  & +e^{-\int_{0}^{t}\lambda(u)\theta(u)\mathrm{d}u}H_{z}(0,\infty,y-t,t)1_{\{y\geq t\}}\nonumber \\
 & = & 1_{\{t>\frac{1}{\mu}\}}\Omega(t,0)+\Omega(t,t-\frac{1}{\mu})1_{\{t>\frac{1}{\mu}\}}-\Omega(t,0)1_{\{t>\frac{1}{\mu}\}}\nonumber \\
 &  & +e^{-\int_{0}^{t}\lambda(u)\theta(u)\mathrm{d}u}H_{z}(0,\infty,y-t,t)1_{\{y\geq t\}}\nonumber \\
 & = & \Omega(t,t-\frac{1}{\mu})1_{\{t>\frac{1}{\mu}\}}+e^{-\int_{0}^{t}\lambda(u)\theta(u)\mathrm{d}u}H_{z}(0,\infty,y-t,t)1_{\{y\geq t\}}.\label{eq:2-2-18}
\end{eqnarray}
For other discrete service time distributions, the computation will
be similar.
\end{rem}

\subsection{Solving Subsystem II}

Having obtained $G_{z}(t,x,0)=H_{z}(t,\infty,x,0)$ by solving Subsystem
III, we next proceed to derive $M(t,x)$ by solving Subsystem II.
We begin by rewriting Equation (\ref{eq:1-7}) into the standard form
for a first-order PDE: 
\begin{equation}
M_{t}(t,x)+M_{x}(t,x)+\lambda(t)M(t,x)=G_{z}(t,x,0).\label{eq:2-3-1}
\end{equation}
We solve this equation using the coordinate method by changing the
variables to $\tau=t$ and $\xi=x-t$. This simplifies Equation (\ref{eq:2-3-1})
into the following first-order ODE along the characteristic curves:
\begin{equation}
\frac{\partial M(\tau,\xi+\tau)}{\partial\tau}+\lambda(\tau)M(\tau,\xi+\tau)=G_{z}(\tau,\xi+\tau,0).\label{eq:2-3-3}
\end{equation}
Similar to our previous subsystems, the solution to Equation (\ref{eq:2-3-3})
is given by: 
\begin{eqnarray}
 &  & M(t,x)=M(\tau,\xi+\tau)\nonumber \\
 & = & e^{-\int_{0}^{\tau}\lambda(u)\mathrm{d}u}\Big(g_{3}(\xi)+\int_{r=0}^{\tau}G_{z}(r,\xi+r,0)e^{\int_{0}^{r}\lambda(u)\mathrm{d}u}\mathrm{d}r\Big)\nonumber \\
 & = & e^{-\int_{0}^{t}\lambda(u)\mathrm{d}u}\Big(g_{3}(x-t)+\int_{r=0}^{t}G_{z}(r,x-t+r,0)e^{\int_{0}^{r}\lambda(u)\mathrm{d}u}\mathrm{d}r\Big),\label{eq:2-3-4}
\end{eqnarray}
where $g_{3}(\cdot)$ is an arbitrary unknown function. We consider
the initial condition at $t=0$, with $g_{3}(x)=M(0,x)$ for $x\geq 0$.
We thus have 
\begin{eqnarray}
M(t,x) & = & e^{-\int_{0}^{t}\lambda(u)\mathrm{d}u}\Big(M(0,x-t)+\int_{r=0}^{t}G_{z}(r,x-t+r,0)e^{\int_{0}^{r}\lambda(u)\mathrm{d}u}\mathrm{d}r\Big).
\end{eqnarray}
When $x<t$, we determine the function $g_{3}(\cdot)$ using the boundary
condition $M(t,0)=0$, which comes from Equation (\ref{eq:1-9}).
Then we obtain 
\begin{equation}
0=M(t,0)=e^{-\int_{0}^{t}\lambda(u)\mathrm{d}u}\Big(g_{3}(-t)+\int_{r=0}^{t}G_{z}(r,-t+r,0)e^{\int_{0}^{r}\lambda(u)\mathrm{d}u}\mathrm{d}r\Big).\label{eq:2-3-5}
\end{equation}
From this, we can solve for the function $g_{3}(\cdot)$ as: 
\begin{equation}
g_{3}(-t)=-\int_{r=0}^{t}G_{z}(r,-t+r,0)e^{\int_{u=0}^{r}\lambda(u)\mathrm{d}u}\mathrm{d}r.\label{eq:2-3-6}
\end{equation}
Substituting Equation (\ref{eq:2-3-6}) back into the general solution
Equation (\ref{eq:2-3-4}) yields the solution for $M(t,x)$: 
\begin{align}
M(t,x) & =e^{-\int_{0}^{t}\lambda(u)\mathrm{d}u}\Big(-\int_{r=0}^{-(x-t)}G_{z}(r,x-t+r,0)e^{\int_{0}^{r}\lambda(u)\mathrm{d}u}\mathrm{d}r+\int_{r=0}^{t}G_{z}(r,x-t+r,0)e^{\int_{0}^{r}\lambda(u)\mathrm{d}u}\mathrm{d}r\Big)\nonumber \\
 & =\int_{r=t-x}^{t}G_{z}(r,x-t+r,0)e^{-\int_{r}^{t}\lambda(u)\mathrm{d}u}\mathrm{d}r.\label{eq:2-3-7}
\end{align}
We therefore summarize the function of $M(t,x)$ as a piecewise function:
\begin{eqnarray}
M(t,x) & = & e^{-\int_{0}^{t}\lambda(u)\mathrm{d}u}H(0,x-t,0,0)1_{\{x\geq t\}}+\int_{r=\max\{0,t-x\}}^{t}G_{z}(r,x-t+r,0)e^{-\int_{r}^{t}\lambda(u)\mathrm{d}u}\mathrm{d}r.\label{eq:2-3-8}
\end{eqnarray}
We have now solved Subsystem II and obtained the complete expression
for $M(t,x)$.

\subsection{Solving Subsystem I}

We finally return to Subsystem I to solve for the AoI distribution
$M(t,x)$. Our strategy is to first simplify the PDE for Subsystem
I. We observe that Subsystem I contains terms such as $D_{z}(t,x,0)$,
$M_{t}(t,x)$, and $M_{x}(t,x)$ that are independent of the variable
$z$. By setting $z=\infty$ in Equation (\ref{eq:2-1}), we can establish
a relationship between these terms and the other functions: 
\begin{align}
0= & -D_{t}(t,x,\infty)+M_{t}(t,x)+\lambda(t)\theta(t)D(t,x,\infty)\nonumber \\
 & +\lambda(t)(1-\theta(t))M(t,x)-\lambda(t)\theta(t)D(t,x,\infty)+\lambda(t)\theta(t)M(t,x)\nonumber \\
 & -D_{x}(t,x,\infty)+D_{x}(t,x,0)-D_{z}(t,x,0).\label{eq:2-4-1}
\end{align}
Substituting this relationship back into the original Equation (\ref{eq:2-1})
and using the fact that $D(t,x,\infty)=\Phi(t,x)$ allow us to eliminate
these terms and arrive at a simplified PDE for $D(t,x,z)$: 
\begin{equation}
\mbox{Subsystem I: }D_{t}(t,x,z)+D_{x}(t,x,z)-D_{z}(t,x,z)+\lambda(t)\theta(t)D(t,x,z)=v_{4}(t,x,z),\label{eq:2-4-2}
\end{equation}
with 
\begin{equation}
v_{4}(t,x,z)=\lambda(t)\theta(t)\Phi(t,x)F(z)+\lambda(t)\big(1-\theta(t)\big)M(t,x)\big(F(z)-1\big)+\Phi_{t}(t,x)+\Phi_{x}(t,x).\label{eq:2-4-3}
\end{equation}
We again solve Equation (\ref{eq:2-4-2}) using the coordinate method.
By changing the variables to $\tau=t$, $\xi=x-t$ and $\eta=z+t$,
the PDE simplifies into the following ODE: 
\begin{equation}
\frac{\partial D(\tau,\xi+\tau,\eta-\tau)}{\partial\tau}+\lambda(\tau)\theta(\tau)D(\tau,\xi+\tau,\eta-\tau)=v_{4}(\tau,\xi+\tau,\eta-\tau),\label{eq:2-4-5}
\end{equation}
Similar to the previous subsystems, the general solution for $D(t,x,z)$
can be found as follows: 
\begin{eqnarray}
 &  & D(t,x,z)=D(\tau,\xi+\tau,\eta-\tau)\nonumber \\
 & = & e^{-\int_{0}^{\tau}\lambda(u)\theta(u)\mathrm{d}u}\Big(\int_{r=0}^{\tau}v_{4}(r,\xi+r,\eta-r)e^{\int_{0}^{r}\lambda(u)\theta(u)\mathrm{d}u}\mathrm{d}r+g_{4}(\xi,\eta)\Big)\nonumber \\
 & = & e^{-\int_{0}^{t}\lambda(u)\theta(u)\mathrm{d}u}\Big(\int_{r=0}^{t}v_{4}(r,x-t+r,z+t-r)e^{\int_{0}^{r}\lambda(u)\theta(u)\mathrm{d}u}\mathrm{d}r+g_{4}(x-t,z+t)\Big),\label{eq:2-4-6}
\end{eqnarray}
where $g_{4}(\cdot,\cdot)$ is an arbitrary real-valued function.
In the following, we determine the function $g_{4}$ for the case
of $x\geq t$ and $x<t$ separately. 

We first determine the function $g_{4}(\cdot,\cdot)$ using the initial
and boundary conditions. Based on the initial condition, we have $g_{4}(x,z)=D(0,x,z)$.
We then have for $x\geq t$ that
\begin{eqnarray}
D(t,x,z) & = & e^{-\int_{0}^{t}\lambda(u)\theta(u)\mathrm{d}u}\bigg\{\int_{r=0}^{t}v_{4}(r,x-t+r,z+t-r)e^{\int_{0}^{r}\lambda(u)\theta(u)\mathrm{d}u}\mathrm{d}r+D(0,x-t,z+t)\bigg\}\nonumber \\
 & = & e^{-\int_{0}^{t}\lambda(u)\theta(u)\mathrm{d}u}\bigg\{\int_{r=0}^{t}\big(\lambda(r)\theta(r)\Phi(r,x-t+r)\big(F(z+t-r)-1\big)\nonumber \\
 &  & +\lambda(r)\big(1-\theta(r)\big)M(r,x-t+r)\big(F(z+t-r)-1\big)\nonumber \\
 &  & +Y(r,x-t+r)\big)e^{\int_{0}^{r}\lambda(u)\theta(u)\mathrm{d}u}\mathrm{d}r+D(0,x-t,z+t)\bigg\}.\label{eq:2-4-7}
\end{eqnarray}
where 
\begin{equation}
Y(t,x)=\Phi_{t}(t,x)+\Phi_{x}(t,x)+\lambda(t)\theta(t)\Phi(t,x).
\end{equation}
 By letting $z=\infty$ in Equation (\ref{eq:2-4-7}), we have 
\begin{eqnarray}
\Phi(t,x) & = & e^{-\int_{0}^{t}\lambda(u)\theta(u)\mathrm{d}u}\Big(\int_{r=0}^{t}Y(r,x-t+r)e^{\int_{0}^{r}\lambda(u)\theta(u)\mathrm{d}u}\mathrm{d}r+\Phi(0,x-t)\Big).\label{eq:2-4-8}
\end{eqnarray}
When letting $z=0$ in Equation (\ref{eq:2-4-7}), combining it with
Equation (\ref{eq:2-4-8}), and using the fact that $D(t,x,0)=M(t,x)$,
we have 
\begin{eqnarray}
M(t,x) & = & e^{-\int_{0}^{t}\lambda(u)\theta(u)\mathrm{d}u}\Big(\int_{r=0}^{t}\big(\lambda(r)\theta(r)\Phi(r,x-t+r)\big(F(t-r)-1\big)\nonumber \\
 &  & +\lambda(r)\big(1-\theta(r)\big)M(r,x-t+r)\big(F(t-r)-1\big)\nonumber \\
 &  & +Y(r,x-t+r)\big)e^{\int_{0}^{r}\lambda(u)\theta(u)\mathrm{d}u}\mathrm{d}r+D(0,x-t,t)\Big)\nonumber \\
 & = & \Phi(t,x)+e^{-\int_{0}^{t}\lambda(u)\theta(u)\mathrm{d}u}\Big(-\Phi(0,x-t)+D(0,x-t,t)\nonumber \\
 &  & +\int_{r=0}^{t}\Big(\lambda(r)\theta(r)\Phi(r,x-t+r)\big(F(t-r)-1\big)\nonumber \\
 &  & +\lambda(r)\big(1-\theta(r)\big)M(r,x-t+r)\big(F(t-r)-1\big)\Big)e^{\int_{0}^{r}\lambda(u)\theta(u)\mathrm{d}u}\mathrm{d}r.\label{eq:2-4-9}
\end{eqnarray}
Rearranging Equation (\ref{eq:2-4-9}), we hence have derived the
AoI distribution for the case of $x\geq t$ as follows: 
\begin{eqnarray}
\Phi(t,x) & = & M(t,x)-e^{-\int_{0}^{t}\lambda(u)\theta(u)\mathrm{d}u}\Big(-\Phi(0,x-t)+D(0,x-t,t)\Big)\nonumber \\
 &  & -\int_{r=0}^{t}\Big(\lambda(r)\theta(r)\Phi(r,x-t+r)\nonumber\\
 & & +\lambda(r)\big(1-\theta(r)\big)M(r,x-t+r)\Big)\Big(F(t-r)-1\Big)e^{-\int_{r}^{t}\lambda(u)\theta(u)\mathrm{d}u}\mathrm{d}r.\label{eq:2-4-10}
\end{eqnarray}

We next derive the function $g_{4}$ for the case of $x<t$. Using
the boundary condition that $D(t,0,z)=0$ (which holds because the
AoI $\Delta(t)$ must be non-negative), we obtain 
\begin{equation}
0=e^{-\int_{0}^{t}\lambda(u)\theta(u)\mathrm{d}u}\Big(\int_{r=0}^{t}v_{4}(r,-t+r,z+t-r)e^{\int_{0}^{r}\lambda(u)\theta(u)\mathrm{d}u}\mathrm{d}r+g_{4}(-t,z+t)\Big),\label{eq:2-4-11}
\end{equation}
from which we determine the function $g_{4}$ as 
\begin{equation}
g_{4}(-t,z+t)=-\int_{r=0}^{t}v_{4}(r,-t+r,z+t-r)e^{\int_{0}^{r}\lambda(u)\theta(u)\mathrm{d}u}\mathrm{d}r.\label{eq:2-4-12}
\end{equation}
We next substitute Equation (\ref{eq:2-4-12}) back into Equation
(\ref{eq:2-4-6}), and obtain the function $D(t,x,z)$ as 
\begin{align}
D(t,x,z) & =e^{-\int_{0}^{t}\lambda(u)\theta(u)\mathrm{d}u}\Big(\int_{r=0}^{t}v_{4}(r,x-t+r,z+t-r)e^{\int_{0}^{r}\lambda(u)\theta(u)\mathrm{d}u}\mathrm{d}r+g_{4}(x-t,z+t)\Big)\nonumber \\
 & =\int_{r=t-x}^{t}v_{4}(r,x-t+r,z+t-r)e^{-\int_{r}^{t}\lambda(u)\theta(u)\mathrm{d}u}\mathrm{d}r\nonumber \\
 & =\int_{r=t-x}^{t}\Big\{Y(r,x-t+r)+\lambda(r)\theta(r)\Phi(r,x-t+r)\big(F(z+t-r)-1\big)\nonumber \\
 & +\lambda(r)\big(1-\theta(r)\big)M(r,x-t+r)\big(F(z+t-r)-1\big)\Big\}e^{-\int_{r}^{t}\lambda(u)\theta(u)\mathrm{d}u}\mathrm{d}r.\label{eq:2-4-13}
\end{align}
By letting $z=\infty$ in Equation (\ref{eq:2-4-13}), we have 
\begin{eqnarray}
\Phi(t,x)=D(t,x,\infty) & = & \int_{r=t-x}^{t}Y(r,x-t+r)e^{-\int_{r}^{t}\lambda(u)\theta(u)\mathrm{d}u}\mathrm{d}r.\label{eq:2-4-14}
\end{eqnarray}
To determine the RHS of Equation (\ref{eq:2-4-14}), we apply the
other boundary condition $D(t,x,0)=M(t,x)$. This identity holds because
the event $\{W(t)=0\}$ (zero remaining workload) is equivalent to
the system being idle, which by definition means $\{A(t)=0,W(t)=0\}$.
Setting $z=0$ in the solution above and substituting the definition
of $v_{4}$, we obtain: 
\begin{align}
M(t,x) & =\int_{r=t-x}^{t}\Big\{ Y(r,x-t+r)+\lambda(r)\theta(r)\Phi(r,x-t+r)\big(F(t-r)-1\big)\nonumber \\
 & +\lambda(r)\big(1-\theta(r)\big)M(r,x-t+r)\big(F(t-r)-1\big)\Big\} e^{-\int_{r}^{t}\lambda(u)\theta(u)\mathrm{d}u}\mathrm{d}r.\label{eq:2-4-15}
\end{align}

This equation provides a relationship between $M(t,x)$, which we
have already solved for in Subsystem II, and $\Phi(t,x)$ (our target).
Finally, we substitute this expression back into Equation (\ref{eq:2-4-14}),
which gives our final result: 
\begin{align}
\Phi(t,x)= & M(t,x)-\int_{r=t-x}^{t}\Big(\lambda(r)\theta(r)\Phi(r,x-t+r)\nonumber \\
 & +\lambda(r)\big(1-\theta(r)\big)M(r,x-t+r)\Big)\Big(F(t-r)-1\Big)e^{-\int_{r}^{t}\lambda(u)\theta(u)\mathrm{d}u}\mathrm{d}r.\label{eq:2-4-16}
\end{align}
Summarizing Equations (\ref{eq:2-4-10}) and (\ref{eq:2-4-16}), we
have 
\begin{eqnarray}
\Phi(t,x) & = & M(t,x)-e^{-\int_{0}^{t}\lambda(u)\theta(u)\mathrm{d}u}\Big(-H(0,x-t,\infty,\infty)+H(0,x-t,\infty,t)\Big)1_{\{x\geq t\}}\nonumber \\
 &  & -\int_{r=\max\{0,t-x\}}^{t}\Big(\lambda(r)\theta(r)\Phi(r,x-t+r)\nonumber \\
 &  & +\lambda(r)\big(1-\theta(r)\big)M(r,x-t+r)\Big)\Big(F(t-r)-1\Big)e^{-\int_{r}^{t}\lambda(u)\theta(u)\mathrm{d}u}\mathrm{d}r.\label{eq:2-4-17}
\end{eqnarray}
This completes the derivations for Theorem \ref{thm:When-assuming-,}.

{
\subsection{Existence of derivatives}\label{subsec:Existence-of-derivatives}

In our early derivations in Section \ref{sec:System-model-and}, we assume the derivatives
of $H$ in each direction exist and then construct the PDE systems. We now show that this assumption and derivation are generally valid in the
sense stated in Theorem 1: $M(t,\infty)$, $M(t,x)$, and $\Phi(t,x)$
are piecewise differentiable in their respective variables, while
$G_z(t,y,0)$ is piecewise continuous and piecewise differentiable with
respect to $y$. We first prove that $B(t,0)=M(t,\infty)$
is differentiable except for a finite number of points. We rewrite
$B(t,0)$ as 
\begin{equation}
B(t,0)=\Omega(t)+\int_{r=0}^{t}B(r,0)L(t,r)\mathrm{d}r,\label{eq:2-6-3}
\end{equation}
where $\Omega(t)=\int_{r=0}^{t}\lambda(r)\theta(r)F(t-r)e^{-\int_{r}^{t}\lambda(u)\theta(u)\mathrm{d}u}\mathrm{d}r+B(0,t)e^{-\int_{0}^{t}\lambda(u)\theta(u)\mathrm{d}u}$
and $L(t,r)=\lambda(r)\big(1-\theta(r)\big)\big(F(t-r)-1\big)e^{-\int_{r}^{t}\lambda(u)\theta(u)\mathrm{d}u}$.
The first term in $\Omega(t)$ is continuous and piecewise
differentiable under Assumption 1. To account for the second term in $\Omega(t)$,  we first have from the absolute continuity in Assumption 2 that
\begin{align}
& H(0,x,y,z) -  H(0,x,0,0) \nonumber\\
& =  \int_{0}^y H_y(0,0, v, 0) \, \mathrm{d}v + \int_{0}^z H_z(0,0, 0, w) \, \mathrm{d}w \nonumber\\
& + \int_{0}^y \int_{0}^x H_{x,y}(0,u, v, 0) \, \mathrm{d}u \, \mathrm{d}v + \int_{0}^z \int_{0}^x H_{x,z}(0,u, 0, w) \, \mathrm{d}u \, \mathrm{d}w + \int_{0}^z \int_{0}^y H_{y,z}(0,0, v, w) \, \mathrm{d}v \, \mathrm{d}w \nonumber\\
& + \int_{0}^z \int_{0}^y \int_{0}^x H_{x,y,z}(0,u, v, w) \, \mathrm{d}u \, \mathrm{d}v \, \mathrm{d}w.
\end{align}

Since the boundary condition in Assumption 2 has $H(0,0,y,z)=0$, we have $H_y(0,0, v, 0)=0$, $H_z(0,0, 0, w)=0$, and $H_{y,z}(0,0, v, w)=0$ in the above equation.  Moreover, since Assumption 2 has $H(0,x,y,0)=H(0,x,0,z)=H(0,x,0,0)$, we can obtain $H_{x,y}(0,u, v, 0)=0$ and $H_{x,z}(0,u, 0, w)=0$. We hence have
\begin{eqnarray}
B(0,t) &=&H(0,\infty,\infty,t) \nonumber \\
&=& H(0,\infty,0,0) 
+\int_{r=0}^{\infty}\int_{u=0}^{\infty}\int_{v=0}^{t}
H_{x,y,z}(0,r,u,v)\,\mathrm{d}v\,\mathrm{d}u\,\mathrm{d}r.
\end{eqnarray}
The above equation shows
that $B(0,t)$ is continuous and differentiable except for a finite
number of points. Then, at every regular point,

\begin{equation}
\frac{d}{dt}\left[
B(0,t)\exp\left(-\int_{u=0}^{t}\lambda(u)\theta(u)\,du\right)
\right]
=\exp\left(-\int_{u=0}^{t}\lambda(u)\theta(u)\,du\right) 
\times\left[
\frac{dB(0,t)}{dt}
-\lambda(t)\theta(t)B(0,t)
\right].
\end{equation}
Hence, $\Omega(t)$ is continuous and differentiable except for a
finite number of points. Moreover, $L(t,r)$ is continuous in the
triangle $r\leq t$ except on a finite number of lines.  According to the main
Theorem in \cite{evans1911volterra}, there exists a unique solution that is continuous
except for a finite number of points.

We next prove that the second term of Equation (\ref{eq:2-6-3}),
i.e., $I(t)=\int_{r=0}^{t}B(r,0)\lambda(r)\big(1-\theta(r)\big)\big(F(t-r)-1\big)e^{-\int_{r}^{t}\lambda(u)\theta(u)\mathrm{d}u}\mathrm{d}r$,
is differentiable except for a finite number of points. To prove the differentiability of $I(t)$,
by definition, we only need to prove that its derivative exists and is
finite. By assumption, on $[0,T]$, the CDF $F$ has finitely many
jumps at points $0<c_{1}<\cdots<c_{m}\le T$, and is absolutely continuous
in each open interval between two consecutive jumps. According to
the Lebesgue decomposition theorem (see \cite{rosenthal2006first}),
$F$ can be decomposed as 
\begin{equation}
F(z)=F_{c}(z)+F_{d}(z),\qquad z\in[0,T], \label{eq:lebesgue_start}
\end{equation}
where $F_{c}$ is absolutely continuous on $[0,T]$, and 
\begin{equation}
F_{d}(z)=\sum_{j=1}^{m}J_{j}\,\mathbf{1}_{\{z\ge c_{j}\}},\qquad J_{j}:=F(c_{j})-F(c_{j}^{-}).
\end{equation}
Moreover, since we assume $F(0)=0$, there exists $f\in L^{1}([0,T])$
such that 
\begin{equation}
F_{c}(z)=\int_{0}^{z}f(u)\,\mathrm{d}u,\qquad z\in[0,T],
\end{equation}
and $F_{c}'(z)=f(z)$ (see Theorem 3.35 of \cite{folland1999real}).
Accordingly, we write $I(t)=I_{c}(t)+I_{d}(t),$ where 
\begin{align}
I_{c}(t) & =\int_{0}^{t}B(r,0)\lambda(r)(1-\theta(r))(F_{c}(t-r)-1)\exp\!\left(-\int_{r}^{t}\lambda(u)\theta(u)\,\mathrm{d}u\right)\mathrm{d}r\nonumber \\
 & =\int_{0}^{t}B(r,0)L_{c}(t,r)\mathrm{d}r
\end{align}
and 
\begin{align}
I_{d}(t) & =\int_{0}^{t}B(r,0)\lambda(r)(1-\theta(r))F_{d}(t-r)\exp\!\left(-\int_{r}^{t}\lambda(u)\theta(u)\,\mathrm{d}u\right)\mathrm{d}r\nonumber \\
 & =\int_{0}^{t}B(r,0)L_{d}(t,r)\mathrm{d}r.
\end{align}
We now prove the differentiability of $I_{c}(t)$ and $I_{d}(t)$
separately. By definition, the derivative of $I_{c}(t)$ is written
as 
\begin{eqnarray}
 &  & \lim_{h\to0}\frac{1}{h}\Big(I_{c}(t+h)-I_{c}(t)\Big)\nonumber \\
 & = & \lim_{h\to0}\frac{1}{h}\int_{r=0}^{t+h}B(r,0)L_{c}(t+h,r)\mathrm{d}r-\lim_{h\to0}\frac{1}{h}\int_{r=0}^{t}B(r,0)L_{c}(t,r)\mathrm{d}r\nonumber \\
 & = & \lim_{h\to0}\frac{1}{h}\int_{r=0}^{t}B(r,0)\Big(L_{c}(t+h,r)-L_{c}(t,r)\Big)\mathrm{d}r+\lim_{h\to0}\frac{1}{h}\int_{r=t}^{t+h}B(r,0)L_{c}(t+h,r)\mathrm{d}r.\label{eq:2-6-4}
\end{eqnarray}
As $B(t,0)$ is piecewise continuous in $t$, we know that the second
term of Equation (\ref{eq:2-6-4}) exists except for a finite number
of points. We then focus on the first term of Equation (\ref{eq:2-6-4}). We let $a_h(t,r) =\frac{1}{h} B(r,0)\big(L_{c}(t+h,r)-L_{c}(t,r)\big)$.
To prove that $\lim_{h\to0} \int_{r=0}^{t} a_{h}(t,r) \mathrm{d}r$ exists, we first find that $L_{c}(t,r)$ is
absolutely continuous in $t$ for any fixed $r\in[0,t]$. By the definition
of absolute continuity, we have 
\begin{align}
\frac{1}{h}\Big(L_{c}(t+h,r)-L_{c}(t,r)\Big) & =\frac{1}{h}\int_{t}^{t+h}L_{c}^{'}(s,r)\mathrm{d}s
\end{align}
with 
\begin{equation}
L_{c}^{'}(s,r)\triangleq\frac{\partial}{\partial s}L_{c}(s,r)=\lambda(r)(1-\theta(r))\exp\!\left(-\int_{r}^{s}\lambda(u)\theta(u)\,\mathrm{d}u\right)\bigg[f(s-r)-\lambda(s)\theta(s)\bigl(F_{c}(s-r)-1\bigr)\bigg],
\end{equation}
where $L_{c}^{'}(s,r)$ exists except for a finite number of points. We next verify the conditions of the general Lebesgue dominated convergence theorem by showing that 
 $a_{h}(t,r) = \frac{1}{h}\int_{t}^{t+h}B(r,0)L_{c}^{'}(s,r)\mathrm{d}s$ is dominated by a sequence of converging and integrable functions. Because $B(r,0)\in[0,1]$, $0\le\theta(t)\le1$,
$0\le F_{c}\le1$, and $\lambda(t)$ is bounded on $[0,T]$, there
exists a constant $C_{T}>0$ such that 
\begin{equation}
\left|B(r,0) \frac{\partial}{\partial s}L_{c}(s,r)\right|\le C_{T}\bigl(|f(s-r)|+1\bigr),\qquad0\le r\le s\le T,
\end{equation}
with 
\begin{equation}
\left|\frac{1}{h}\int_{t}^{t+h}B(r,0)L_{c}^{'}(s,r)\mathrm{d}s\right|\leq\frac{1}{h}\int_{t}^{t+h}C_{T}\bigl(|f(s-r)|+1\bigr)\mathrm{d}s\triangleq b_{h}(t-r).
\end{equation}
As $f$ is integrable, by the Lebesgue differentiation theorem (Theorem 14 on Page 126 of \cite{royden2010real}), for almost every $u=t-r$, we have $b_h(u) \to C_T(|f(u)| + 1) \triangleq b(u)$  as $h\to0$. As $f(s-r)$ is a non-negative and integrable function over the area $(s,r)\in [t,t+h]\times[0,t]$, then by the Fubini-Tonelli theorem (see \cite{folland1999real}), we have
\begin{eqnarray}
\lim_{h\to0}\int_{r=0}^{t}b_{h}(t-r)\mathrm{d}r &=& \lim_{h\to0}\frac{1}{h}\int_{r=0}^{t}\int_{s=t}^{t+h}C_{T}\bigl(|f(s-r)|+1\bigr)\mathrm{d}s\mathrm{d}r\nonumber \\ & = & \lim_{h\to 0}\frac{1}{h}\int_{s=t}^{t+h}\int_{r=0}^{t}C_{T}|f(s-r)|\mathrm{d}r\mathrm{d}s+C_{T}t\nonumber \\
 & = & \lim_{h\to0} \frac{1}{h}\int_{s=t}^{t+h}C_{T}\big(F_{c}(s) -F_{c}(s-t) \big)\mathrm{d}s+C_{T}t \nonumber \\
 & = & C_{T}\big(F_{c}(t) -F_{c}(0) \big)+C_{T}t\nonumber \\
 & = & C_{T}F_{c}(t)+C_{T}t \quad \mbox{since }F_{c}(0)=0.   
\end{eqnarray}
Notice that the last term of the above equation also equals $\int_{r=0}^{t}b(t-r) \mathrm{d}r \leq C_{T}(1+t) < \infty$. Applying the general Lebesgue dominated
convergence theorem (Page 89 of \cite{royden2010real}), we have 
\begin{equation}
\lim_{h\to0}\frac{1}{h}\int_{r=0}^{t}B(r,0)\Big(L_{c}(t+h,r)-L_{c}(t,r)\Big)\mathrm{d}r=\int_{0}^{t}B(r,0)\frac{\partial}{\partial t}L_{c}(t,r)\mathrm{d}r.
\end{equation}
As both parts in Equation (\ref{eq:2-6-4}) exist except for a finite
number of points, we conclude that $I_{c}(t)$ is differentiable except
for a finite number of points. We next consider $I_{d}(t)$. Since
\begin{equation}
F_{d}(t-r)=\sum_{j=1}^{m}J_{j}\,\mathbf{1}_{\{t-r\ge c_{j}\}}=\sum_{j=1}^{m}J_{j}\,\mathbf{1}_{\{r\le t-c_{j}\}},
\end{equation}
we may write 
\begin{eqnarray}
I_{d}(t) & = & \int_{0}^{t}B(r,0)\lambda(r)(1-\theta(r))\sum_{j=1}^{m}J_{j}\,\mathbf{1}_{\{r\le t-c_{j}\}}\exp\!\left(-\int_{r}^{t}\lambda(u)\theta(u)\,\mathrm{d}u\right)\mathrm{d}r\nonumber \\
 & = & \sum_{j=1}^{m}J_{j}\int_{0}^{(t-c_{j})^{+}}B(r,0)\lambda(r)(1-\theta(r))\exp\!\left(-\int_{r}^{t}\lambda(u)\theta(u)\,\mathrm{d}u\right)\mathrm{d}r.
\end{eqnarray}
Let $R(t,r)=B(r,0)\lambda(r)(1-\theta(r))\exp\!\left(-\int_{r}^{t}\lambda(u)\theta(u)\,\mathrm{d}u\right)$.
We have 
\begin{eqnarray}
 &  & \lim_{h\to0}\frac{1}{h}\Big(I_{d}(t+h)-I_{d}(t)\Big)\nonumber \\
 & = & \lim_{h\to0}\frac{1}{h}\sum_{j=1}^{m}J_{j}\Big(\int_{0}^{(t+h-c_{j})_{+}}R(t+h,r)\mathrm{d}r-\int_{0}^{(t-c_{j})_{+}}R(t,r)\mathrm{d}r\Big)\nonumber \\
 & = & \lim_{h\to0}\frac{1}{h}\sum_{j=1}^{m}J_{j}\Big(\int_{0}^{(t-c_{j})_{+}}\big(R(t+h,r)-R(t,r)\big)\mathrm{d}r+\int_{(t-c_{j})_{+}}^{(t+h-c_{j})_{+}}R(t+h,r)\mathrm{d}r\Big).\label{eq:2-6-5}
\end{eqnarray}
It is easy to observe that $\lim_{h\to0}\frac{1}{h}\int_{(t-c_{j})_{+}}^{(t+h-c_{j})_{+}}R(t+h,r)\mathrm{d}r$
exists except for a finite number of points. To prove the existence
of $\lim_{h\to0}\frac{1}{h}\int_{0}^{(t-c_{j})_{+}}\big(R(t+h,r)-R(t,r)\big)\mathrm{d}r$,
we notice that $R(t,r)$ is also an absolutely continuous function
in $t$ with derivative $\frac{\partial}{\partial t}R(t,r)=-\lambda(t)\theta(t)R(t,r)$,
which is dominated by an integrable function. Similar to our discussion
above, we can again apply the Lebesgue dominated convergence theorem
and have 
\begin{equation}
\lim_{h\to0}\frac{1}{h}\int_{0}^{(t-c_{j})_{+}}\big(R(t+h,r)-R(t,r)\big)\mathrm{d}r=-\lambda(t)\theta(t)\int_{0}^{(t-c_{j})_{+}}R(t,r)\,\mathrm{d}r, \label{eq:lebesgue_end}
\end{equation}

The derivative of $I_{d}(t)$ thus exists except for a finite number
of points. Combining the two parts, we conclude that $I'(t)=I_{c}'(t)+I_{d}'(t)$
exists for $t\in[0,T]$ except for a finite number of points. Hence
we can show that $B(t,0)$ is continuous and differentiable except
for a finite number of points.

Furthermore, we have 
\begin{eqnarray}
G_{z}(t,y,0) & = & \int_{r=\max\{0,t-y\}}^{t}\Big(\lambda(r)\theta(r)+\lambda(r)\big(1-\theta(r)\big)B(r,0)\Big)e^{-\int_{r}^{t}\lambda(u)\theta(u)\mathrm{d}u}\mathrm{d}F(t-r)\nonumber \\
 &  & +e^{-\int_{0}^{t}\lambda(u)\theta(u)\mathrm{d}u}H_{z}(0,\infty,y-t,t)1_{\{y\geq t\}}.\label{eq:2-6-9}
\end{eqnarray}

For $y\ge t$, Assumption \ref{assum2}(1), \ref{assum2}(3), and \ref{assum2}(4) give
$$
H_z(0,\infty,y-t,t)
=
\int_{r=0}^{\infty}\int_{u=0}^{y-t}
H_{x,y,z}(0,r,u,t)\,\mathrm{d}u\,\mathrm{d}r.
$$
Hence, $H_z(0,\infty,y-t,t)$ is piecewise continuous and
piecewise differentiable with respect to $y$. By applying the same
Lebesgue decomposition of $F$ as in Equation (\ref{eq:lebesgue_start})--(\ref{eq:lebesgue_end}), the integral term
in Equation (\ref{eq:2-6-9}) is also piecewise continuous and piecewise differentiable
with respect to $y$. Therefore, $G_z(t,y,0)$ is piecewise
continuous and piecewise differentiable with respect to $y$. We do
not require differentiability of $G_z(t,y,0)$ with respect to $t$ in the main theorem or subsequent derivations.
We next consider $M(t,x)$. For $x\ge t$, the initial term in Equation (\ref{eq:2-3-8}) is
\begin{equation}
\exp\left(-\int_{u=0}^{t}\lambda(u)\,du\right)
H(0,x-t,0,0).
\end{equation}
By Assumption \ref{assum2}(5), its derivatives with respect to $x$ and $t$
exist except at a finite number of points. In particular,
\begin{equation}
\frac{\partial}{\partial x}
\left[
\exp\left(-\int_{u=0}^{t}\lambda(u)\,du\right)
H(0,x-t,0,0)
\right]
=
\exp\left(-\int_{u=0}^{t}\lambda(u)\,du\right)
H_x(0,x-t,0,0),
\end{equation}
and
\begin{align}
& \frac{\partial}{\partial t}
\left[
\exp\left(-\int_{u=0}^{t}\lambda(u)\,du\right)
H(0,x-t,0,0)
\right] \nonumber \\
 = &
-\exp\left(-\int_{u=0}^{t}\lambda(u)\,du\right) \times
\left[
\lambda(t)H(0,x-t,0,0)
+H_x(0,x-t,0,0)
\right].
\end{align}
The remaining term in (\ref{eq:2-3-8}) is piecewise differentiable by the
Leibniz rule and the piecewise differentiability of $G_z(t,y,0)$
with respect to $y$. Hence, $M(t,x)$ is piecewise differentiable
with respect to $t$ and $x$.

Finally, to analyze $\Phi(t,x)$ for $x\ge t$, the initial-distribution terms in (\ref{eq:2-4-17})
satisfy
\begin{equation}
-H(0,x-t,\infty,\infty)
+H(0,x-t,\infty,t) =
-\int_{r=0}^{x-t}\int_{u=0}^{\infty}\int_{v=t}^{\infty}
H_{x,y,z}(0,r,u,v)\,\mathrm{d}v\,\mathrm{d}u\,\mathrm{d}r. 
\end{equation}
Thus, the initial-distribution terms in (\ref{eq:2-4-17}) are piecewise
differentiable under Assumption \ref{assum2}(4). It remains to verify the regularity of the
Volterra equation in (\ref{eq:2-4-22}). Let
$$
u=t-x,\qquad \widehat{\Phi}(u,x)=\Phi(u+x,x),\qquad
a(u)=\max\{0,-u\}.
$$
Then Equation (\ref{eq:2-4-22}) can be written as
\begin{equation}
\widehat{\Phi}(u,x)
=
Q(u,x)+\int_{\tau=a(u)}^x
\rho(u,x,\tau)\widehat{\Phi}(u,\tau)\,\mathrm{d}\tau ,
\end{equation}
where
$$
\rho(u,x,\tau)
=
-\lambda(\tau+u)\theta(\tau+u)
\big(F(x-\tau)-1\big)
\exp\left(-\int_{s=u+\tau}^{u+x}\lambda(s)\theta(s)\,\mathrm{d}s\right),
$$
and $Q(u,x)$ denotes the remaining non-integral terms and the integral
term involving $M(\tau+u,\tau)$. By Assumption 1, Assumption 2, and the
piecewise differentiability of $M(t,x)$ proved above, $Q(u,x)$ and
$\rho(u,x,\tau)$ are piecewise differentiable in $x$ and in the parameter
$u$, except on a finite number of discontinuity or characteristic lines. 

For every fixed $u$, the equation above is a one-dimensional Volterra
integral equation in $x$. Hence, by the same argument used for
$B(t,0)$, $\widehat{\Phi}(u,x)$ is piecewise differentiable with
respect to $x$. Moreover, on each region $u>0$ or $u<0$, the lower limit
$a(u)=\max\{0,-u\}$ is differentiable, and both $Q(u,x)$ and the Volterra kernel $\rho(u,x,\tau)$ above are piecewise differentiable on the parameter $u$. Therefore, by
Theorem 1.2 in Chapter 13 of \cite{gripenberg1990volterra}, $\widehat{\Phi}(u,x)$ is piecewise
differentiable with respect to $u$. At every regular point,
$\widehat{\Phi}_u(u,x)$ satisfies
\begin{eqnarray}
\widehat{\Phi}_u(u,x)
&=&
Q_u(u,x)
-a'(u)\rho(u,x,a(u))\widehat{\Phi}(u,a(u))  \nonumber \\
& &+
\int_{\tau=a(u)}^x
\rho_u(u,x,\tau)\widehat{\Phi}(u,\tau)\,\mathrm{d}\tau
+
\int_{\tau=a(u)}^x
\rho(u,x,\tau)\widehat{\Phi}_u(u,\tau)\,\mathrm{d}\tau .
\end{eqnarray}

Here $a'(u)=0$ for $u>0$ and $a'(u)=-1$ for $u<0$, while the line
$u=0$ is one of the characteristic lines excluded from the regularity
claim. Since $\Phi(t,x)=\widehat{\Phi}(t-x,x)$, the chain rule gives the piecewise differentiability of $\Phi(t,x)$ with respect to $t$ and $x$.
}

\section{Proof of Theorem \ref{thm:The-LST-of}}

\label{subsec:Solution-to-subsystems-1}

Similar to our approach for the time-varying system, we solve the
stationary system by analyzing its subsystems. We begin by constructing
the Subsystem V, by setting $y=\infty$ in Equation (\ref{eq:3-0-2}).
We define $D(x,z)\triangleq H(x,\infty,z)$ and $M(x)\triangleq H(x,0,0)$,
which leads to the following formulation of Subsystem V: 
\begin{align}
\mbox{Subsystem V: } & \lambda\theta D(x,\infty)F(z)+\lambda(1-\theta)M(x)F(z)-\lambda\theta D(x,z)\nonumber \\
 & +\lambda\theta M(x)-D_{x}(x,z)+D_{z}(x,z)+D_{x}(x,0)-D_{z}(x,0)=0.\label{eq:3-0-4}
\end{align}
We observe that Subsystem V is not self-contained, since the $M(x)$
function remains undetermined. To solve for the marginal AoI distribution
$\Phi(x)=H(x,\infty,\infty)$, it is necessary to first characterize
$M(x)$. Recall that the evolution of $M(x)$ is captured by Equation
(\ref{eq:3-0-1}). Its solution can be expressed as 
\begin{equation}
M(x)=e^{-\lambda x}\Big(\int_{s=0}^{x}G_{z}(s,0)e^{\lambda s}\mathrm{d}s\Big),\label{eq:3-0-5}
\end{equation}
where $G(y,z)\triangleq H(\infty,y,z)$. To obtain $M(x)$, we must
first derive the function $G_{z}(x,0)$. To this end, we set $x=\infty$
in Equation (\ref{eq:3-0-2}), which yields Subsystem VI: 
\begin{align}
\mbox{Subsystem VI: } & \lambda\theta F(z)+\lambda(1-\theta)M(\infty)F(z)-\lambda\theta G(y,z)\nonumber \\
 & +\lambda\theta M(\infty)-G_{y}(y,z)+G_{z}(y,z)-G_{z}(y,0)=0.\label{eq:3-0-6}
\end{align}
Using the fact that $\lambda M(\infty)=G_{z}(\infty,0)$ (obtained
from setting $x=\infty$ in Equation (\ref{eq:3-0-1})), we find that
Subsystem VI in Equation (\ref{eq:3-0-6}) is self-contained. Therefore,
our strategy is to first solve the Subsystem VI to obtain $G_{z}(y,0)$
and $M(x)$, and then solve the Subsystem V to obtain the AoI distribution
$\Phi(x)=D(x,\infty)$.

\subsection{Solution to Subsystem VI}

To solve Subsystem VI, we use the method of characteristics by changing
the variables to $y'=y+z$ and $z'=y$. By letting $v_{7}(y,z)=\lambda\theta F(z)+\lambda\big((1-\theta)F(z)+\theta\big)M(\infty)-G_{z}(y,0)$,
the solution to Equation (\ref{eq:3-0-6}) is given as 
\begin{eqnarray}
G(y,z) & = & e^{-\lambda\theta z'}\Big(\int_{s=0}^{z'}v_{7}(s,y'-s)e^{\lambda\theta s}\mathrm{d}s+g_{7}(y')\Big)\nonumber \\
 & = & e^{-\lambda\theta y}\Big(\int_{s=0}^{y}v_{7}(s,y+z-s)e^{\lambda\theta s}\mathrm{d}s+g_{7}(y+z)\Big),\label{eq:3-1-1}
\end{eqnarray}
where $g_{7}(\cdot)$ is an arbitrary real-valued function. From the
boundary conditions that $G(y,0)=G(0,z)=M(\infty)$, we obtain $g_{7}(z)=M(\infty).$
This implies that $g_{7}(\cdot)$ is a constant function. Consequently,
we have 
\begin{equation}
G(y,0)=e^{-\lambda\theta y}\Big(\int_{s=0}^{y}v_{7}(s,y-s)e^{\lambda\theta s}\mathrm{d}s+M(\infty)\Big)=M(\infty).\label{eq:3-1-2}
\end{equation}
Reorganizing Equation (\ref{eq:3-1-2}), we have 
\begin{eqnarray}
 &  & M(\infty)(e^{\lambda\theta y}-1)=\int_{s=0}^{y}v_{7}(s,y-s)e^{\lambda\theta s}\mathrm{d}s\nonumber \\
 & = & \int_{s=0}^{y}\Big\{\lambda\theta F(y-s)+\lambda\big((1-\theta)F(y-s)+\theta\big)M(\infty)-G_{z}(s,0)\Big\} e^{\lambda\theta s}\mathrm{d}s,\label{eq:3-1-3}
\end{eqnarray}
from which we obtain 
\begin{equation}
\int_{s=0}^{y}G_{z}(s,0)e^{\lambda\theta s}\mathrm{d}s=\int_{s=0}^{y}\Big(\lambda\theta+\lambda(1-\theta)M(\infty)\Big)F(y-s)e^{\lambda\theta s}\mathrm{d}s.\label{eq:3-1-4}
\end{equation}
To further solve the integral equation of (\ref{eq:3-1-4}), we take
the derivative with respect to $y$ to solve for the function explicitly:
\begin{equation}
G_{z}(y,0)e^{\lambda\theta y}=\lambda\Big(\theta+(1-\theta)M(\infty)\Big)\Big(\int_{s=0}^{y}e^{\lambda\theta s}\mathrm{d}F(y-s)\Big).\label{eq:3-1-5}
\end{equation}
Reorganizing Equation (\ref{eq:3-1-5}), we simplify the result as
\begin{align}
G_{z}(y,0) & =\lambda\Big(\theta+(1-\theta)M(\infty)\Big)\Big(\int_{s=0}^{y}e^{-\lambda\theta(y-s)}\mathrm{d}F(y-s)\Big)\nonumber \\
 & =\lambda\Big(\theta+(1-\theta)M(\infty)\Big)\Big(\int_{s=0}^{y}e^{-\lambda\theta s}\mathrm{d}F(s)\Big).\label{eq:3-1-6}
\end{align}
Note that the constant $M(\infty)$ in Equation (\ref{eq:3-1-6})
still remains unknown. To solve for $M(\infty)$, we set $y=\infty$
in Equation (\ref{eq:3-1-6}) and obtain 
\begin{equation}
G_{z}(\infty,0)=\lambda\Big(\theta+(1-\theta)M(\infty)\Big)\tilde{F}(\lambda\theta).\label{eq:3-1-7}
\end{equation}
Setting $x=\infty$ in Equation (\ref{eq:3-0-1}), we obtain $G_{z}(\infty,0)=\lambda M(\infty)$.
Combining this condition with Equation (\ref{eq:3-1-7}) yields the
following result: 
\begin{equation}
M(\infty)=\frac{\theta\tilde{F}(\lambda\theta)}{1-(1-\theta)\tilde{F}(\lambda\theta)}.\label{eq:3-1-8}
\end{equation}

We have thus far solved Subsystem VI and determined the closed-form
expression for $G_{z}(y,0)$. We substitute Equation (\ref{eq:3-1-6})
into Equation (\ref{eq:3-0-5}), and derive the expression of the
function $M(x)$ as: 
\begin{align}
M(x) & =e^{-\lambda x}\int_{s=0}^{x}G_{z}(s,0)e^{\lambda s}\mathrm{d}s\nonumber \\
 & =\lambda\Big(\theta+(1-\theta)M(\infty)\Big)e^{-\lambda x}\int_{y=0}^{x}\left[\int_{s=0}^{y}e^{-\lambda\theta s}\mathrm{d}F(s)\right]e^{\lambda y}\mathrm{d}y\nonumber \\
 & =\lambda\Big(\theta+(1-\theta)M(\infty)\Big)e^{-\lambda x}\int_{s=0}^{x}\left[\int_{y=s}^{x}e^{\lambda y}\mathrm{d}y\right]e^{-\lambda\theta s}\mathrm{d}F(s)\nonumber \\
 & =\Big(\theta+(1-\theta)M(\infty)\Big)\int_{s=0}^{x}\left(e^{-\lambda\theta s}-e^{\lambda(s-\theta s-x)}\right)\mathrm{d}F(s).\label{eq:3-1-9}
\end{align}
We will then substitute $M(x)$ into Equation (\ref{eq:3-0-4}) and
solve Subsystem V.

\subsection{Solution to Subsystem V}

We now proceed to solve Subsystem V. To simplify the analysis, we
set $z=\infty$ in Equation (\ref{eq:3-0-4}), which yields: 
\begin{equation}
\lambda M(x)-D_{x}(x,\infty)+D_{x}(x,0)-D_{z}(x,0)=0.\label{eq:3-2-1}
\end{equation}
We substitute Equation (\ref{eq:3-2-1}) into Equation (\ref{eq:3-0-4}),
and simplify Subsystem V as 
\begin{align}
 & \lambda\theta\Phi(x)F(z)+\Phi_{x}(x)+\lambda(1-\theta)M(x)\big(F(z)-1\big)\nonumber \\
 & -\lambda\theta D(x,z)-D_{x}(x,z)+D_{z}(x,z)=0.\label{eq:3-2-2}
\end{align}
To solve Equation (\ref{eq:3-2-2}), we again change the coordinates
by letting $x'=x+z$ and $z'=x$. By letting $v_{8}(x,z)=\lambda\theta\Phi(x)F(z)+\Phi_{x}(x)+\lambda(1-\theta)M(x)\big(F(z)-1\big)$,
the solution to differential equation (\ref{eq:3-2-2}) can be easily
derived as follows: 
\begin{align}
D(x,z) & =e^{-\lambda\theta z'}\Big(\int_{s=0}^{z'}v_{8}(s,x'-s)e^{\lambda\theta s}\mathrm{d}s+g_{8}(x')\Big)\nonumber \\
 & =e^{-\lambda\theta x}\Big(\int_{s=0}^{x}v_{8}(s,x+z-s)e^{\lambda\theta s}\mathrm{d}s+g_{8}(x+z)\Big),\label{eq:3-2-3}
\end{align}
where $g_{8}(\cdot)$ is a real-valued function. From the boundary
conditions that $D(0,z)=0$ and $D(x,0)=M(x)=H(x,\infty,0)$, we have
$g_{8}(z)=0$ and 
\begin{eqnarray}
M(x) & = & e^{-\lambda\theta x}\int_{s=0}^{x}\Big(\lambda\theta\Phi(s)F(x-s)+\lambda(1-\theta)M(s)(F(x-s)-1)+\Phi_{x}(s)\Big)e^{\lambda\theta s}\mathrm{d}s\nonumber \\
 & = & \Phi(x)+\int_{s=0}^{x}\Phi(s)\lambda\theta\left(F(x-s)-1\right)e^{-\lambda\theta(x-s)}\mathrm{d}s\nonumber \\
 &  & +\int_{s=0}^{x}\lambda(1-\theta)M(s)\left(F(x-s)-1\right)e^{-\lambda\theta(x-s)}\mathrm{d}s.\label{eq:3-2-4}
\end{eqnarray}
By reorganizing Equation (\ref{eq:3-2-4}), we have 
\begin{equation}
\Phi(x)=M(x)+\theta\int_{s=0}^{x}\Phi(s)K(x-s)\mathrm{d}s+(1-\theta)\int_{s=0}^{x}M(s)K(x-s)\mathrm{d}s,\label{eq:3-2-5}
\end{equation}
where $K(x)=\lambda\left(1-F(x)\right)e^{-\lambda\theta x}$. We observe
that Equation (\ref{eq:3-2-5}) is a Volterra integral equation of
the second kind \cite{polyanin2008handbook} when $\theta>0$. Note
that we can also apply a similar argument in Appendix \ref{subsec:Existence-of-derivatives} to show that the derivative
of $\Phi(x)$ exists except for a finite number of points. 

We now apply Laplace Transform to both sides of Equation (\ref{eq:3-2-5})
and use Equation (\ref{eq:3-1-9}) for $M(x)$, which yields the following:
\begin{equation}
\Phi^{*}(s)\triangleq\int_{x=0}^{\infty}e^{-sx}\Phi(x)dx=M^{*}(s)+\theta\Phi^{*}(s)K^{*}(s)+(1-\theta)M^{*}(s)K^{*}(s),\label{eq:3-2-6}
\end{equation}
where 
\begin{align}
M^{*}(s) & =\int_{x=0}^{\infty}e^{-sx}M(x)dx=\frac{\lambda}{(\lambda+s)s}\frac{\theta}{1-(1-\theta)\tilde{F}(\lambda\theta)}\tilde{F}(\lambda\theta+s)\label{eq:3-2-7}
\end{align}
and 
\begin{align}
K^{*}(s) & =\int_{x=0}^{\infty}e^{-sx}K(x)dx=\frac{\lambda}{\lambda\theta+s}\left(1-\tilde{F}(\lambda\theta+s)\right),\label{eq:3-2-8}
\end{align}
where $\Phi^{*}(s)$, $M^{*}(s)$ and $K^{*}(s)$ are the Laplace
Transforms of $\Phi(x)$, $M(x)$ and $K(x)$. Solving Equation (\ref{eq:3-2-6}),
we have 
\begin{equation}
\Phi^{*}(s)=M^{*}(s)\frac{1+(1-\theta)K^{*}(s)}{1-\theta K^{*}(s)}.\label{eq:3-2-10}
\end{equation}
The LST of the AoI distribution is thus given as 
\begin{align}
\tilde{\Phi}(s) & \triangleq\int_{x=0}^{\infty}e^{-sx}\Phi_{x}(x)\mathrm{d}x\nonumber \\
 & =\Phi(x)e^{-sx}\big|_{x=0}^{\infty}+s\int_{x=0}^{\infty}e^{-sx}\Phi(x)\mathrm{d}x\nonumber \\
 & =s\cdot M^{*}(s)\frac{1+(1-\theta)K^{*}(s)}{1-\theta K^{*}(s)}.\label{eq:3-2-11}
\end{align}
We therefore have derived the LST of the AoI for the system with probabilistic
preemption, as summarized in the first part of Theorem \ref{thm:The-LST-of}.

For the system with no preemption ($\theta=0$), Equation (\ref{eq:3-2-7})
simplifies to: 
\begin{equation}
\Phi(x)=M(x)+\int_{s=0}^{x}M(s)K(x-s)\mathrm{d}s.\label{eq:3-3-1}
\end{equation}
$M(x)$ for this case is found by setting $\theta=0$ in Equation
(\ref{eq:3-1-9}) and performing an integration by parts: 
\begin{eqnarray}
M(x) & = & M(\infty)\int_{s=0}^{x}(1-e^{\lambda(s-x)})\mathrm{d}F(s)\nonumber \\
 & = & M(\infty)\lambda\int_{s=0}^{x}F(s)e^{-\lambda(x-s)}\mathrm{d}s.\label{eq:3-3-2}
\end{eqnarray}
The value of $M(\infty)$ is obtained by applying the L'Hospital rule
to Equation (\ref{eq:3-1-8}), yielding: 
\begin{eqnarray}
M(\infty) & = & \lim_{\theta\to0}\frac{\theta\tilde{F}(\lambda\theta)}{1-(1-\theta)\tilde{F}(\lambda\theta)}\nonumber \\
 & = & \lim_{\theta\to0}\frac{\tilde{F}(\lambda\theta)+\lambda\theta\tilde{F}^{\prime}(\lambda\theta)}{\tilde{F}(\lambda\theta)-\lambda(1-\theta)\tilde{F}^{\prime}(\lambda\theta)}\nonumber \\
 & = & \frac{1}{1-\lambda\tilde{F}^{\prime}(0)}.\label{eq:3-3-4}
\end{eqnarray}
This completes the derivation of the AoI distribution for the system
without preemption, as summarized in the second part of Theorem \ref{thm:The-LST-of}.

\section{Proof of Corollary \ref{cor:When-the-processing}}\label{sec:Proof-of-Corollary}
\begin{IEEEproof}
When $\lambda\neq\mu$ and $\mu\neq\lambda(1-\theta)$, the AoI distribution
can be derived by substituting the properties of the exponential distribution
$\tilde{F}(s)=\mu/(s+\mu)$ into the formulas in Theorem 2. We then
have 
\begin{equation}
\widetilde{\Phi}(s)=\frac{\lambda\mu(\mu+\lambda\theta)(s+\mu+\lambda)}{(\lambda+s)(\mu+s)(\mu+\lambda)(\mu+\lambda\theta+s)}.
\end{equation}
We can write the partial fraction expansion as 
\begin{equation}
\widetilde{\Phi}(s)=\frac{a}{s+\lambda}+\frac{b}{s+\mu}+\frac{c}{s+\mu+\lambda\theta}.\label{eq:pf_form}
\end{equation}
The coefficients are residues: 
\begin{align}
a & =\lim_{s\to-\lambda}(s+\lambda)\widetilde{\Phi}(s)=-\frac{\lambda\mu^{2}(\mu+\lambda\theta)}{(\lambda-\mu)(\lambda+\mu)\bigl(\mu-\lambda(1-\theta)\bigr)},\label{eq:res_a}\\
b & =\lim_{s\to-\mu}(s+\mu)\widetilde{\Phi}(s)=\frac{\lambda\mu(\mu+\lambda\theta)}{\theta(\lambda-\mu)(\lambda+\mu)},\label{eq:res_b}\\
c & =\lim_{s\to-(\mu+\lambda\theta)}(s+\mu+\lambda\theta)\widetilde{\Phi}(s)=\frac{\lambda\mu(\theta-1)(\mu+\lambda\theta)}{\theta(\lambda+\mu)\bigl(\mu-\lambda(1-\theta)\bigr)}.\label{eq:res_c}
\end{align}
Taking the inverse LST, we can obtain Equation (\ref{eq:5-2-1}):
\begin{align}
\Phi(x) & =1+\frac{\mu^{2}(\mu+\lambda\theta)}{(\lambda-\mu)(\lambda+\mu)\bigl(\mu-\lambda(1-\theta)\bigr)}e^{-\lambda x}\nonumber \\
 & -\frac{\lambda(\mu+\lambda\theta)}{\theta(\lambda-\mu)(\lambda+\mu)}e^{-\mu x}-\frac{\lambda\mu(1-\theta)}{\theta(\lambda+\mu)\bigl(\mu-\lambda(1-\theta)\bigr)}e^{-(\mu+\lambda\theta)x}.
\end{align}
When $\theta=0$, the result from Theorem \ref{thm:The-LST-of} is consistent with the findings in \cite{champati2019distribution},
where a sample path analysis was used to derive the same expression.
When $\lambda=\mu,$ there are repeated poles. We can thus write the
LST as 
\begin{eqnarray}
\widetilde{\Phi}(s) & = & \frac{a}{(s+\mu)}+\frac{b}{(s+\mu)^{2}}+\frac{c}{s+\mu+\mu\theta}.
\end{eqnarray}
Using the technique above, we can obtain that 
\begin{eqnarray}
\Phi(x) & = & \frac{\mu(1+\theta)}{2\theta}\Big(\frac{1}{\mu}-(x+\frac{1}{\mu})e^{-\mu x}\Big)+\frac{(1+\theta)(1-\theta)}{2\theta^{2}}\Big(e^{-\mu x}-1\Big)\nonumber \\
 &  & +\frac{1-\theta}{2\theta^{2}}\Big(1-e^{-\mu(1+\theta)x}\Big).
\end{eqnarray}
When $\mu=\lambda(1-\theta),$ there are also repeated poles. Similarly,
we write the LST as 
\begin{eqnarray}
\widetilde{\Phi}(s) & = & \frac{a}{(s+\lambda)}+\frac{b}{(s+\lambda)^{2}}+\frac{c}{s+\lambda(1-\theta)}.
\end{eqnarray}
Using the technique above, we can obtain that 
\begin{eqnarray}
\Phi(x) & = & \frac{\lambda(1-\theta)^{2}}{\theta(2-\theta)}xe^{-\lambda x}+\frac{(1-\theta)(1+\theta-\theta^{2})}{\theta^{2}(2-\theta)}\Big(e^{-\lambda x}-1\Big)\nonumber \\
 &  & +\frac{1}{\theta^{2}(2-\theta)}\Big(1-e^{-\lambda(1-\theta)x}\Big).
\end{eqnarray}
We next prove that the preemption with probability $\theta=1$ is optimal with $\lambda\neq\mu$ and $\mu\neq\lambda(1-\theta)$.
Note that we have
\begin{eqnarray}
 &  & \lim_{t\to\infty}\boldsymbol{P}(\Delta_{1}(t)\leq x)-\lim_{t\to\infty}\boldsymbol{P}(\Delta_{\theta}(t)\leq x)\nonumber \\
 & = & \frac{\lambda^{2}\mu(1-\theta)}{\lambda+\mu}e^{-(\mu+\lambda\theta)x}\frac{\dfrac{e^{\lambda\theta x}-1}{\lambda\theta}-\dfrac{e^{(\mu-\lambda+\lambda\theta)x}-1}{\mu-\lambda+\lambda\theta}}{\lambda-\mu}\nonumber \\
 & = & \frac{\lambda^{2}\mu(1-\theta)}{\lambda+\mu}e^{-(\mu+\lambda\theta)x}\frac{\int_{0}^{x}e^{\lambda\theta s}ds-\int_{0}^{x}e^{(\mu-\lambda+\lambda\theta)s}ds}{\lambda-\mu}\nonumber \\
 & = & \frac{\lambda^{2}\mu(1-\theta)}{\lambda+\mu}e^{-\mu x}J(x;\theta),
\end{eqnarray}
where $J(x;\theta)=e^{-\lambda\theta x}\frac{\int_{0}^{x}e^{\lambda\theta s}ds-\int_{0}^{x}e^{(\mu-\lambda+\lambda\theta)s}ds}{\lambda-\mu}.$
From the fact that 
\begin{align}
\frac{\partial J(x;\theta)}{\partial\theta} & =-\lambda xe^{-\lambda\theta x}\frac{\int_{0}^{x}e^{\lambda\theta s}ds-\int_{0}^{x}e^{(\mu-\lambda+\lambda\theta)s}ds}{\lambda-\mu}+e^{-\lambda\theta x}\frac{\int_{0}^{x}\lambda se^{\lambda\theta s}ds-\int_{0}^{x}\lambda se^{(\mu-\lambda+\lambda\theta)s}ds}{\lambda-\mu}\nonumber \\
 & =e^{-\lambda\theta x}\lambda\frac{\int_{0}^{x}e^{\lambda\theta s}(s-x)ds-\int_{0}^{x}e^{(\mu-\lambda+\lambda\theta)s}(s-x)ds}{\lambda-\mu}\leq0,
\end{align}
we have $J(x;\theta)\geq J(x;1)=e^{-\lambda x}\frac{\int_{0}^{x}e^{\lambda s}ds-\int_{0}^{x}e^{\mu s}ds}{\lambda-\mu}\geq0.$
The discussions on the cases of $\lambda=\mu$ and $\mu=\lambda(1-\theta)$ are similar, so we omit the details here.
\end{IEEEproof}

\section{Proof of Corollary \ref{cor:MD11}}\label{sec:proof-of-cor2}
\begin{proof} For a deterministic processing time of $1/\mu$, we
have $-\tilde{F}'(0)=1/\mu$, which gives $M(\infty)=\frac{\mu}{\lambda+\mu}$
and 
\begin{eqnarray}
M(x)=\begin{cases}
0, & \mbox{if }x\leq\frac{1}{\mu};\\
\frac{\mu}{\lambda+\mu}(1-e^{-\lambda x+\frac{\lambda}{\mu}}), & \mbox{if }x>\frac{1}{\mu}.
\end{cases}
\end{eqnarray}
We then evaluate the convolution integral $\Phi(x)=M(x)+\lambda\int_{0}^{x}M(s)(1-F(x-s))\mathrm{d}s$.
If $x<\frac{1}{\mu}$, then $M(x)=0$ and the integral is zero, so
$\Phi(x)=0$; If $\frac{1}{\mu}\leq x<\frac{2}{\mu}$, the integral
becomes 
\begin{equation}
\Phi(x)=M(x)+\lambda\int_{1/\mu}^{x}M(s)\mathrm{d}s=\frac{\lambda\mu}{\lambda+\mu}x-\frac{\lambda}{\lambda+\mu};
\end{equation}
If $x\geq\frac{2}{\mu}$, then 
\begin{equation}
\Phi(x)=M(x)+\lambda\int_{x-1/\mu}^{x}M(s)\mathrm{d}s=1-\frac{\mu}{\lambda+\mu}e^{-\lambda x+2\frac{\lambda}{\mu}}.
\end{equation}
We thus derived the results of this corollary. \end{proof}

\section{Pseudocode for the heuristic approach}\label{sec:pseudocode}
Algorithm \ref{alg:Heuristic-approach-for} provides the pseudocode of the heuristic approach for selecting the sampling and preemption rates in Section \ref{subsec:Optimal-design-for}.
\begin{algorithm}[h!]
\begin{algorithmic}[1] \State{Initialize: time
points $\boldsymbol{t}=(t_{0},...,t_{n})$, AoI threshold $\boldsymbol{x}=(x_{0},...,x_{n-1})$,
probability thresholds $\boldsymbol{p}=(p_{0},...,p_{n-1})$, evaluation
grid $\boldsymbol{\eta}=(\eta_{1},...,\eta_{m})$, maximum number
of iterations $ite_{max}$, preemption probability grid $\boldsymbol{\theta}_{grid}$, and rate increment $\delta>0$.
} \State{Construct the sub-interval time points $\boldsymbol{\tilde{t}}=(t_{0},t_{1}-x_{1},t_{1},...,t_{n-1}-x_{n-1},t_{n-1},t_{n})$.
Set $ite\gets0$ and $violation\leftarrow True$. } \While{$violation==True$
and $ite<ite_{max}$} \State{$violation\gets False$} \If{$ite==0$}
\Comment{Initial pass: search over $\theta$ and $\lambda$ per
sub-interval} \For{$i=0,...,n-2$} \State{Perform a grid search
over $\boldsymbol{\theta}_{grid}$ to find the minimum $\lambda_{2i}$
and corresponding $\theta_{2i}$ such that $\Phi(x_{i};\lambda_{2i},\theta_{2i})\geq p_{i}$
using stationary approximation} \State{Perform a grid search over
$\boldsymbol{\theta}_{grid}$ to find the minimum $\lambda_{2i+1}$
and corresponding $\theta_{2i+1}$ such that $\Phi(x_{i};\lambda_{2i+1},\theta_{2i+1})\geq p_{i}$
and $\Phi(x_{i+1};\lambda_{2i+1},\theta_{2i+1})\geq p_{i+1}$ using
stationary approximation} \EndFor \State{Perform a grid search
over $\boldsymbol{\theta}_{grid}$ to find the minimum $\lambda_{2n-2}$
and corresponding $\theta_{2n-2}$ such that $\Phi(x_{n-1};\lambda_{2n-2},\theta_{2n-2})\geq p_{n-1}$
using stationary approximation} \EndIf \For{$i=1,...,m$} \State{Let
$k$ be the index such that $\eta_{i}\in[t_{k},t_{k+1})$} \State{Evaluate
$\Phi(\eta_{i},x_{k})$ at $\eta_{i}$ using the piecewise constant
sampling rate $\lambda(t)$ and preemption probability $\theta(t)$
defined over $[t_{0},t_{n})$, where $\lambda(t)=\lambda_{j}$ and
$\theta(t)=\theta_{j}$ for $t\in[\tilde{t}_{j},\tilde{t}_{j+1})$}
\If{$\Phi(\eta_{i},x_{k})<p_{k}$} \State{Find slot $s$ such
that $\eta_{i}\in[\tilde{t}_{s},\tilde{t}_{s+1})$} \State{Increase
$\lambda_{s}\gets\lambda_{s}+\delta$, $violation\leftarrow True$}
\EndIf \EndFor \If{$violation==False$} \State{\textbf{break}}
\EndIf \State{$ite\leftarrow ite+1$} \EndWhile \If{$violation==True$}
\State{\textbf{return} No feasible solution found} \Else \State{\textbf{return}
Piecewise constant functions $\lambda(t)$ and $\theta(t)$ with values
$(\lambda_{0},...,\lambda_{2n-2})$ and $(\theta_{0},...,\theta_{2n-2})$
over intervals defined by $\boldsymbol{\tilde{t}}$} \EndIf  \end{algorithmic}
\caption{Heuristic approach for system parameter design}
\label{alg:Heuristic-approach-for}
\end{algorithm}

\end{document}